\documentclass[twocolumn,
 amsmath,amssymb,
 aps,
prb,
]{revtex4-2}
\usepackage{caption}
\usepackage{bm}
\usepackage[normalem]{ulem}
\usepackage{subcaption}
\usepackage[colorinlistoftodos, color=green!40, prependcaption]{todonotes}

\usepackage{amsthm}
\usepackage{mathtools}
\usepackage{dsfont}
\usepackage{physics}
\usepackage{xcolor}
\usepackage{graphicx}
\usepackage[left=23mm,right=13mm,top=35mm,columnsep=15pt]{geometry} 
\usepackage{adjustbox}
\usepackage{placeins}
\usepackage[T1]{fontenc}
\usepackage{lipsum}
\usepackage{csquotes}
\usepackage{stmaryrd}
\usepackage{tikz-3dplot}

\usepackage[pdftex, pdftitle={Article}, pdfauthor={Author}]{hyperref} % For hyperlinks in the PDF
\usepackage{hyperref}
\newcommand{\Harvard}{Department of Physics, Harvard University, Cambridge, Massachusetts 02138, USA}

\begin{document}
\title{Understanding Stabilizer Codes Under Local Decoherence Through A General Statistical Mechanics Mapping}

\author{Anasuya Lyons}
\affiliation{\Harvard}

%\author{Authors}

\date{\today} % Leave empty to omit a date

%TC:ignore
\begin{abstract}
We consider the problem of a generic stabilizer Hamiltonian under local, incoherent Pauli errors. Using two different approaches--- (i) Haah's polynomial formalism \cite{haahCommutingPauliHamiltonians2013} and (ii) the CSS-to-homology correspondence--- we construct a mapping from the $n$th moment of the decohered ground state density matrix to a classical statistical mechanics model. We demonstrate that various measures of information capacity-- (i) quantum relative entropy, (ii) coherent information, and (iii) entanglement negativity--- map to thermodynamic quantities in the statistical mechanics model and can be used to characterize the decoding phase transition. As examples, we analyze the 3D toric code and X-cube model, deriving bounds on their optimal decoding thresholds and gaining insight into their information properties under decoherence. Additionally, we demonstrate that the SM mapping acts an an ``ungauging'' map; the classical models that describe a given code under decoherence also can be gauged to obtain the same code. Finally, we comment on correlated errors and non-CSS stabilizer codes.
\end{abstract}

%\keywords{first keyword, second keyword, third keyword}

\maketitle
%TC:endignore

\tableofcontents

\section{Introduction}
% Notes: expand the intro with more details about the results; in particular the last paragraph. end by giving a summary of the organization of the paper

% results to highlight: able to construct order parameters for stat mech models that weren't used before

% explain polynomial formalism and homological formalism--- add introductory appendix about both 

% add table summarizing how different information measures look on either side of the phase transition

Over the past 20 years, there has been a growing understanding of the deep connections between condensed matter physics and quantum information. Information-theoretic tools have become invaluable for analyzing quantum matter; concepts like entanglement and topological order have become foundational to our understanding of exotic quantum phases. In the other direction, a wealth of condensed matter systems have emerged as promising platforms for quantum computing. For example, Abelian topological phases can act as robust quantum memories \cite{Kitaev_2003, Dennis_2002}, while non-Abelian phases can host anyons suitable for quantum computation \cite{Mochon_2003, Das_Sarma_2005}. The notion of a phase of matter and a quantum code have become deeply intertwined. 

In the era of noisy-intermediate scale quantum (NISQ) devices, any quantum computer or quantum simulator will have to contend with noise. Recently, experimental advances have meant that many topological codes can be realized in quantum simulators \cite{satzingerRealizingTopologicallyOrdered2021, Semeghini_2021, andersen2023nonabelian, iqbal2023topological, Iqbal_2024}; an understanding of how errors affect these phases is crucial for analyzing and pushing these experiments forwards. On the fundamental side, we want to better understand novel-decoherence induced phenomena in topological quantum phases.

An important class of quantum codes are stabilizer codes \cite{gottesman1997stabilizer}; they are exactly solvable, classically-simulable models, which nonetheless describe a wide array of exotic quantum phases. For instance, many Abelian topological phases \cite{Bomb_n_2014, haah_2021,Ellison_2022} and fracton models \cite{chamonQuantumGlassiness2005, Haah_2011, Yoshida_2013, vijayNewKindTopological2015, vijayFractonTopologicalOrder2016} have fixed-point stabilizer descriptions. Stabilizer codes are also ubiquitous in the quantum computing world; much attention has been paid to quantum low-density parity check (qLDPC) codes, as they have provided the first examples of truly ``good'' quantum codes \cite{panteleevAsymptoticallyGoodQuantum2022, panteleevQuantumLDPCCodes2022}. Since they are so widespread and are very tractable to study analytically, stabilizer models are a good starting point for investigating mixed state topological order. 

In this work, we construct a general toolkit for analyzing stabilizer codes under local Pauli decoherence by mapping information quantities to thermodynamic ones in a classical statistical mechanics model. The information transition in the quantum code can be understood via comparison with the classical finite-temperature transition (see Table \ref{tab:info-SM-mapping} for a summary of of the mapping). Statistical mechanics mappings have been widely used for analyzing the error-correcting properties of stabilizer codes \cite{Dennis_2002, wangConfinementHiggsTransitionDisordered2003, Bombin_2012, kovalevNumericalAnalyticalBounds2018, chubbStatisticalMechanicalModels2021, chen2023separability}. These works have used the statistical mechanics models to understand how well particular decoders can withstand errors. Recently, an SM mapping was demonstrated for the decohered density matrix itself in the specific case of the 2D toric code \cite{fanDiagnosticsMixedstateTopological2023}. We construct a generic version of this mapping, applicable to any Pauli stabilizer code. We use this mapping to analyze the 3D toric code and the X-Cube model under decoherence in detail--- we are able to extract coding thresholds and calculate information quantities on either side of the transition. Additionally, we are able to use our knowledge of these stabilizer codes to construct novel order parameters for their corresponding classical models, demonstrating our methods are useful for better understanding classical statistical mechanics models as well as quantum codes. 

We construct the general mapping using Haah's polynomial formalism for translation-invariant stabilizer codes \cite{haahCommutingPauliHamiltonians2013}. The basic idea of the formalism to to express the geometry of stabilizers in the form of Laurent polynomials, which can then be analyzed using the tools of commutative algebra. The form of interactions in the statistical mechanics model can be related to the \emph{excitation map}, which gives a systematic way to determine the syndrome for a given error pattern. We show additionally that an equivalent statistical mechanics mapping can be constructed using the CSS-to-homology correspondence. Any CSS code can be viewed as a chain complex of length three \cite{kitaevQuantumComputationsAlgorithms1997a, Bombin_2007, bravyiHomologicalProductCodes2013}; the boundary maps belonging to the chain complex describe the statistical mechanics models for $X$ and $Z$ errors.

The rest of the paper is organized as follows: in section \ref{sec:warm-up}, we use the 3D toric code as a playground to explain the set-up of the problem and give an explicit example of the statistical-mechanics mapping first introduced in \cite{fanDiagnosticsMixedstateTopological2023}. Then, in section \ref{sec:generic-mapping}, we construct the mapping for a generic CSS stabilizer code. Section \ref{sec:info-quantities} details the mapping for the Réyni relative entropy, coherent information, and entanglement negativity. In section \ref{sec:ungauging}, we give a short proof that the general SM mapping presented can be viewed as an ungauging procedure, and that, for CSS codes, the SM models that describe bit-flip and phase errors will be Kramers-Wannier dual. We present an alternate method of deriving the SM models in \ref{sec:homology}, using the homological description of CSS codes; in \ref{sec:replica} we consider taking the replica limit $n \rightarrow 1$ and show the connection between our general mapping and the SM models defined by the Kitaev-Preskill decoder \cite{Dennis_2002}. 

In Section \ref{sec:examples}, we apply the general results derived in section \ref{sec:generic-mapping} to the 3D toric code and the X-cube model. We derive their corresponding SM models, obtain bounds on their decoding thresholds, and analyze in detail the behavior of the various information measures on either side of the decoding transition. The statistical mechanics models for the 3D toric code are well studied, the ones for the X-cube model are less so: using knowledge of the X-cube model, we are able to construct novel order parameters for the ferromagnetic-to-paramagnetic transition. 

Sections \ref{sec:corr-errs} and \ref{sec:non-CSS} apply our formalism to correlated errors and non-CSS codes. We discuss several examples of correlated error channels in the 2D toric code, including single-site $Y$ errors and $\psi$-errors (which create fermions). We comment on the generalization of our mapping to non-CSS codes, deriving the SM model for the CBLT code as an example. We conclude with a discussion of the results and possible future directions in section \ref{sec:discussion}.

\section{Warm-Up: 3D Toric Code under Local Decoherence}
\label{sec:warm-up}

We begin by reviewing the methods developed in \cite{fanDiagnosticsMixedstateTopological2023} by analyzing the 3D toric code \cite{Dennis_2002, Hamma_2005, Castelnovo_2008, michnicki3DTopologicalQuantum2014} under local decoherence. We aim to introduce the motivations and conceptual basis of the paper in a concrete way through this example, so that the general mapping explained in the next section is maximally clear.

\begin{figure}
    %\centering
    \includegraphics[width=\linewidth]{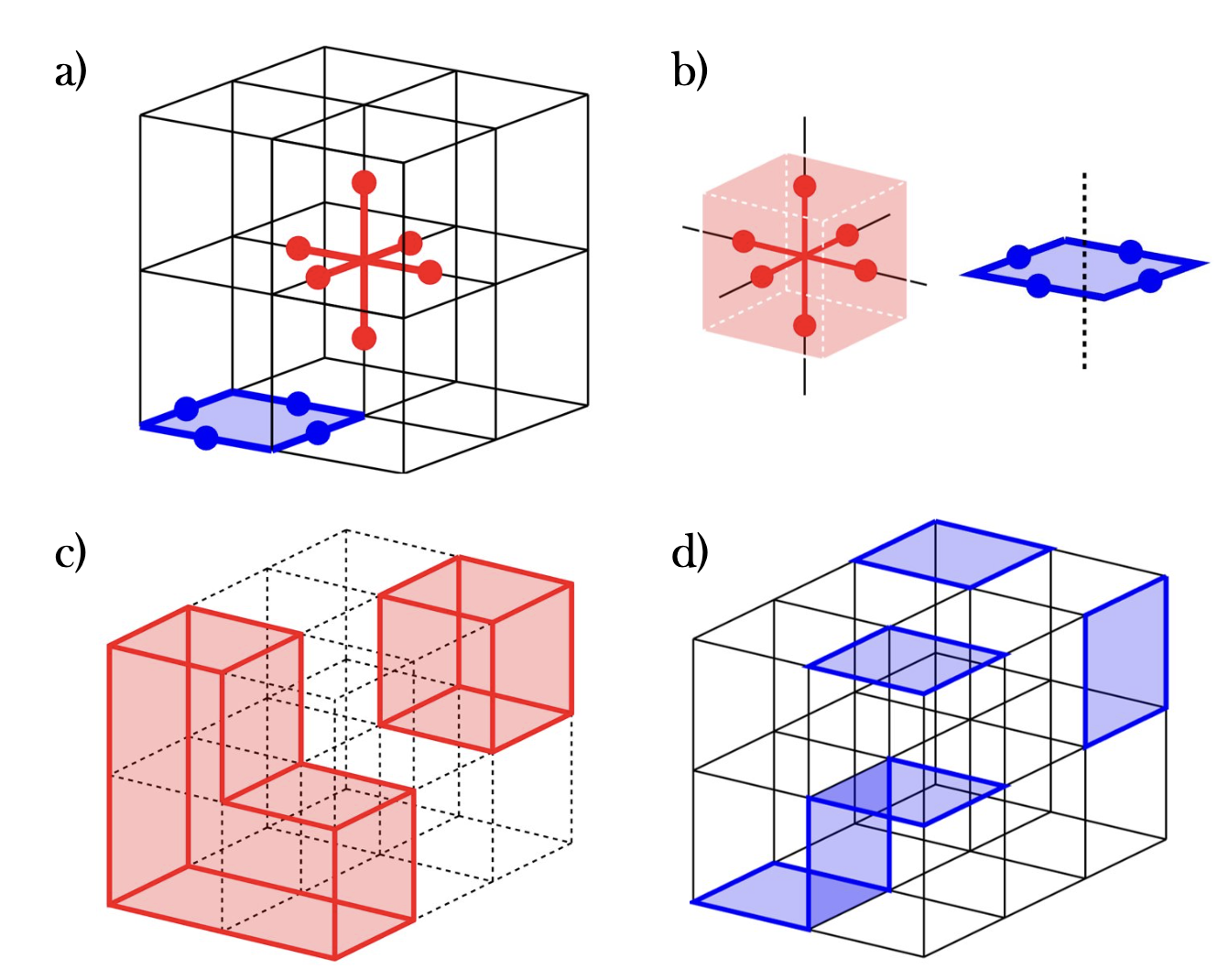}
    \caption{\textbf{3D Toric Code:} a) The stabilizers as defined on the cubic lattice for the 3D toric code. $A_v$ is pictured in red, while $B_p$ for an example plaquette is pictured in blue. b) The stabilizers and their counterparts on the dual lattice; the vertex term is mapped to a cube on the dual lattice, while the plaquette terms are mapped to dual edges (indicated by the dashed line). c) An example stabilizer configuration $g_x$, which is a closed  ``wireframe'' configuration on the dual cubic lattice. d) An example stabilizer configuration $g_z$, which is a set of closed loops on the direct lattice.}
    \label{fig:3dtc}
\end{figure}

We will consider the 3D toric code on the cubic lattice. The Hamiltonian is given by:
\begin{equation}
    H_{3DTC} = -\sum_{v \in V} A_v - \sum_{p \in P} B_p
\end{equation}
with the following stabilizers:
\begin{equation}
\begin{aligned}
    A_v &= \prod_{\partial e \ni v} X_e \\
    B_p &= \prod_{e \in \partial p} Z_e   
\end{aligned}
\end{equation}
Here, $v \in \partial e$ indicates that the edge $e$ has an end on the vertex $v$, and $e \in \partial p$ indicates the edge $e$ belongs to the boundary of plaquette $p$. See Fig. \ref{fig:3dtc}a for an illustration. 

%\begin{tikzpicture}
% \draw[thick] (0, 0) -- (0, 0.5);
% \draw[thick] (0, 0) -- (0, -0.5);
% \draw[thick] (0, 0) -- (0.5, 0);
% \draw[thick] (0, 0) -- (-0.5, 0);
% \draw[thick] (0, 0) -- (-0.4, -0.3);
% \draw[thick] (0, 0) -- (0.4, 0.3);
% \filldraw[black] (0, 0) circle (2pt) node[anchor=east] {$v$};
% \node [fill=white,inner sep=1pt] at (0,0.25) {$X$}
% \end{tikzpicture}

We now consider the ground state density matrix, given by the product of projectors onto the $+1$ eigenstate of all the stabilizers $A_v$ and $B_p$:
\begin{equation}
\begin{aligned}
    \rho_0 &= \frac{1}{2^3} \prod_v \frac{1}{2}(1+A_v) \prod_p \frac{1}{2}(1 + B_p)\\
    &= \frac{1}{2^{N}} \sum_{g_x} g_x \sum_{g_z} g_z
\end{aligned}
\label{eq:3dtc-rho0}
\end{equation}
where $g_x$ are products of $A_v$ stabilizers, which can be visualized as closed ``wireframes'' on the dual cubic lattice (see Fig. \ref{fig:3dtc}c). The $g_z$ are products of plaquette terms, which form closed loops on the direct lattice (see Fig. \ref{fig:3dtc}d). Here, $N$ is the number of physical qubits (equivalently, the number of edges of the cubic lattice).

We would like to examine the effects of local Pauli errors on the properties of this ground state density matrix; in particular, is there a critical error rate such that we lose the ability to store any information in the ground state manifold? We will focus on local Pauli error channels, where there is some probability $p$ to apply a single-qubit Pauli $P_j$ to our density matrix $\rho_0$:
\begin{equation}
    \begin{aligned}
        \mathcal{E}_{X_j}[\rho_0] &= (1-p_x)\rho + p_x X_j \rho X_j \\
        \mathcal{E}_{Z_j}[\rho_0] &= (1-p_z)\rho + p_z Z_j \rho Z_j \\
    \end{aligned}
\label{eq:channels}
\end{equation}
Here, $j = 1, \dots, N$ indexes the physical qubits living on the edges of the cubic lattice. The benefits of expressing $\rho_0$ as a sum over stabilizer configurations, as in Eq. \ref{eq:3dtc-rho0}, is that the action of the above error channels take a simple form. For example, consider the bit-flip channel $\mathcal{E}_{X_j}$: 
\begin{equation}
    \begin{aligned}
        \mathcal{E}_{X_j}[\rho_0] &= \frac{1}{2^{N}} \sum_{g_x} \mathcal{E}_{X_j}[g_x] \sum_{g_z} \mathcal{E}_{X_j}[g_z] \\
        &= \frac{1}{2^{N}} \sum_{g_x} g_x \sum_{g_z} (1-p_x) g_z + p_x X_j g_z X_j \\
        &= \frac{1}{2^{N}} \sum_{g_x} g_x \sum_{g_z} (1-2p_x)^{\abs{g_{z, j}}} g_z
    \end{aligned}
\end{equation}
where $\abs{g_{z, j}}=1$ if $Z_j$ is in the support of $g_z$, and is $0$ otherwise. The effect of the phase error channel $\mathcal{E}_{Z_j}$ will be the same, but with the roles of $g_x$ and $g_z$ swapped.

%\zx{[} The $g_x$ configurations are untouched by the bit-flip channel since they commute with $X_j$. The $g_z$ configurations with no support on the site $j$ are also untouched, since they also commute with $X_j$.  The $g_z$ configurations with support on site $j$ will pick up a factor of $1-2p_x$ because of the anti-commutation relations between $Z_j$ and $X_j$. \zx{] The texts in the square bracket may not be necessary} 

Applying the channel $\mathcal{E}_{X_j}$ for all $j = 1, \dots, N$ yields:
\begin{equation}
    \begin{aligned}
        \circ_j \mathcal{E}_{X_j}[\rho_0] &= \frac{1}{2^{N}} \sum_{g_x} g_x \sum_{g_z} (1-2p_x)^{\sum_{j=1}^N \abs{g_{z, j}}} g_z \\
        &= \frac{1}{2^{\mathcal{N}}} \sum_{g_x} g_x \sum_{g_z} (1-2p_x)^{\abs{g_{z}}} g_z 
    \end{aligned}
\end{equation}
We will call $\abs{g_z} = \sum_{j=1}^N \abs{g_{z, j}}$ the \emph{weight with respect to $X$} of the configuration $g_z$. In this case, this is equivalent to the number of edges on which $g_z$ has nontrivial support--- however, we will see in sections \ref{sec:corr-errs} and \ref{sec:non-CSS} that when we generalize to correlated errors and non-CSS codes, this will not necessarily be the case. If we now apply $\mathcal{E}_{Z_j}$ as well, we obtain the ``error-corrupted density matrix'' $\rho$:
\begin{equation}
    \rho = \frac{1}{2^{N}} \sum_{g_x} (1-2p_z)^{\abs{g_x}} g_x \sum_{g_z} (1-2p_x)^{\abs{g_{z}}} g_z 
\end{equation}

Our goal is to identify and understand any phase transitions in the information properties of $\rho$ as we tune the channel error rates, $p_x$ and $p_z$. Crucially, any order parameter of such a transition must be a nonlinear function of $\rho$ --- physical observables (i.e. linear functions of $\rho$) can only be smooth functions of error rate \cite{fanDiagnosticsMixedstateTopological2023}. This stems from the fact that a local quantum channel can be purified through a finite-depth unitary circuit, and this finite-depth circuit cannot introduce any non-analyticities into the functional form of an expectation value. As they probe correlations between multiple copies of the density matrix, non-linear quantities are not physical observables and as such are not required to be smooth functions. 
These non-linear quantities still carry meaningful knowledge about $\rho$, though; they probe the \emph{information} properties of our state and can characterize phase transitions - in this case the loss of quantum memory.

The simplest non-linear quantity one can consider is the $n$th moment of the density matrix, $\tr \rho^n$, for $n \geq 2$. As demonstrated in \cite{fanDiagnosticsMixedstateTopological2023}, $\tr \rho^n$ can be mapped onto the partition function for a pair of classical statistical mechanics models (SM). The critical properties of these SM models can in turn be mapped back onto information quantities in the quantum code. Let's calculate the $n$th moment of the corrupted density matrix of the 3D toric code:
\begin{equation}
    \tr \rho^n = \frac{1}{2^{nN}} \prod_{\alpha=x,z} \sum_{\{g^{(m)}_{\alpha}\}}e^{-\mu_{\bar{\alpha}} \sum_{m=1}^n \abs{g^{(m)}_\alpha}} \tr \left ( \prod_{m=1}^n g^{(m)}_\alpha \right)
\end{equation}
For convience, we introduce the notation $\bar{\alpha}$ which denotes the opposite Pauli to $\alpha$: for instance, $\bar{z} = x$. Additionally, we have defined $\mu_{\bar{\alpha}} = -\log(1-2p_{\bar{\alpha}})$. Since we have $n$ copies of $\rho$, we have $n$ flavors of stabilizers, indexed by $m = 1, \dots, n$. All terms with residual Pauli matrices will disappear from the trace: only configurations where the $n$th copy is equal to the product of the previous $n-1$ copies contribute to the sum. This constrains the $n$ copies as follows: 
\begin{equation}
\prod_{m=1}^{n-1} g_\alpha^{(m)} = g_x^{(n)}
\end{equation}

Plugging back into the expression for the $n$th moment yields:
\begin{equation}
    \mathrm{tr} \rho^n = \frac{1}{2^{(n-1)N}} \prod_{\alpha = x, z} \sum_{\{g^{(m)}_{\alpha}\}} e^{-\mu_{\bar{\alpha}} \left (\sum\limits_{m=1}^{n-1} \abs{g^{(m)}_{\alpha}} + \abs{\prod\limits_{m=1}^{n-1} g^{(m)}_{\alpha}}\right)} 
\label{eq:3dtc-nth-moment}
\end{equation}

We would like to map this expression onto the partition function for a classical SM model, where the exponent can be interpreted as the Hamiltonian, and the sum over stabilizer configurations $\{g_\alpha\}$ can be interpreted as a sum over classical spin configurations. The key step is to calculate the weight of a stabilizer configuration $\abs{g_\alpha}$, since this forms the building block of the above SM model.

\begin{figure}
    \centering
    \includegraphics[width=\linewidth]{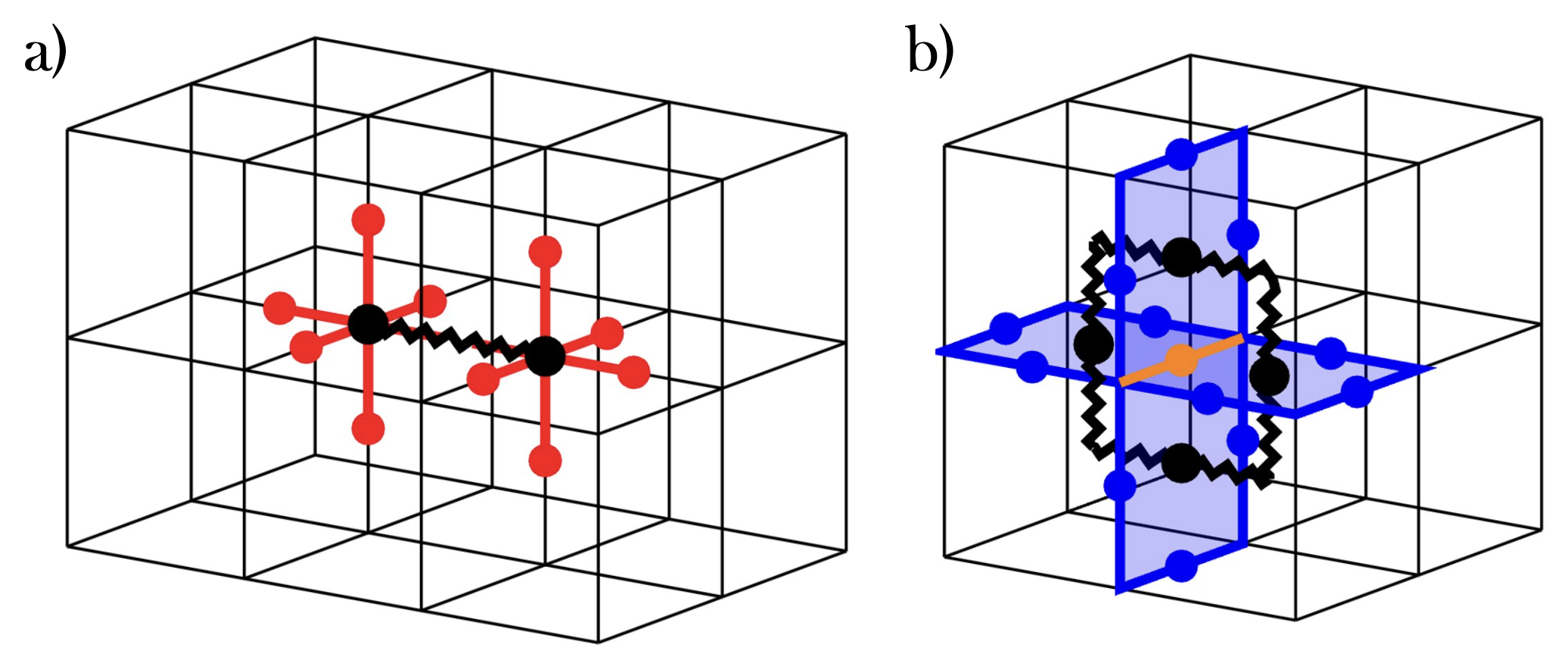}
    \caption{\textbf{SM Models for 3D Toric Code:} a) The Ising-type interaction induced by phase errors is pictured. The classical spins corresponding to the two vertex terms are denoted by black circles. The error-affected edge, and the interaction induced between the classical spins, is marked by the black zig-zag line. b) A single Pauli $X$ error (the orange zig-zag line) produces a different set of excitations, denoted by the shaded plaquettes. Assigning classical spins to the centers of the plaquettes (black circles), we see that the induced interaction is exactly given by 3D $\mathbb{Z}_2$ lattice gauge theory, which is dual to the 3D Ising model.}
    \label{fig:3dtc-SM}
\end{figure}

First, we consider the effect of phase errors on the error-corrupted density matrix. We only need to consider the $g_x$ configurations: to determine the form of interactions in our SM model, we place classical spins at each vertex on the cubic lattice. Each possible $g_x$ is assigned a corresponding configuration of classical spins -- if a vertex term is included in $g_x$, the corresponding classical spin has $\sigma^z = -1$, and if a vertex term is not included in $g_x$, its classical spin has $\sigma^z = +1$. A single $Z_e$ anti-commutes with the two vertex terms $A_v$ that share the edge $e$; $g_x$ will have support on $e$ if it includes one and only one of these vertex terms. This corresponds to the scenario where the two classical spins on the endpoints of the edge are pointing in opposite directions. Nearest-neighbor spins pointing in the same direction indicate that $g_x$ has no support on the link connecting them. Our SM model should count all of the domain walls in the classical spin configuration for $g_x$, which means it must be the 3D Ising model on the cubic lattice (see Fig. \ref{fig:3dtc-SM}a). Put more concretely, we can map $\abs{g_{x, i}}$ onto:
\begin{equation}
    \abs{g_{x, i}} = \frac{1}{2}(1 + \sigma^z_j \sigma^z_k)
\end{equation}
where $j, k$ denote the classical spins separated by the edge $i$ (see Fig. \ref{fig:3dtc-SM}a). Plugging into Eq. \ref{eq:3dtc-nth-moment} gives us the following SM model:
\begin{equation}
        \mathcal{H}^{(n)}_z
        =  - \frac{\mu_z}{2} \sum_{\langle jk \rangle} \left( \sum_{m=1}^{n-1} \sigma^{(m), z}_j \sigma^{(m), z}_k + \prod_{m=1}^{n-1} \sigma^{(m), z}_j \sigma^{(m), z}_k\right)
\end{equation}
where we have dropped the constant terms, which only shift the spectrum. The sum over $\langle jk\rangle$ denotes a sum over nearest neighbor vertices on the 3D cubic lattice. This SM model is a $(n-1)$-flavor 3D Ising model with a flavor-coupling term; for $n=2$, it reduces to the usual 3D Ising model.

We can apply the same procedure to analyze the bit-flip error channel. Our classical spins for a configuration $g_z$ live at the centers of plaquettes, or equivalently, edges of the dual lattice. A single $X_e$ acting on some edge of the toric code ground state will excite the four plaquettes that share the edge $e$; accordingly, we can determine whether a given edge has support in $g_z$ or not based on the parity of these four classical spins. The correct interaction term in our SM model is then $\sigma^z_i \sigma^z_j \sigma^z_k \sigma^z_l$, where $ijkl$ denotes the four plaquettes sharing the edge $e$ (see Fig. \ref{fig:3dtc-SM}b). Alternately, we can view them as the dual edges forming the boundary of a plaquette on the dual lattice. Plugging into Eq. \ref{eq:3dtc-nth-moment}, we find the $n$th SM model for bit-flip errors:
\begin{equation}
    \begin{aligned}
        \mathcal{H}^{(n)}_x &=  - \frac{\mu_x}{2} \sum_{\langle ijkl \rangle} \left( \sum_{m=1}^{n-1} \sigma^{(m), z}_i \sigma^{(m), z}_j \sigma^{(m), z}_k \sigma^{(m), z}_l \right.\\
        &\qquad\qquad\qquad\left.+ \prod_{m=1}^{n-1} \sigma^{(m), z}_i \sigma^{(m), z}_j \sigma^{(m), z}_k \sigma^{(m), z}_l \right)
    \end{aligned}
\end{equation}
where $\langle ijkl \rangle$ denotes all sets of dual edges $i,j,k,l$ that form the boundary of a plaquette on the dual cubic lattice. For $n=2$, this Hamiltonian describes $\mathbb{Z}_2$ lattice gauge theory \cite{wegnerDualityGeneralizedIsing1971, kogutIntroductionLatticeGauge1979}. We have now mapped the $n$th moment of the 3D toric code density matrix onto two partition functions; one for $Z$ errors and one for $X$ errors:
\begin{equation}
\begin{aligned}
    \tr \rho^n &= \frac{1}{2^{(n-1)N}} \prod_{\alpha = x, z} \mathcal{Z}^{(n)}_\alpha\\
    \mathcal{Z}^{(n)}_\alpha &= \sum_{\{g_\alpha\}} e^{-\mathcal{H}^{(n)}_\alpha}
\end{aligned}
\end{equation}

If the classical model $\mathcal{H}^{(n)}_\alpha$ has a finite-temperature phase transition, any good order parameters for this transition on the classical side should map back to the information quantities on the quantum side. We will investigate some candidate information quantities in section \ref{sec:generic-mapping} which are sensitive to these decoherence-induced phase transitions. The two SM models we have derived for the 3D toric code do have finite-temperature transitions, which we will examine closer in \ref{sec:3dtc}.

\section{Generic Statistical Mechanics Mapping}
\label{sec:generic-mapping}

Now that we have worked through the example of the 3D toric code, we focus on developing a generic mapping from stabilizer codes under decoherence to their corresponding statistical mechanics models. We present the generic mapping for the $n$th moment of the density matrix $\rho$ and three relevant information quantities: (i) the relative entropy, (ii) the coherent information, and (iii) the entanglement negativity. Our mapping generalizes the one introduced in \cite{fanDiagnosticsMixedstateTopological2023} for the 2D toric code. We restrict ourselves to CSS codes for the moment, and will come back to the non-CSS case in section \ref{sec:non-CSS}.

Consider a generic translation-invariant stabilizer Hamiltonian of the form:
\begin{equation}
    H = -\sum_{\alpha = x, z}\sum_{i_{\alpha}} S^{(\alpha)}_{i_{\alpha}}
    \label{eq:stabilizer-Hamiltonian}
\end{equation}
where $S^{(\alpha)}_{i_{\alpha}}$ are local Pauli operators with finite support in some region centered at $i_\alpha$ of the lattice, and $\alpha$ indicates whether $S$ is a $Z$ or $X$ check. 

%\zx{\sout{and $i_{\alpha}$ indexes physical locations where stabilizers are centered. Note that these locations depend on the type of stabilizer in question: for instance, the 2D toric code Hamiltonian would have $i_{X}$ indexing vertices of the square lattice, while $i_{Z}$ indexes plaquettes.}}

As all $S_{i_\alpha}^{(\alpha)}$'s mutually commute, the ground state manifold for $H$ is given by the projection into their simultaneous $+1$ eigenstate. The completely mixed state in the ground state manifold is given by the following:
\begin{equation}
    \rho_0 = \frac{1}{2^{\mathcal{N}}}\prod_{\alpha} \prod_{i_{\alpha}} \frac{1}{2}(1 + S_{i_\alpha}^{(\alpha)})
    \label{eq:density-matrix}
\end{equation}
where $\mathcal{N}$ is the dimension of the ground state manifold. If $N$ is the number of physical qubits in the system, $N_s$ is the number of stabilizers, and $N_c$ is the number of independent constraints on the stabilizers, $\mathcal{N}$ will be given by:
\begin{equation}
    \mathcal{N} = {N - N_s + N_c}
\end{equation}

In the vein of the analysis in section \ref{sec:warm-up}, we want to express the density matrix as a sum over different configurations of stabilizers, denoted $g_{\alpha}$. When we expand the product over $i_{\alpha}$ in Eq. \ref{eq:density-matrix}, we find terms corresponding to all possible products of the stabilizers $S^{(\alpha)}_{i_{\alpha}}$.
\begin{equation}
\begin{aligned}
    \rho_0 &= \frac{1}{2^{N-N_s+N_c}} \frac{1}{2^{N_s - N_c}} \prod_{\alpha = x, z} \sum_{g_{\alpha}}g_{\alpha}\\
    &= \frac{1}{2^N}\prod_{\alpha = x, z} \sum_{g_{\alpha}}g_{\alpha}
\end{aligned}
\end{equation}

The factor of $1/2^{N_s - N_c}$ in the first line comes from the fact there are $N_s$ factors of $1/2$ from the projectors, and $2^{N_c}$ of the resulting stabilizer products in the expansion are equal to the identity, stemming from the stabilizer constraints. 

Now we consider applying a local decoherence channel to this density matrix. For simplicity, we will restrict ourselves to the single site $X$ (bit-flip) or $Z$ (phase) errors we considered for the 3D toric code, as defined in Eq. \ref{eq:channels}. More general Pauli errors will be considered in section \ref{sec:corr-errs}. 

Given that our density matrix $\rho_0$ is a sum over Pauli operators, both $X$ and $Z$ errors will introduce some relative phases between terms in the sum, depending on which stabilizer configurations commute or anti-commute with $X$ or $Z$ at a given site. For convenience, we introduce the so-called ``scalar commutator'' \cite{chubbStatisticalMechanicalModels2021}:  When two operators commute up to a number, the scalar commutator $\llbracket \cdot, \cdot \rrbracket$ returns this number. For example, $\llbracket X_i, Z_j \rrbracket = (-1)^{\delta_{ij}}$. We can recast the action of the error channels into one general formula for single-site Pauli errors $P^{(\alpha)}_j = X, Z$ if $\alpha = x, z$:
\begin{equation}
    \mathcal{E}_{P^{(\alpha)}_j}[\rho] = (1 - (1 + \llbracket P^{(\alpha)}_j, \rho \rrbracket ) p_\alpha)\rho\\
\end{equation}
If we apply the channel $\mathcal{E}_\alpha = \circ_{j=1}^N \mathcal{E}_{P^{(\alpha)}_j}$ we find:
\begin{equation}
\begin{aligned}
    \mathcal{E}_\beta[\rho_0] &= \frac{1}{2^N} \sum_{g_{\alpha}} g_\alpha \sum_{g_{\bar{\alpha}}} \left(\prod_{j=1}^N (1 + (\llbracket P^{(\alpha)}_j, g_{\bar{\alpha}} \rrbracket - 1)p_{\alpha}) \right ) g_{\bar{\alpha}} \\
    &=\frac{1}{2^N} \sum_{g_{\alpha}} g_\alpha \sum_{g_{\bar{\alpha}}} e^{-\mathrm{log}(1+2p_P)\sum_{j=1}^N |g_{\bar{\alpha}, j}|} g_{\bar{\alpha}}\\
    &= \frac{1}{2^N} \sum_{g_{\alpha}} g_\alpha \sum_{g_{\bar{\alpha}}} e^{-\mu_\alpha \abs{g_{\bar{\alpha}}}} g_{\bar{\alpha}}
\end{aligned}
\end{equation}
where we have defined $\abs{g_{\bar{\alpha}}}$ as:
\begin{equation}
    \begin{aligned}
    \abs{g_{\bar{\alpha}}} &= \sum_{j = 1}^N \abs{g_{\bar{\alpha}, j}} \\
    &= \sum_{i = j}^N \frac{1}{2} (1 - \llbracket P^{(\alpha)}_j, g_{\bar{\alpha}, j} \rrbracket) \\
\end{aligned}
\end{equation}

As before, $\abs{g_{\bar{\alpha}}}$ is the ``weight'' of $g_{\bar{\alpha}}$ with respect to the Pauli error $P^{(\alpha)}$--- it gives the number of sites on which the support of $g_{\bar{\alpha}}$ anti-commutes with $P^{(\alpha)}$. Here, $g_{\bar{\alpha}, j}$ is the support of $g_{\bar{\alpha}}$ on physical site $j$.

Consider the $n$th moment of the density matrix $\rho_\alpha = \mathcal{E}_\alpha[\rho_0]$:
\begin{equation}
        \mathrm{tr} \rho_\alpha^n =  \frac{1}{2^{nN}} \sum_{\{g^{(m)}_\alpha\}} \sum_{\{g^{(m)}_{\bar{\alpha}}\}} e^{-\mu_\alpha \sum\limits_{m=1}^n \abs{g^{(m)}_{\bar{\alpha}}}} \mathrm{~tr} \left (\prod_{m=1}^n g^{(m)}_\alpha g^{(m)}_{\bar{\alpha}}\right) 
\end{equation}
where $m = 1, \dots, n$ indexes copies of the density matrix and $\{g^{(m)}_{\alpha, \bar{\alpha}}\}$ denotes a choice of possible $g_{\alpha, \bar{\alpha}}$ configurations for each copy. All terms in the sum will vanish except those whose product $\prod_{m=1}^n g^{(m)}_\alpha g^{(m)}_{\bar{\alpha}}$ is the identity, so we can write $\tr \rho^n $ in terms of the first $n-1$ copies:
\begin{equation}
     \mathrm{tr} \rho_\alpha^n = \frac{1}{2^{(n-1)N}} \sum_{\{g^{(m)}_{\bar{\alpha}}\}} e^{-\mu_\alpha \sum\limits_{m=1}^{n-1} \abs{g^{(m)}_{\bar{\alpha}}} + \abs{\prod\limits_{m=1}^{n-1} g^{(m)}_{\bar{\alpha}}}}
\label{eq:nth-before-poly}
\end{equation}

We now have something that looks like the partition function for a statistical mechanics model--- to complete the mapping, we need to find the SM model that properly calculates the value of $\abs{g_{\bar{\alpha}}}$. To do this, we will utilize the commutative algebra approach to analyzing stabilizer Hamiltonians, first introduced by Haah \cite{haahCommutingPauliHamiltonians2013}. We only touch on the elements of the formalism necessary for the general mapping here--- we direct the interested reader to \cite{haahCommutingPauliHamiltonians2013, vijayFractonTopologicalOrder2016, Haah_2017} for a deeper introduction. 

The idea is to encode the stabilizer geometry into a vector of Laurent polynomials, which allow us to systematically understand commutation relations. Suppose our stabilizer model lives on a $d$-dimensional Bravais lattice with $l$ qubits per unit cell. Given some choice of unit cell $s_0$, we express every stabilizer $S$ acting on $s_0$ solely in terms of Pauli $Z$ and $X$ operators. Each spin where $S$ has $Z$ support is assigned a monomial $x_1^a x_2^b \cdots x_d^c$ where $(a, b, \dots, c)$ is the coordinate of the spin's unit cell relative to $s_0$. The monomials corresponding to the $n$th spin in each unit cell are summed together, and form the $n$th component of the vector of polynomials corresponding to $S$. In the same way, the $n$th spins in each unit cell where $S$ has $X$ support are assigned monomials and summed together to give the $(n + l)$th component. We denote the map from a Pauli operator to its corresponding vector of polynomials as $\mathsf{\Omega}: S \rightarrow \mathsf{\Omega}(S)$. The vector of polynomials $\mathsf{\Omega}(S) \in \mathbb{F}_2^{2 l}$ has binary coefficients $\{0, 1\}$, and polynomial addition is defined (mod 2) going forwards.

We collect the polynomial vector for each distinct stabilizer type and concatentate them to form the so-called \emph{stabilizer map}, $\mathsf{S}$. This is a $2l \times m$ matrix, where $m$ is the number of distinct stabilizer types. We can think of $\mathsf{S}$ as a map from the space of stabilizer types to the bits on the physical lattice that form the given stabilizer.
\begin{equation}
    \mathsf{S} = \begin{pmatrix}
        \mathsf{\Omega}(S^{(1)}) & \mathsf{\Omega}(S^{(2)}) & \cdots & \mathsf{\Omega}(S^{(m)})
    \end{pmatrix}
\end{equation}
%\zx{[How about the font $\mathsf{S}, \mathsf{E}, \mathsf{\Omega}$? ]}
There is a corresponding \emph{excitation map}, $\mathsf{E}$, which maps from the space of physical bits to the space of stabilizer types. $\mathsf{E}$ takes as input a set of bits where errors have occurred, and outputs the stabilizers that are violated by the given error pattern. 
\begin{equation}
    \mathsf{E} = \mathsf{S}^{\dagger}\lambda_l
\end{equation}
where $\mathsf{S}^{\dagger} = (\bar{\mathsf{S}})^T$, and $\bar{\mathsf{S}}$ indicates every polynomial entry $f(x_1, x_2, \hdots, x_d)$ in $S$ is mapped to $f(x_1^{-1}, x_2^{-1}, \hdots, x_d^{-1})$. The matrix $\lambda_l$ is defined as:
\begin{equation}
    \lambda_l = \begin{pmatrix}
        0 & \mathbb{I}_{l\times l} \\
        \mathbb{I}_{l\times l} & 0
    \end{pmatrix}
\end{equation}

We now have the tools necessary to write down the SM model for our stabilizer Hamiltonian. The weight of each configuration $g_{\alpha}$ is the number of physical sites on which it anti-commutes with the Pauli error under consideration. Each $g_{\alpha}$ is some product of stabilizers:
\begin{equation}
    g_{\alpha} = \prod_{u} (S^{(\alpha)}_u)^{\mathbf{c}^{\alpha}_u}
\end{equation}
where $\mathbf{c}^{\alpha}_u$ is a binary vector containing the information on the support of $g_{\alpha}$. We assign a classical spin variable $s_u$ to the center of each stabilizer $S^{(\alpha)}_{u}$, which will take value $s_u = +1$ if that stabilizer is not involved in the product $g_{\alpha}$ and $s_u = -1$ if it is involved in $g_{\alpha}$. In other words, $s_u = 2\mathbf{c}^{(\alpha)}_u - 1$. For a configuration $g_{\alpha}$ to have support on a site $j$, it must involve an odd number of stabilizers $S^{(\alpha)}_u$ that themselves have support on $j$, as otherwise the Pauli operators will multiply to the identity. We can count the parity of stabilizers acting on $j$ by looking at the product of classical spins $s_u$ corresponding to the stabilizers with support on $j$. This set of stabilizers is precisely the output of the excitation map, $\mathsf{E}$, when we input an error occurring on site $j$:
\begin{equation}
    \abs{g_\alpha} = \sum_{j=1}^N \frac{1}{2}(1 - \mathsf{E} \cdot \mathsf{\Omega}(P_j)). 
\label{eq:weight}
\end{equation}
Here, $j$ indexes a physical site, $P_j$ is the Pauli error under consideration, and we have taken the output of $\mathsf{E}$ to be in the space of classical spins rather than stabilizers (they are isomorphic given we have assigned a classical spin to the center of each stabilizer). 
%\zx{[Maybe we can move (27)(28) to right after (21), and then say we can use the polynomial formalism to simplify (28) and then introduce details of the formalism? ]}
Using this expression derived for $\abs{g_{\alpha}}$ in Eq. \ref{eq:weight} and plugging back into Eq. \ref{eq:nth-before-poly}, we arrive at the desired SM model:
\begin{equation}
    \begin{aligned}
        \mathrm{tr} \rho_\alpha^n &= \frac{1}{2^{(n-1)N}} \sum_{\{g^{(m)}_{\bar{\alpha}}\}} e^{- \mathcal{H}^{(n)}_\alpha} = \frac{1}{2^{(n-1)N}} \mathcal{Z}^{(n)}_\alpha \\
        \mathcal{H}^{(n)}_\alpha &= - \frac{\mu_\alpha}{2}\sum_{j=1}^N \left (\sum_{m=1}^{n-1} \mathsf{E}^{(m)} \cdot \mathsf{\Omega}(P^{(\alpha, m)}_j) \right. \\
        &\qquad\qquad\qquad\qquad+ \left. \prod_{m=1}^{n-1} \mathsf{E}^{(m)} \cdot \mathsf{\Omega}(P^{(\alpha, m)}_j)\right)
    \end{aligned}
\label{eq:nth-SM-model}
\end{equation}
Here, $\mathsf{E}^{(m)}$ maps an error pattern $P^{(\alpha, m)}$ on the $m$th copy to the corresponding classical spins of flavor $m = 1, \dots, n-1$.

It will be useful to have the error-corrupted density matrix for the case that both kinds of Pauli errors occur with probabilities $p_x$ and $p_z$: $\rho = \mathcal{E}_x \circ \mathcal{E}_z[\rho_0]$. As we are working with CSS codes, the SM models decouple:
\begin{subequations}
\begin{align}
&\rho =  \frac{1}{2^N} \sum_{g_x}e^{-\mu_z \abs{g_x}} g_x \sum_{g_x} e^{-\mu_x \abs{g_x}} g_x \\
&\tr \rho^n  = \frac{1}{2^{(n-1)N}} \mathcal{Z}^{(n)}_z \mathcal{Z}^{(n)}_x
\end{align}
\label{eq:both-errors-rho}
\end{subequations}

We are now able to derive the family of SM models describing the $n$th moment of the ground state density matrix for \emph{any} stabilizer code, as long as we have the excitation map, $\mathsf{E}$.

\subsection{SM Mapping for Information Quantities}
\label{sec:info-quantities}

\renewcommand{\arraystretch}{2}
\begin{table*}[t]
    \centering
    \begin{tabular}{|c|c|c|c|}
        \hline
        \hline
         \textbf{Information Measure} & \textbf{SM Model Counterpart} & \textbf{PM} & \textbf{FM} \\
         \hline
         Entropy & Free energy, $\mathcal{F}^{(n)}$ & $O(L^3)$ & $0$\\
         \hline
         Relative Entropy & Logarithm of generalized spin-spin correlator, $\left \langle \mathsf{E}^{(n)}\cdot\mathsf{\Omega}(P_k)\right\rangle$ & $O(\abs{i-j})$ & $0$\\
         \hline
         Coherent Information & Related to excess free energy of inserting defects along logicals & $\mathcal{N} \log 2$& 0\\
         \hline
         Entanglement Negativity & Excess free energy of pinning spins along boundary $\partial A$ & $\abs{\partial A}/\xi - \gamma$ & $\abs{\partial A}/\xi - \gamma'$\\
         \hline
         \hline
    \end{tabular}
    \caption{\textbf{SM Mapping for Information Measures:} A table summarizing different information measures and their corresponding quantities in the classical SM model. The behavior of the information quantity in the paramagnetic (PM) and ferromagnetic (FM) phases is also listed. We call the general high-temperature phase ``paramagnetic'' and the low-temperature phase ``ferromagnetic'' for consistency, even if a specific model (like $\mathbb{Z}_2$ gauge theory) doesn't have a conventional paramagnetic or ferromagnetic phase.}
    \label{tab:info-SM-mapping}
\end{table*}
\renewcommand{\arraystretch}{1}

The previous section derived the SM model corresponding to the $n$th moment of the density matrix, $\mathrm{tr} \rho^n$ under incoherent Pauli errors, $P$. We will now consider other useful information quantities and their SM counterparts. In particular, we focus on the quantities introduced in \cite{fanDiagnosticsMixedstateTopological2023} in order to diagnose the transition out of the topologically ordered phase. See Table \ref{tab:info-SM-mapping} for a summary of the different information quantities and their corresponding quantities in the SM models. 

\subsubsection{Réyni Entropy}

Now that we have the SM models for $n$th moments of the density matrix, the most basic information measures to consider are the Réyni entropies $S^{(n)}(\rho)$ \cite{renyi1961measures}. Recall that the $n$th Réyni entropy is given by:
\begin{equation}
    S^{(n)} = \frac{1}{1-n} \mathrm{log}(\mathrm{tr}\rho^n). 
\end{equation}
In the limit $n \rightarrow 1$, this formula recovers the von Neumann entropy, $S(\rho) = \tr \rho \log \rho$. We can plug in Eq. \eqref{eq:both-errors-rho}b to find:
\begin{equation}
    \begin{aligned}
        S^{(n)} &= \frac{1}{1-n} \mathrm{log}\left(\frac{1}{2^{(n-1)N}}\mathcal{Z}^{(n)}_x \mathcal{Z}^{(n)}_z\right)\\
        &= \frac{1}{n-1}\left(\mathcal{F}^{(n)}_x + \mathcal{F}^{(n)}_z \right) + N \log 2,
    \end{aligned}
\end{equation}
where we have defined the free energy densities $\mathcal{F}_\alpha^{(n)}$. So the Réyni entropy will have a transition wherever the stat-mech model undergoes an ferromagnetic-to-paramagnetic transition. This result tells us that the critical error rate for the $n$th moment of the density matrix will provide an upper bound for the critical error rate in the replica limit $n \rightarrow 1$. This is due to the non-increasing behavior of the Réyni entropies \cite{Muller_Lennert_2013}:
$S^{(n)} \leq S^{(m)}$ if $n > m$. If $S^{(n)}$ undergoes a transition from ferromagnetic (zero free energy) to paramagnetic (extensive free energy), it bounds from below the critical temperature $T_c$ of $S^{(m)}$, and consequently bounds from above the inverse critical temperature $\beta_c$. This is an upper bound on the critical error rate since we have a monotonically increasing relationship between $\beta_c$ and $p_c$. In particular, we can use the critical error rates extracted for Réyni-2 quantities to place the tightest bounds on the replica limit.

\subsubsection{Relative Entropy}
\label{sec:relative-entropy}

The error channels $\mathcal{E}_x$ and $\mathcal{E}_z$ create excitations on top of the ground state. Our ability to retrieve any initial information encoded in the state hinges upon our ability to detect and correct these errors. The \emph{quantum relative entropy} \cite{ume1962, Vedral_2002} provides a way to quantify our ability to detect errors, as it is a measure of the distinguishability of two density matrices, $\rho$ and $\rho_\epsilon$.
\begin{equation}
    D(\rho \vert \vert \rho_\epsilon)  = \tr \rho \log \rho - \tr \rho \log \rho_{\epsilon}
\end{equation}
When $\rho$ and $\rho_\epsilon$ are identical, $D^{(n)} = 0$, and when $\rho$ and $\rho_\epsilon$ are orthogonal, $D^{(n)} \rightarrow \infty$. Consider the case when $\rho_\epsilon$ is equal to the error-corrupted density matrix $\rho$, but with one additional pair of excitations added. If we cannot distinguish the two density matrices, we have no chance of decoding any information that might have originally been stored in $\rho_0$, since we cannot accurately determine the error syndrome.
The Réyni versions of the relative entropy \cite{PETZ198657, Muller_Lennert_2013} are given by:
\begin{equation}
    D^{(n)}(\rho \vert \vert \rho_\epsilon)  = \frac{1}{1-n}\log \frac{\tr\rho \rho_{\epsilon}^{n-1}}{\tr\rho^n}
\label{eq:rel-entropy-original}
\end{equation}

We consider the case where the additional error is another single-site Pauli, $P^{(\alpha)}_k$, for some physical site $k$. Then $\rho_{\epsilon}$ is given by:
\begin{equation}
    \begin{aligned}
        \rho_{\epsilon} &= P^{(\alpha)}_k \rho P^{(\alpha)}_k \\
        &= \frac{1}{2^N} \sum_{g_{\alpha}}  e^{-\mu_{\bar{\alpha}} \abs{g_{\alpha}}} g_\alpha \sum_{g_{\bar{\alpha}}} e^{-\mu_\alpha \abs{g_{\bar{\alpha}}}} P^{(\alpha)}_k g_{\bar{\alpha}} P^{(\alpha)}_k \\
        &= \frac{1}{2^N} \sum_{g_{\alpha}}  e^{-\mu_{\bar{\alpha}} \abs{g_{\alpha}}} g_\alpha \sum_{g_{\bar{\alpha}}} \llbracket g_{\bar{\alpha}}, P^{(\alpha)}_k \rrbracket e^{-\mu_\alpha \abs{g_{\bar{\alpha}}}}  \\
        &= \frac{1}{2^N} \sum_{g_{\alpha}} e^{-\mu_{\bar{\alpha}} \abs{g_{\alpha}}} g_\alpha \sum_{g_{\bar{\alpha}}} \mathsf{E} \cdot\mathsf{\Omega}(P^{(\alpha)}_k)\vert_{g_{\bar{\alpha}}} e^{-\mu_\alpha \abs{g_{\bar{\alpha}}} g_{\bar{\alpha}}}\\
    \end{aligned}
\end{equation} 
where $\mathsf{E} \cdot\mathsf{\Omega}(P^{(\alpha)}_k)\vert_{g_{\bar{\alpha}}}$ denotes the evalutation of the interaction term given by $\mathsf{E} \cdot\mathsf{\Omega}(P^{(\alpha)}_k)$ with respect to the classical spin configuration corresponding to $g_{\bar{\alpha}}$.
\begin{equation}
    \begin{aligned}
        \tr \rho &\rho_{\epsilon}^{n-1} = \frac{1}{2^{nN}} \sum_{\{g^{(m)}_{\alpha,\bar{\alpha}}\}}
        \prod_{m=1}^{n-1} \left (\mathsf{E}^{(m)}\cdot\mathsf{\Omega}(P^{(\alpha, m)}_k)\vert_{g^{(m)}_{\bar{\alpha}}} \right)\\
        & \times e^{-\mu_{\bar{\alpha}} \sum_{m=1}^n \abs{g^{(m)}_{\alpha}}-\mu_\alpha \sum_{m=1}^n \abs{g^{(m)}_{\bar{\alpha}}}} \tr \left ( \prod_{m=1}^n g^{(m)}_\alpha g^{(m)}_{\bar{\alpha}} \right )\\
        &=  \frac{1}{2^{(n-1)N}} \sum_{\{g^{(m)}_{\alpha,\bar{\alpha}}\}} 
        \left (\mathsf{E}^{(n)}\cdot\mathsf{\Omega}(P^{(\alpha, n)}_k)\vert_{g^{(n)}_{\bar{\alpha}}}\right) e^{-\mathcal{H}^{(n)}_\alpha - \mathcal{H}^{(n)}_{\bar{\alpha}}}\\
        &= \frac{1}{2^{(n-1)N}} \mathcal{Z}^{(n)}_\alpha  \mathcal{Z}^{(n)}_{\bar{\alpha}} \left \langle \mathsf{E}^{(n)}\cdot\mathsf{\Omega}(P^{(\alpha, n)}_k)\right\rangle_{\alpha}
    \end{aligned}
\end{equation}

So the relative entropy $D^{(n)}$ is related to the expectation value of $\left \langle \mathsf{E}^{(n)}\cdot\mathsf{\Omega}(P^{(\alpha, n)}_k)\right\rangle$ in the theory $\mathcal{H}^{(n)}_{\alpha}$:
\begin{equation}
    D^{(n)}(\rho \vert \vert \rho_\epsilon) = \frac{1}{1-n}\log \left \langle \mathsf{E}^{(n)}\cdot\mathsf{\Omega}(P^{(\alpha, n)}_k)\right\rangle_{\alpha}
\label{eq:rel-entropy}
\end{equation}

We can generalize the applied error $P^{(\alpha)}_k$ to any set of Pauli operators that creates a minimal set of excitations in the quantum theory: for example, an anyon string operator in a topological phase. Then the relative entropy will be related to a generalized spin-spin correlator, where the classical spins are located at the same places as the additional excitations. If the SM model does have a finite-temperature transition, the ferromagnetic phase will be characterized by $D^{(n)} \approx 0$--- the ground state of the ordered phase should have long-range order, indicating $\left\langle \mathsf{E}^{(n)}\cdot\mathsf{\Omega}(P^{(n)}_k)\right\rangle \approx 1$. In the paramagnetic phase, $D^{(n)} \rightarrow \infty$ since $\left\langle \mathsf{E}^{(n)}\cdot\mathsf{\Omega}(P{(n)}_k)\right\rangle \approx 0$ due to the larger number of fluctuations. So we expect this quantity to be a good witness for the information transition in the quantum model, since it displays markedly different behavior in each classical phase. We will present examples in section \ref{sec:examples}. 

\subsubsection{Coherent Information}
\label{sec:coherent-info}

Another key measure of the information capacity of our quantum code is the ground state degeneracy, %\zx{[Later when we include the deocherence from cluster state, we may want to modify this sentence.]} \zx{[btw, another motivation why we may want to start from a cluster state and do the fine tuned decoherence, is because, this can possibly give rise to a phase which exhibits strong-to-weak SSB of 1-form symmetry. ]} 
which determines the number of logical qubits that can be stored. We want to track how the logical information is degraded when we apply an error channel. We can do this by entangling the logical qubits of our code with ancilla qubits, and then measuring the \emph{coherent information} \cite{Schumacher_1996, Lloyd_1997} between the ancillas and our system after the decoherence has occurred. The coherent information is a standard measure for the information that survives in a state after it passes through a quantum channel:
\begin{equation}
    I_c(R ~\rangle~ Q) = S_Q - S_{RQ}
\end{equation}
%\zx{[this $\rangle$ looks weird...]}
where $Q$ is our original state, $R$ consists of the ancilla qubits, and $S_{Q}$ and $S_{RQ}$ are the von Neumann entropies of the respective systems. The Rényi versions of $I_c$ are given by:
\begin{equation}
    I_c^{(n)}(R ~\rangle~ Q) = \frac{1}{n-1} \text{log} \left(\frac{\text{tr}\rho_{RQ}^n}{\text{tr}\rho^n_Q}\right)
\end{equation} 
What does $\rho_{RQ}$ look like? We have assumed that our code has ground state degeneracy $2^{\mathcal{N}}$, so there must be $\mathcal{N}$ pairs of logical operators corresponding to logical $X$ and logical $Z$ for each logical qubit $\gamma = 1, \dots, \mathcal{N}$. We denote the logical operators for $\gamma$ as $\hat{g}_{x,\gamma}$ and $\hat{g}_{z,\gamma}$. The density matrix (without errors) for $RQ$ with ancillas completely entangled with the logical qubits is given by:
\begin{equation}
\begin{aligned}
    \rho_{0, RQ} &= \frac{1}{2^N} \prod_{\alpha=x, z} \prod_{\gamma = 1}^{\mathcal{N}}\frac{1}{2}(1 + K^{(\alpha)}_\gamma \hat{g}_{\alpha, \gamma}) \sum_{g_\alpha} g_\alpha \\
    &= \frac{1}{2^{\mathcal{N} +N}} \prod_{\alpha=x, z} \sum_{g_\alpha} \sum_{d_\gamma^\alpha} g_\alpha \prod_{\gamma=1}^{\mathcal{N}}(K_\gamma^{(\alpha)} \hat{g}_{\alpha, \gamma})^{d_\gamma^\alpha}
\end{aligned}
\end{equation}
where $K^{(\alpha)}_\gamma = X_\gamma, Z_\gamma$ are Pauli operators acting on the ancilla entangled with logical $\gamma$, and $d_\gamma^\alpha$ is a binary vector with $\mathcal{N}$ components indicating whether $\hat{g}_{\alpha, \gamma}$ is included in the given term or not. The sum is over all possible binary vectors of length $\mathcal{N}$. The mathematical rewriting of projectors $\frac{1}{2}(1+K_{\gamma}^{(\alpha)} \hat{g}_{\alpha,\gamma})$ in terms of $\sum_{d_{\gamma}^{\alpha} }(K_{\gamma}^{(\alpha)} \hat{g}_{\alpha,\gamma})^{d_{\gamma}^{\alpha}} $ is for later convenience. If we apply the error channel $\mathcal{E} = \mathcal{E}_x \circ \mathcal{E}_z$, the resulting density matrix is given by:
\begin{equation}
    \begin{aligned}
        \rho_{RQ} &= \frac{1}{2^{\mathcal{N}+N}} \prod_{\alpha = x, z} \sum_{g_\alpha, \mathbf{d}^\alpha} e^{-\mu_{\bar{\alpha}} \abs{g_\alpha \hat{g}_{\alpha}^{\mathbf{d}^\alpha}}} g_\alpha (K^{(\alpha)} \hat{g}_{\alpha})^{\mathbf{d}^\alpha}
    \end{aligned}
\end{equation}
%\zx{[The notation is heavy. If we write the exponential as $\exp (...)$, will it be better? Or maybe we can omit some indices?]}
where there is an implied summation over all logicals $\gamma$; we have suppressed the $\gamma$ index to avoid clutter. The $n$th moment of this density matrix is given by:
\begin{equation}
    \begin{aligned}
        \tr \rho^n_{RQ} &= \frac{1}{2^{(n-1)(\mathcal{N}+N)}} \prod_{\alpha = x, z} \sum_{\{\mathbf{d}^{(\alpha, m)}\}} \sum_{\{g^{(m)}_{\alpha}\}}  e^{-\mathcal{H}^{(n)}_{\bar{\alpha}, \mathbf{d}^{\alpha}}}\\
        \mathcal{H}^{(n)}_{\bar{\alpha}, \mathbf{d}^{\alpha}} &= - \frac{\mu_{\bar{\alpha}}}{2} \sum_{j = 1}^N \left (\sum_{m=1}^{n-1} (-1)^{\abs{(\hat{g}^{(m)}_{\alpha, j})^{\mathbf{d}^{(\alpha, m)}}}} \mathsf{E}^{(m)} \cdot \mathsf{\Omega}(P^{(\alpha, m)}_j) \right.\\
        &\left. \qquad+ (-1)^{\abs{\prod_{m=1}^{n-1} (\hat{g}^{(m)}_{\alpha, j})^{\mathbf{d}^{(\alpha, m)}}}}\prod_{m=1}^{n-1} \mathsf{E}^{(m)} \cdot \mathsf{\Omega}(P^{(\alpha, m)}_j)\right)
    \end{aligned}
\end{equation} 
where we have used the following identity:
% \begin{equation}
%     \prod_{m=1}^{n-1} g^{(m)}_\alpha (\hat{g}^{(m)}_{\alpha} K^{(\alpha, m)})^{\mathbf{d}^{\alpha, m}} = 
%     g^{(n)}_\alpha (\hat{g}^{(n)}_{\alpha} K^{(\alpha, n)})^{\mathbf{d}^{\alpha, n}}
% \end{equation}
\begin{equation}
    \abs{g_\alpha \hat{g}^{\mathbf{d}^{\alpha}}_\alpha} = \sum_{j=1}^N \frac{1}{2}\left(1 - (-1)^{\abs{\hat{g}_\alpha^{\mathbf{d}^{\alpha}}}} \mathsf{E} \cdot \mathsf{\Omega}(P^{(\alpha)}_j)\vert_g\right)
\end{equation}

The resulting SM model is different from that described by $\mathcal{H}_{\bar{\alpha}}^{(n)}$; the signs of interactions along the logicals $\gamma$ have been flipped to compensate for the added weight of $\hat{g}_{\alpha, \gamma}$. The vectors $\{\mathbf{d}^{(\alpha, m)}\}$ denote the logicals that are present in each replica. $\mathcal{H}_{\bar{\alpha}}^{(n)}$, as defined in Eq. \ref{eq:nth-SM-model}, corresponds to $\mathbf{d}^{(\alpha, m)} = \mathbf{0}$ for all $m$.

Plugging back into the full formula for the Réyni coherent information, we find:
\begin{equation}
\begin{aligned}
    I_c^{(n)}(R \rangle Q) &= \frac{1}{n-1}\log \left (\frac{1}{2^{(n-1)\mathcal{N}}} \prod_{\alpha = x, z}  \frac{\sum_{\{\mathbf{d}^{(\alpha, m)}\}}\mathcal{Z}^{(n)}_{\bar{\alpha}, \mathbf{d}^{\alpha}}}{\mathcal{Z}^{(n)}_{\bar{\alpha}, \{\mathbf{0}\}}} \right) \\
    &= \frac{1}{n-1} \sum_{\alpha = x, z} \log \left(\sum_{\{\mathbf{d}^{(\alpha, m)}\}} e^{-\Delta \mathcal{F}^{(n)}_{\bar{\alpha}, \mathbf{d}^{\alpha}}}\right) \\
    &\qquad \qquad\qquad \qquad \qquad \qquad \qquad - \mathcal{N}\log 2
\end{aligned} 
\label{eq:coherent-info}
\end{equation}
where $\Delta \mathcal{F}^{(n)}_{\bar{\alpha}, \mathbf{d}^{\alpha}}$ is the free-energy cost of inserting defects along bonds corresponding the support of the type-$\alpha$ logical operators in the theory $\mathcal{H}^{(n)}_{\bar{\alpha}}$. In the paramagnetic phase (or analogue to the paramagnetic phase) of both SM models, we expect $\Delta \mathcal{F} \sim 0$, leading to $I_c^{(n)} = \mathcal{N}\log 2$. This indicates we have the full ground state manifold available to store information. In the ordered phase of both models, $\Delta \mathcal{F}$ will scale with the size of the logical operators, which is extensive for a topologically ordered phase. This leads to $I^{(n)}_c = - \mathcal{N} \log 2$, indicative of a trivial memory.

\subsubsection{Entanglement Negativity}
\label{sec:entanglement-negativity}

The \emph{entanglement negativity} \cite{Peres_1996, Horodecki_1996, horodecki1998, Vidal_2002} is a measure of mixed state entanglement; it is a diagnostic for the separability of a mixed-state density matrix. It has been demonstrated that gapped, topological phases of matter have a universal $O(1)$ correction to their ground state entanglement negativity, called the \emph{topological entanglement negativity} \cite{lee2013}. It has also been shown that type-I fracton phases have in general linear, subextensive corrections to their entanglement \emph{entropy} \cite{Ma_2018, Shi_2018, shirleyUniversalEntanglementSignatures2019}. However, it remains to be demonstrated that the entanglement negativity generally behaves the same way for fractonic order, type-I or otherwise. We note that recently, the authors of \cite{chen2023separability} demonstrated that the 3D toric code and X-Cube models indeed undergo ``separability'' transitions at the decoding threshold.

The Réyni versions of the entanglement negativity are given by \cite{calabrese2012}:
\begin{equation}
    \mathcal{E}^{(2n)}_A(\rho) = \frac{1}{2 - 2n} \text{log}\frac{\text{tr}(\rho^{T_A})^{2n}}{\text{tr}\rho^{2n}}
\end{equation}
where $A$ is some subset of physical spins. $T_A$ denotes the partial transpose of $\rho$ on $A$. Let us recall our expression for the error-corrupted ground state density matrix, $\rho$:
\begin{equation}
    \rho =  \frac{1}{2^N} \sum_{g_{x}, g_x}e^{-\mu_z \abs{g_{x}} - \mu_x \abs{g_z}} g_x g_{z}
\end{equation}
The stabilizer configuration $g_x g_z$ is some tensor product of Pauli operators; it will be invariant under the partial transpose $T_A$ unless $g_x$ and $g_z$ overlap on $A$ such that there is a Pauli $Y$; $Y^T = -Y$. So $g_x g_z$ picks up a factor of $(-1)$ for every such overlap within $A$:
\begin{equation}
    \rho^{T_A} =  \frac{1}{2^N} \sum_{g} (-1)^{Y_A(g)} e^{-\mu_z \abs{g_{x}} - \mu_x \abs{g_z}}  g
\end{equation}
where $g = g_x g_z$ and $Y_A$ is the number of Pauli $Y$ within the region $A$. The $n$th moment of this density matrix will be the expectation value of $\mathcal{O}^{(n)}_N = \prod_{m=1}^n (-1)^{Y_A(g^{(m)})}$:
\begin{equation}
    \begin{aligned}
        \tr (\rho^{T_A})^n &= \frac{1}{2^{(n-1)N}} \sum_{\{g^{(m)}\}} \mathcal{O}^{(n)}_N  e^{-\mathcal{H}^{(n)}_z - \mathcal{H}^{(n)}_x} 
    \end{aligned}
\end{equation}
and the entanglement negativity is given by:
\begin{equation}
    \mathcal{E}_A^{(2n)}(\rho) = \frac{1}{2-2n}\log \left \langle \mathcal{O}^{(2n)}_N \right \rangle
\end{equation}
We have used the same constraint as before: $\prod_{m=1}^{n-1} g^{(m)} = g^{(n)}$. This constraint allows us to rewrite $\mathcal{O}^{(n)}$:
\begin{equation}
    \mathcal{O}^{(n)}_N = (-1)^{Y_A\left(\prod_{m=1}^{n-1} g^{(m)}\right)}\prod_{m=1}^{n-1}(-1)^{Y_A(g^{(m)})}
\end{equation}
In \cite{fanDiagnosticsMixedstateTopological2023}, the authors showed that this observable can be rewritten in the following way:
\begin{equation}
    \mathcal{O}^{(n)}_N = \prod_{m, l = 1, m\neq l}^{n-1} \text{sgn}_A (g_x^{(m)}, g_z^{(l)})
\end{equation}
where
\begin{equation}
    \mathrm{sgn}_A ({g^{(m)}_x}, {g^{(l)}_z}) = \begin{cases}
        1 & \text{if } [g^{(m)}_{x,A}, g^{(l)}_{z, A}] = 0 \\
        -1 & \text{if } \{g^{(m)}_{x,A}, g^{(l)}_{z, A}\} = 0
    \end{cases}
\end{equation}
If we assume that there are only $X$ errors, for instance, then we can simplify the evaluation of $\langle \mathcal{O}^{(n)}_N \rangle$:
\begin{equation}
\begin{aligned}
    \tr (\rho^{T_A})^n &= \frac{1}{2^{(n-1)N}} \sum_{\{g_z^{(m)}\}} \mathcal{O}^{(n)}_{N,z}  e^{-\mathcal{H}^{(n)}_z} \\
    \mathcal{O}^{(n)}_{N,z} &= \sum_{\{g^{(m)}_x\}} \mathcal{O}^{(n)}_N 
\end{aligned}
\end{equation}
For a fixed set of $\{g^{(l)}_z\}$, the sum over all $\{g^{(m)}_x\}$ will vanish unless the $\mathcal{O}^{(n)}_N$ terms add constructively. This requires that $\prod_{l \neq m} \text{sgn}_A(g^{(m)}_x, g^{(l)}_z) = 1$ for all $g^{(m)}_x$. We can satisfy this requirement if $G_z^{(m)} = \prod_{l \neq m} g^{(l)}_z$ is ``$A$-separable''. Here, we define a configuration $g$ to be $A$-separable if the restriction of $g$ to $A$ is itself a stabilizer configuration, meaning $g$ contains no stabilizers that live on the boundary $\partial A$. Then, the terms in the sum that survive are $\{g_x^{(m)}\}$ and $\{G_z^{(m)}\}$ containing only $A$-separable configurations. We can re-write $\mathcal{O}^{(n)}_{N,z}$ as follows:
\begin{equation}
    \mathcal{O}^{(n)}_{N,z} = \prod_{m = 1}^{n-1} N_x \delta_{G_z^{(m)}}
\end{equation}
where $\delta_{G_z^{(m)}}$ is zero unless $G_z^{(m)}$ is $A$-separable. $N_x$ is a numerical prefactor from the sum over $A$-separable $g^{(m)}_x$ configurations. 

The requirement of $A$-separability means that the classical spin configurations corresponding to $G^{(m)}_z$ cannot have any energetic contribution from interactions lying on the boundary $\partial A$; the expectation value of $\mathcal{O}^{(n)}_N$ is the free energy for the SM model with these interactions pinned. So we see that the Réyni entanglement negativity $\mathcal{E}^{(2n)}_A$ is connected to the excess free energy of forcing spins along the boundary $\partial A$ to be `aligned'' with respect to the SM model interaction.

% \zx{[Is any part of the derivations in this subsection different from those in Ruihua's appendix? If everything is the same or if there is only one or two changes, then we can just cite their work and highlight the difference. ]}

\subsection{SM Mapping as Ungauging}
\label{sec:ungauging}

The gauging map, first introduced by Wegner \cite{wegnerDualityGeneralizedIsing1971}, is a duality transformation from a model with some given global symmetry to a model with a corresponding gauge symmetry. For example, intrinsically topological order can be obtained by gauging a symmetry-protected topological order \cite{Levin_2012, Haegeman_2015}, and fracton phases can be constructed by gauging classical spin models with subsystem symmetries \cite{vijayFractonTopologicalOrder2016, Williamson_2016}. Here, we demonstrate that the SM mapping developed in the previous section acts as an ``un-gauging'' map for $n=2$ -- when we input a stabilizer model, the mapping returns its ungauged counterpart.

We start with a classical spin model, with a corresponding classical stabilizer map, $\mathsf{S}_c$. As the model is classical, $\mathsf{S}_c$ will have only $Z$-type stabilizers. We can view $\mathsf{S}_c$ as a map from the space of stabilizers to the space of spins on which the stabilizers have support. The quantum model that results from gauging the symmetries of $\mathsf{S}_c$ will have the following stabilizer map \cite{vijayFractonTopologicalOrder2016}:
\begin{equation}
    \mathsf{S} = \begin{pmatrix}
        \mathsf{S}^{\dagger}_c & 0 \\
        0 & \mathsf{G}
    \end{pmatrix}
\end{equation}
where $\mathsf{G}$ is a matrix whose columns are the generators of $\ker(\mathsf{S}_c)$ \footnote{Intuitively, $\mathsf{G}$ generates the collection of subsets of classical spins that can be simultaneously flipped without incurring an energy penalty, i.e., symmetries of the classical model. When we gauge the classical model to obtain the quantum one, these symmetries become allowed gauge transformations}. $\mathsf{G}$ is a map from the space of spins; it takes a spin to the element of $\ker \mathsf{S_c}$ it belongs to. The excitation map for the quantum model is given by:
\begin{equation}
        E = \begin{pmatrix}
            0 & \mathsf{S}_c \\
            \mathsf{G}^{\dagger} & 0
        \end{pmatrix}
\end{equation}

We see that for $n=2$, $X$ errors must be described by the theory $\mathsf{S}_c$, and $Z$ errors are described by $\mathsf{G}^{\dagger} = \bar{\mathsf{G}}^T$. Recall that $\bar f(x_1, \cdots, x_d) = f(x_1^{-1}, \cdots, x_d^{-1})$. We can show that $\mathsf{S}_c$ and $\mathsf{G}^{\dagger}$ are in general Kramers-Wannier dual \cite{kramers-wannier} SM models by considering a high-temperature expansion \cite{Kardar_2007} of the partition function for $\mathsf{S}_c$. Let $\{s_i\}$ be a classical spin configuration, where $i = 1, \dots, N$ indexes the sites: 
\begin{equation}
    \begin{aligned}
    \mathcal{Z}_c &= \sum_{\{s_i\}} e^{-\beta \sum_{j=1}^M \sum_{i=1}^N (\sigma_i)^{\mathsf{S}^j_c}}\\
    &= (\cosh\beta)^{MN} \sum_{\{s_i\}} \prod_{i=1}^N \prod_{j=1}^M \left ( 1 + (\sigma_i)^{\mathsf{S}^j_c} \tanh\beta \right) \\
    &= (\cosh\beta)^{MN} \sum_{\{s_i\}} \sum_{I} \sum_{\{J_i\}} (\tanh\beta)^{\sum_{i \in I} \abs{J_i}}  \prod_{i \in I}(\sigma_i)^{\sum_{j \in J_i} \mathsf{S}^j_c}
    \end{aligned}
\end{equation}
Here, $(\sigma_i)^{\mathsf{S}^j_c}$ denotes the the interaction term corresponding to the $j = 1, \dots, M$th column of $\mathsf{S}_c$, centered at site $i$. $I$ is a subset of spins, and the sum is over all possible subsets; $J_i$ is a subset of stabilizer types assigned to spin $i$, and $\{J_i\}$ is a choice of subset for each $i \in I$. Since we are summing over all possible spin configurations, the only terms that will end up contributing to the sum are those for which all $\sigma_i$ are raised to an even power (mod 2). In other words, we require that
\begin{equation}
    \sum_{i \in I} \sum_{j \in J_i} \mathsf{\Omega}(i)\mathsf{S}^j_c  = 0
\end{equation}
where $\mathsf{\Omega}(i)$ is the polynomial representing the location of $i$ with respect to some chosen origin $i_0$. We define a vector $\mathbf{g}_I$ with $M$ components:
\begin{equation}
    (\mathbf{g}_I)_j = \sum_{i \in I \mathrm{~s.t.~} j \in J_i} \mathsf{\Omega}(i)
\end{equation}
Then we have:
\begin{equation}
    \sum_j (\mathbf{g}_I)_j \mathsf{S}_c^j = 0
\end{equation}
By definition, $ \mathbf{g}_I \in \ker \mathsf{S}_c$, and so is in the image of $\mathsf{G}$. We can write the high-temperature expansion as follows:
\begin{equation}
    \mathcal{Z}_c = 2^N (\cosh\beta)^{MN} \sum_{\mathbf{g}_I}(\tanh\beta)^{\abs{\mathbf{g}_I}} 
\end{equation}
where $\abs{\mathbf{g}_I} = \sum_{i \in I}\abs{J_i}$. 

We want to find the classical model with a low-temperature expansion that also sums over $\mathbf{g}_I$. Consider the classical theory defined by $\mathsf{G}^{\dagger}$; its low-temperature expansion will involve a sum over configurations with small numbers of flipped interactions. If we flip a single spin in $\mathsf{G}^{\dagger}$, any stabilizers with support on that spin will be flipped as well. $\mathsf{G}$ takes spins as input and returns the stabilizers that they belong to--- $\mathsf{G}$ gives us the information needed for the low-temperature expansion. The image of $\mathsf{G}$ is exactly the set of all $\mathbf{g}_I$, and so we see that the low-temperature expansion of $\mathsf{G}^{\dagger}$ is the high-temperature expansion of $\mathsf{S}_c$.

We note further that in the special case $\mathsf{S}_c = \mathsf{G}^{\dagger}$, i.e. the classical spin model $\mathsf{S}_c$ is self-dual, that the quantum model obtained upon gauging will have $em$ duality. Conversely, a given stabilizer model with $em$ duality will map to a statistical mechanics model that is its own Kramers-Wannier dual. The $em$ duality is clearly seen from the fact $X$ and $Z$ stabilizers will have the exact same geometry, as they are described by the same model.

\subsection{SM Mapping Through A Homological Lens}
\label{sec:homology}

Using the homological perspective on CSS codes, we can construct the general SM mapping in a different, but equivalent way. We assume the reader is familiar with homology theory and its applications to understanding stabilizer codes--- see any of the following for a more thorough introduction: \cite{browne-notes, mittal-notes, rakovszkyPhysicsGoodLDPC2023, palPhysicistIntroductionAlgebraic2019}.

A given CSS code can be described as a $2$-chain complex $\mathcal{C}$, built from two underlying classical codes $H_Z$ and $H_X$ \cite{Bombin_2007, bravyiHomologicalProductCodes2013, kitaevQuantumComputationsAlgorithms1997a}:
\begin{equation}
    \mathcal{C} = C_2 \xrightarrow{H_Z^T} C_1 \xrightarrow{H_X} C_0
\end{equation} 
where $C_2 \sim \mathbb{F}^m_2$ is the space of $Z$ checks, $C_1 \sim \mathbb{F}^n_2$ is the space of physical bits, and $C_0 \sim \mathbb{F}^k_2$ is the space of $X$ checks \footnote{ Here, $\mathbb{F}_2 = \{0, 1\}$ is the field over two elements.}. The map $H_Z^T : C_2 \rightarrow C_1$ takes a $Z$ stabilizer $S^Z_i$ to the subset of physical bits it acts on:
\begin{equation}
    S^Z_i = \prod_{j \in H_Z^T(i)} Z_j
\end{equation}

The map $H_X : C_1 \rightarrow C_0$ takes a physical bit to the set of $X$ checks $S^X_l$ with support on that bit:
\begin{equation}
    S^X_l = \prod_{j ~st.~ l \in H_X(j)} X_j
\end{equation}

The condition that $H_Z^T$ and $H_X$ be valid boundary maps is equivalent to the condition that $X$ and $Z$ checks must commute: $H_X H_Z^T = 0$.

The parent SM models derived in section \ref{sec:generic-mapping} can also be obtained from $H_X$ and $H_Z^T$, since they also contain the geometric information about stabilizers. We re-interpret $C_2$ and $C_0$ as the spaces corresponding to our classical spins. $X$ errors on a link $j$ will generate an interaction between all classical spins assigned to $Z$ stabilizers with support on $j$. This is described simply by the map $H_Z : C_1 \rightarrow C_2$. Similarly, $Z$ errors on a link $j$ will generate an interaction between all classical spins assigned to $X$ stabilizers with support on $j$. This is described by the map $H_X : C_1 \rightarrow C_0$. Thus, we have discovered that the two underlying classical codes for a CSS model describe its behavior under single-site Pauli decoherence. This parallels the view of the SM mapping as ungauging; here, the SM model unzips the chain complex and returns the constituent classical codes.

\subsection{Taking the Replica Limit and the Connection to Maximum-Likelihood Decoding}
\label{sec:replica}

So far, we have focused on deriving SM models that describe the critical behavior of various Réyni information quantities. While these provide bounds on the actual threshold of the code, they are not themselves physical. We need to take the replica limit $n \rightarrow 1$ to get the tightest bounds on decoding thresholds. It is not clear how to take this limit using the formalism developed so far. However, we can take an alternate approach inspired by the Kitaev-Preskill decoder to get a handle on the replica limit, which the authors of \cite{fanDiagnosticsMixedstateTopological2023} call the ``error-string'' picture. 

Instead of constructing the ground state density matrix by applying the error channel to each edge individually, we apply all possible error strings to the clean ground state, weighted by their probability of occurring. Consider some pattern of $X$ errors, $\mathcal{C}_x$ supported on a subset of the physical bits; the effect of all such error strings on the clean density matrix $\rho_0$ will be:
\begin{equation}
    \rho = \sum_{\mathcal{C}_x} \mathcal{P}(\mathcal{C}_x) {\mathcal{C}_x} \rho_0 {\mathcal{C}_x},
\end{equation}
where $\mathcal{P}(\mathcal{C}_x) = p_x^{\abs{\mathcal{C}_x}}(1-p_x)^{N -\abs{\mathcal{C}_x}}$. Again, we want to find a SM mapping for the $n$th moment of the density matrix:
\begin{equation}
    \mathrm{tr}\rho^n = \sum_{\mathcal{C}_x^{(m)}} \bigg(\prod_{m=1}^n \mathcal{P} (\mathcal{C}_x^{(m)})\bigg)\mathrm{ tr}\left( \prod_{m=1}^n {\mathcal{C}_x^{(m)}} \rho_0 {\mathcal{C}_x^{(m)}} \right) 
\end{equation}

Let $\rho_0 = \ket{\Psi_0}\bra{\Psi_0}$ be a pure eigenstate of the logical operators rather than the completely mixed density matrix. Then:
\begin{equation}
\begin{aligned}
    \mathrm{ tr}\left( \prod_{m=1}^n {\mathcal{C}_x^{(m)}} \rho_0 {\mathcal{C}_x^{(m)}} \right) &= \mathrm{ tr}\left( \prod_{m=1}^n {\mathcal{C}_x^{(m)}} \ket{\Psi_0}\bra{\Psi_0} {\mathcal{C}_x^{(m)}} \right)\\
    &= \prod_{m=1}^{n-1} \bra{\Psi_0} {\mathcal{C}_x^{(m)}} {\mathcal{C}_x^{(m+1)}} \ket{\Psi_0}
\end{aligned}
\end{equation}

This will only be nonzero if the operators inside do not create any excitations or any logical operators; the error strings $\mathcal{C}_x^{(m)}$ and $\mathcal{C}_x^{(m+1)}$ must multiply to an $X$ stabilizer configuration $g_x$. We can then express the error string on each copy in terms of the first copy and the stabilizer configuration $g^{(m)}$ needed to create $X^{\mathcal{C}_x^{(m)}}$:
\begin{equation}
    {\mathcal{C}_x^{(m)}} = {\mathcal{C}_x^{(1)}}g_x^{(m)}
\end{equation}
The $n$th moment of the density matrix now becomes:
\begin{equation}
\begin{aligned}
    \mathrm{tr}\rho^n &= \sum_{\mathcal{C}_x^{(1)}} \mathcal{P}(\mathcal{C}_x^{(1)}) \sum_{\{g_x^{(m)}\}} \prod_{m=1}^{n-1} \mathcal{P}(\mathcal{C}_x^{(1)}g_x^{(m)}) \\
    &= (1-p_x)^N  \sum_{\mathcal{C}_x^{(1)}} \mathcal{P}(\mathcal{C}_x^{(1)}) \sum_{\{g_x^{(m)}\}} e^{-2J \sum_{m=1}^{n-1} \abs{\mathcal{C}_x^{(1)}g_x^{(m)}}}
\end{aligned}
\end{equation}
where $e^{-2J} = p_x/(1-p_x)$. As with the first mapping, our goal is to write down an SM model Hamiltonian that properly calculates the value of $\abs{\mathcal{C}_x^{(1)} g_x^{(m)}}$--- this will allow us to interpret $\tr \rho^n$ as a partition function. We already know the SM model for $\abs{g_x}$, we just need to modify it to account for the additional factor of $\mathcal{C}^{(1)}_x$. Where $\mathcal{C}^{(1)}_x$ and $g^{(m)}_x$ overlap, the weight of $g^{(m)}_x$ will be removed, but where $\mathcal{C}^{(1)}_x$ is supported and $g^{(m)}_x$ is not, there will be weight added. We can account for this change by flipping the sign of the classical spin interactions from ferromagnetic to antiferromagnetic along the links spanned by $\mathcal{C}_x^{(1)}$:
\begin{equation}
\begin{aligned}
\tr \rho^n &= (1-p_x)^N  \sum_{\mathcal{C}_x^{(1)}} \mathcal{P}(\mathcal{C}_x^{(1)}) \sum_{\{g_x^{(m)}\}} e^{-\mathcal{H}^{(n)}_{\mathcal{C}^{(1)}_x}}\\
\mathcal{H}^{(n)}_{\mathcal{C}^{(1)}_x} &=  -J \sum_{j=1}^N (-1)^{\abs{\mathcal{C}^{(1)}_{x,j}}}\left (\sum_{m=1}^{n-1} \mathsf{E}^{(m)} \cdot \mathsf{\Omega}(Z^{(m)}_j) \right.\\
&\left.\qquad\qquad\qquad\qquad\qquad+ \prod_{m=1}^{n-1} \mathsf{E}^{(m)} \cdot \mathsf{\Omega}(Z^{(m)}_j)\right)
\end{aligned}
\end{equation}
We are counting the weight of $g_x$ stabilizers, so the interaction term is determined by where $Z_j$ anti-commutes with the given configuration. So the $n$th moment of $\rho$ is described by the following partition function:
\begin{equation}
\begin{aligned}
    \tr \rho^n &= (1-p_x)^N \sum_{\mathcal{C}_x^{(1)}} \mathcal{P}(\mathcal{C}_x^{(1)}) \mathcal{Z}^{(n)}_{\mathcal{C}_x^{(1)}} \\
    &\equiv\ (1-p_x)^N \bar{\mathcal{Z}}^{(n)}.
\end{aligned}
\end{equation}
$\bar{\mathcal{Z}}^{(n)}$ looks like a partition function for a random bond version of the SM model defined in Eq. \ref{eq:nth-SM-model}, averaged over all possible disordered realizations of random bonds. Taking the replica limit is clearer in this case than the previous; it leads to a random-bond SM model for a single copy, with quenched disorder (no disorder-averaging). As pointed out in \cite{fanDiagnosticsMixedstateTopological2023}, this is exactly the SM model obtained via the Kitaev-Preskill maximum-likelihood decoder \cite{Dennis_2002}, indicating this is indeed the optimal decoder.

We note that the SM model for $X$ errors obtained in the error-string picture will be the random-bond version of the SM model obtained for $Z$ errors in the ``stabilizer'' picture considered before; in the limit of zero disorder, these models should be dual to each other (as shown in section \ref{sec:ungauging}). Additionally, in the replica limit, both models should exhibit transitions at the same point, since they describe the same density matrix. 

\section{Examples}
\label{sec:examples}

\begin{table*}[t]
    \centering
    \begin{tabular}{|c|c|c|c|c|}
    \hline
    \hline
        \textbf{Stabilizer Model} & $\mathcal{H}^{(2)}_X$ & $p^{(2)}_{X, c}$ & $\mathcal{H}^{(2)}_Z$ & $p^{(2)}_{Z, c}$ \\
        \hline
        \hline
        2D toric code \cite{Kitaev_2003, fanDiagnosticsMixedstateTopological2023} & 2D Ising Model & 0.178 & 2D Ising Model & 0.178 \\
        \hline
        3D toric code & 3D $\mathbb{Z}_2$ Gauge Theory \cite{wegnerDualityGeneralizedIsing1971} & 0.266 & 3D Ising model \cite{HASENBUSCHMARTIN2001MCSO} & 0.099\\
        \hline
        X-Cube model \cite{vijayFractonTopologicalOrder2016} & Anisotropically-Coupled Ashkin-Teller model \cite{johnstonDualGonihedric3D2011} & 0.336 & Plaquette Ising model \cite{johnstonGonihedric3DIsing1996} & 0.213\\
        \hline 
        Cubic code \cite{Haah_2011} & Fractal Ising model \cite{vijayFractonTopologicalOrder2016} & 0.178 & Fractal Ising model & 0.178 \\
         \hline
         \hline
    \end{tabular}
    \caption{\textbf{Summary of SM mappings for $n =2$:} We list various CSS stabilizer models considered in the text, along with their corresponding $n=2$ SM models and thresholds for $X$ and $Z$ errors. Where necessary, the sources for numerical data on the different SM models are cited.}
    \label{tab:SM_summary}
\end{table*}

In this section, we work out some representative examples of the mapping outlined in the section \ref{sec:generic-mapping}. We return to the 3D toric code, and also consider the X-Cube model \cite{vijayFractonTopologicalOrder2016}. We derive the SM models for both $X$ and $Z$ errors, and examine in depth the behavior of the different information quantities for each model. 

\subsection{3D Toric Code: Redux}
\label{sec:3dtc}
We have already analyzed the 3D toric code in section \ref{sec:warm-up}. Now we can return to this model armed with the general SM mapping toolkit we developed in section \ref{sec:generic-mapping}. The stabilizer and excitation maps for the 3D toric code are:

\begin{equation}
\begin{aligned}
    \mathsf{S}_{3DTC} &= \begin{pmatrix}
        1 + y & 1 + z & 0 & 0\\
        1 + x & 0 & 1+ z & 0\\
        0 & 1+x & 1 + y & 0\\
        0 & 0 & 0 & 1 + \bar{x}\\
        0 & 0 & 0 & 1 + \bar{y}\\
        0 & 0 & 0 & 1 + \bar{z} \\
        \end{pmatrix}\\
    \mathsf{E}_{3DTC} &= \begin{pmatrix}
        0 & 0 & 0 & 1 + x & 1 + y & 1+ z \\
        1 + \bar{y} & 1 + \bar{x} & 0 & 0 & 0 & 0\\
        1 + \bar{z} & 0 & 1 + \bar{y} & 0 & 0 & 0 \\
        0 & 1+ \bar{z} & 1 + \bar{y} & 0 & 0 & 0
    \end{pmatrix}
\end{aligned}
\label{eq:3dtc-polynomial}
\end{equation}

Consider the single site $Z$ error corresponding to the vector $\mathbf{\omega}_z = \begin{pmatrix} 0 & 0 & 0 & 1 & 0 & 0 \end{pmatrix}^T$, which produces the excitation pattern $\mathsf{E}_{3DTC} \cdot \mathbf{\omega}_z = \begin{pmatrix} 1 + x & 0 & 0 & 0 \end{pmatrix}^T$. This is exactly an Ising interaction between the vertex $v$ and the vertex $v + \hat{x}$. $Z$ errors acting on the other two edges in the unit cell will product the interactions between $v$ and its other nearest neighbors in the $\hat{y}$, $\hat{z}$ directions.

 Now, consider the $X$ error given by $\mathbf{\omega}_x = \begin{pmatrix} 1 & 0 & 0 & 0 & 0 & 0 \end{pmatrix}^T$. This corresponds to interaction term $E_{3DTC} \cdot \mathbf{\omega}_x = \begin{pmatrix} 0 & 1+\bar{y} & 1+\bar{z} & 0 \end{pmatrix}^T$, which can be identified as the four-body interaction in $\mathbb{Z}_2$ gauge theory on the cubic lattice. 

Numerical studies place the inverse critical temperature of the 3D Ising model at $\beta_c = 0.2217$ \cite{HASENBUSCHMARTIN2001MCSO}. Using $-\ln(1-2p_c) = 2 \beta_c$, we find that the critical error rate for the Réyni-2 quantities is $p^{(2)}_{c, Z} = 0.099$. By Kramers-Wannier duality, the critical inverse temperature for $\mathbb{Z}_2$ lattice gauge theory is $\beta_c = 0.76$, which gives a threshold of $p^{(2)}_{c, X} = 0.266$. These are upper bounds \cite{Muller_Lennert_2013} for the optimal thresholds $p_{c, X}$ and $p_{c, Z}$, which are connected to the critical temperatures of the random-bond 3D Ising model (RBIM) and random-plaquette 3D $\mathbb{Z}_2$ gauge theory (RPGM) along the Nishimori line, respectively (as the roles of 3D Ising model and gauge theory switch when going to the error-string picture). The 3D RBIM has a transition at $p_{c, X} = 0.233$ \cite{Ozeki1987PhaseDA, Hasenbusch_2007}, while the RPGM has a transition at $p_{c, Z} = 0.033$ \cite{wangConfinementHiggsTransitionDisordered2003, ohnoPhaseStructureRandomPlaquette2004}. 

For $n=3$, the SM model for phase errors will look like a two-flavor Ising model with an inter-flavor coupling term:
\begin{equation}
    \mathcal{H}^{(3)}_z = - \frac{\mu_z}{2} \sum_{\langle ij \rangle} \sigma^{(1)}_i  \sigma^{(1)}_j + \sigma^{(2)}_i \sigma^{(2)}_j + \sigma^{(1)}_i  \sigma^{(1)}_j \sigma^{(2)}_i \sigma^{(2)}_j
\end{equation}
where $\langle ij \rangle$ denotes nearest-neighbors on the cubic lattice. This is the Hamiltonian for the 3D isotropic Ashkin-Teller model \cite{ashkin-teller, FAN1972136}, which is equivalent to the 3D four-state Potts model.

For bit-flip errors, the $n=3$ model is a variation of $\mathbb{Z}_2$ gauge theory:
\begin{equation}
    \mathcal{H}^{(3)}_x = - \frac{\mu_z}{2} \sum_{j=1}^N \prod_{j \in p} \sigma^{(1)}_j  + \prod_{j \in p} \sigma^{(2)}_j + \prod_{j \in p} \sigma^{(1)}_j \sigma^{(2)}_j
\end{equation}
where $p$ denotes a plaquette of the cubic lattice. As far as we are aware, this model has not yet been studied.

 \subsubsection{Relative Entropy}

 We showed in section \ref{sec:relative-entropy} that the Rényi relative entropy, $D^{(n)}$ (see Eq. \ref{eq:rel-entropy-original}), maps onto a generalized spin-spin correlator in the SM models $\mathcal{H}^{(n)}_{x, z}$. We can read off the form of this correlator from the excitation map $\mathsf{E}$ (see Eq. \ref{eq:rel-entropy}). More intuitively, 
 the correlator will include all the classical spins corresponding to stabilizers excited by the additional error string in $\rho_\epsilon$. 
 
 For simplicity, we will focus on Réyni-2 quantities. We first consider the 3D toric code under $Z$ errors; a $Z$ string creates two $e$ anyons at either end. As the corresponding SM model is the 3D Ising model, the relative entropy will be related to the usual 3D Ising two-spin correlator.
 \begin{equation}
     D^{(2)} = \log \langle \sigma_i \sigma_j \rangle 
 \end{equation}
 where $i, j$ index the vertices at either end of the $Z$ string. In the paramagnetic phase of the 3D Ising model, this correlator decays exponentially with $\abs{i-j}$, and so $D^{(2)} \sim \abs{i - j}$. By comparison, in the ferromagnetic phase of the 3D Ising model, this correlator is $O(1)$, so loss of the topological order can be detected by the significant decrease of $D^{(2)}$ when $i, j$ are far apart.  

 For $X$ errors, we need to consider the 3D $\mathbb{Z}_2$ lattice gauge theory. An $X$ string will create an $m$-loop; there will be an extensive number of excited plaquettes along the string. If we map these plaquettes to their corresponding dual edges, we see that the necessary correlator is actually a Wilson loop, $W$:
 \begin{equation}
     D^{(2)} = \log \left \langle \prod_{i \in W} \sigma_i \right \rangle 
 \end{equation}
At low temperatures, this has a perimeter-law scaling \cite{wegnerDualityGeneralizedIsing1971}, so $D^{(2)} \sim |W|$. At high temperatures, the Wilson loop correlation function has area-law scaling, meaning $D^{(2)} \sim \abs{A(W)}$. On either side of the transition, $\rho$ and $\rho_\epsilon$ are distinguishable; the only thing that changes is the scaling of the relative entropy with $m$-loop size. This connected to the fact $m$ anyons are confined in the 3D toric code, which is a self-correcting memory under $X$ errors by themselves.

 \subsubsection{Coherent Information}

 \begin{figure}
     \centering
     \includegraphics[width=0.75\linewidth]{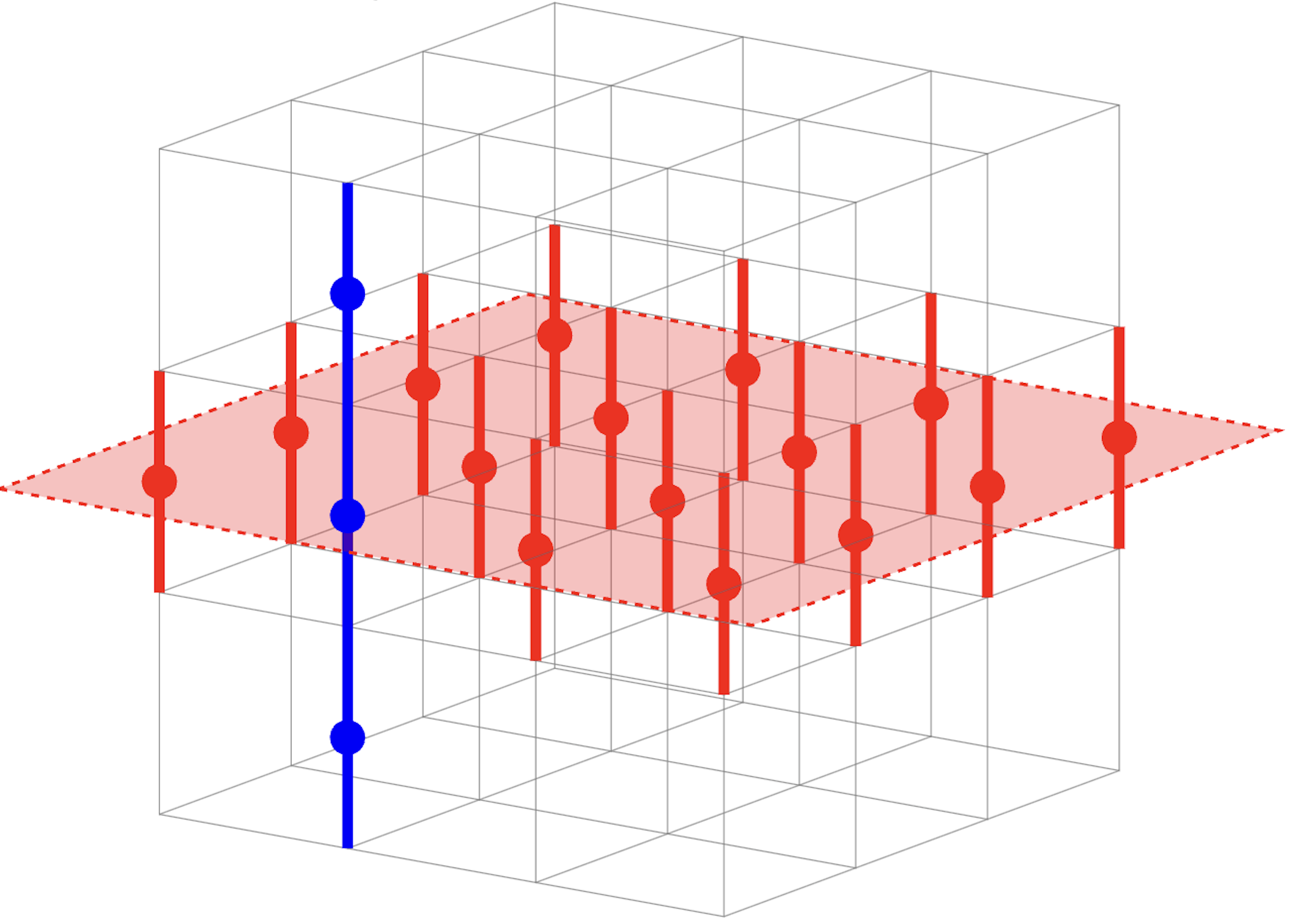}
     \caption{\textbf{3D Toric Code Logicals: } The two kinds of logical operators for the 3D toric code are pictured; the string-like logical $Z$ is in blue, while the membrane-like logical $X$ is pictured in red. These correspond to defects inserted into the 3D Ising model and $\mathbb{Z}_2$ lattice gauge theory, respectively. In the 3D Ising model, the coherent information is connected to the excess free energy of inserting a line of antiferromagnetic interactions along the logical $Z$ string. In the Ising gauge theory, we instead flip the sign of gauge plaquette interactions in the shaded red plane (one plaquette per edge in the logical).}
     \label{fig:3dtc-info}
 \end{figure}

 In section \ref{sec:coherent-info}, we demonstrated that the Réyni coherent information $I_c^{(n)}$ is related to the excess free energy associated with flipping the sign of interactions in the SM model $\mathcal{H}^{(n)}_\alpha$ along logicals of the original quantum code (see Eq. \ref{eq:coherent-info}). Again, we will focus on $I_c^{(2)}$ for simplicity. 

We note that the $Z$ logicals for the 3D toric code are string-like, while the $X$ logicals are membrane-like and extend along 2D planes of the lattice (see Fig. \ref{fig:3dtc-info}). First, we consider the case where both the 3D Ising model and $\mathbb{Z}_2$ gauge theory are in their high-temperature phases. For the 3D Ising model in the paramagnetic phase, the spins are completely disordered, and there is no cost to inserting a domain wall anywhere: $\Delta \mathcal{F}_Z = 0$. For the $\mathbb{Z}_2$ gauge theory, a ``domain'' wall corresponds to a plane of plaquettes with flipped interaction signs. Similarly to the paramagnetic Ising phase, $\Delta \mathcal{F}_X \sim 0$ in the area-law (high-temperature) phase, due to the large number of fluctuations. Essentially, the flipped interactions are irrelevant. Plugging into Eq. \ref{eq:coherent-info}, we find $I^{(2)}_c = 3\log 2$, indicating we have lost no information capacity (recall $\mathcal{N}=3$ is the number of logical qubits that can be encoded in the ground state manifold). 

Now consider the scenario where both models are instead in their low-temperature phases. For the 3D Ising model in the ferromagnetic phase, the free energy cost of a domain wall scales linearly with its length, since there is no entropy gain at all due to the low number of fluctuations. For $\mathbb{Z}_2$ lattice gauge theory in the perimeter law phase, the free energy cost of a flipped plane of plaquettes will scale with the area of the plane, for the same reason. In this case, we have $\Delta \mathcal{F}_Z \sim N^{\frac{1}{3}} \rightarrow \infty$ and $\Delta \mathcal{F}_X \sim N^{\frac{2}{3}} \rightarrow \infty$ in the thermodynamic limit. Plugging in, we find $I^{(2)}_c = -3 \log 2$, indicating we have lost all quantum information in the error channel.

What if we are at error rates $p_x$ and $p_z$ such that the Ising model is in the paramagnetic phase, while the $\mathbb{Z}_2$ gauge theory is still in the perimeter law phase? This is possible, since the 3D Ising model has a lower critical temperature. In this case, the free energy contribution from the Ising model is 0, while the contribution from the gauge theory goes to $\infty$ in the thermodynamic limit. We end up with $I^{(n)}_c = 0$, indicating we can no longer send quantum information through the error channel, but classical information can still be preserved.

 \subsubsection{Entanglement Negativity}

 \begin{figure}
     \centering
     \includegraphics[width=\linewidth]{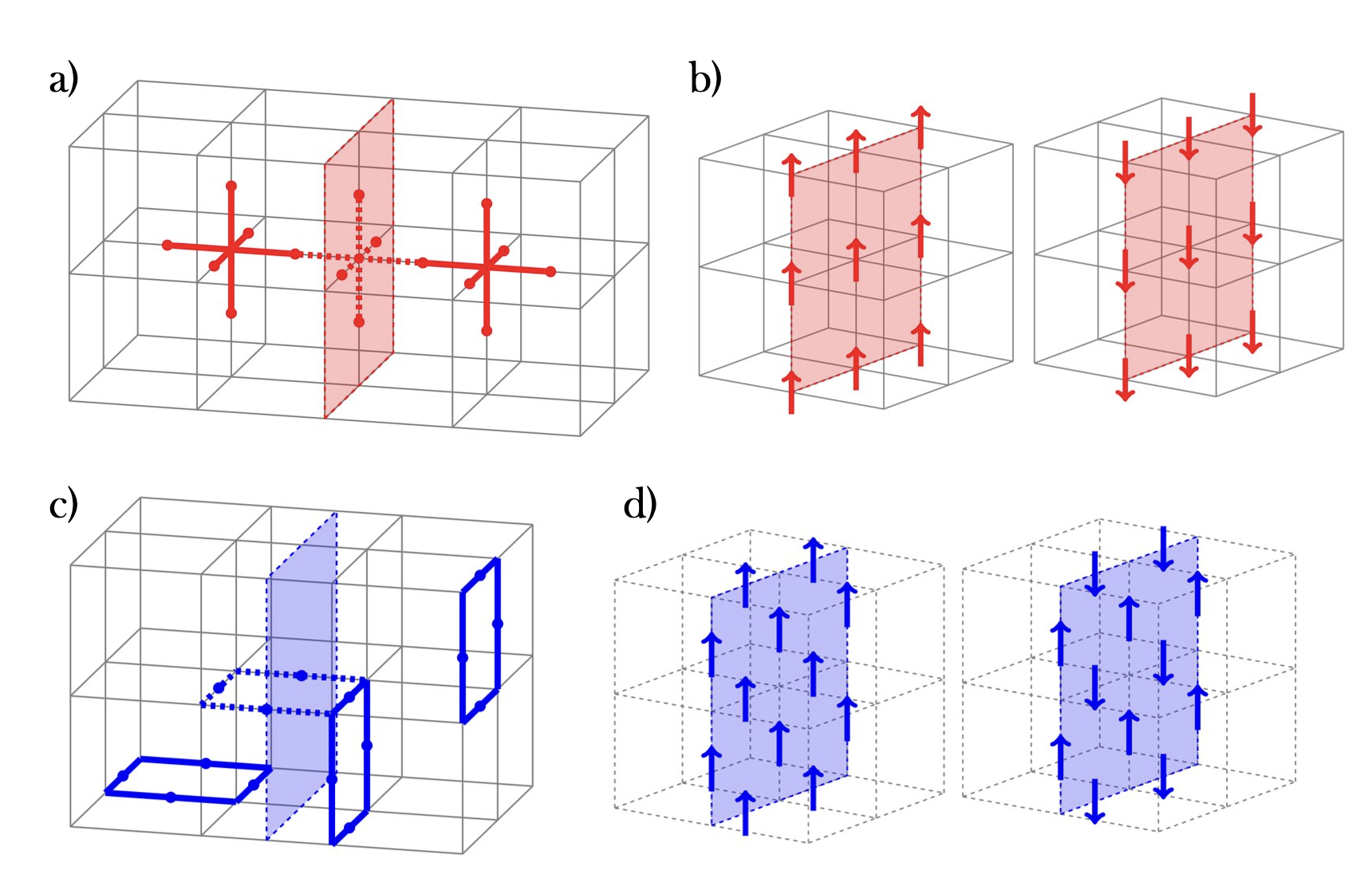}
     \caption{\textbf{Entanglement Negativity in the 3D Toric Code: }a) For $Z$ errors, the boundary $\partial A$ is shaded in red. Allowable vertex terms in $g_x$ are depicted with solid lines, while forbidden vertex terms that cross $\partial A$ are drawn with dashed lines. b) The corresponding allowable classical spin configuration are drawn; the spins along $\partial A$ must all be aligned. In the paramagnetic phase, the spins can fluctuate collectively. b) For $X$ errors, a rough boundary, or a boundary on the dual lattice is more convenient-- an example is shown shaded in blue. Allowed $g_x$ are drawn in solid lines, while $g_x$ that cross the boundary are in dashed lines. This constraint corresponds to pinning the gauge interactions on the shaded plane to be +1. d) Two allowable spin configurations are shown on the right; these represent the two available states in the paramagnetic phase (after gauge-fixing).}
     \label{fig:3dtc-negativity}
 \end{figure}

 The Réyni entanglement negativity $\mathcal{E}^{(2n)}_A$ is connected to the excess free energy of fixing the sign of interactions on the boundary of some region $A$. For simplicity, we consider the case where $\partial A$ is a bipartition of the cubic lattice. The case of the 3D toric code is similar to the 2D case, and we use the same scaling arguments that can be found in \cite{fanDiagnosticsMixedstateTopological2023}. For phase errors, we want to enforce the constraint that no $g_x$ (wireframe) configurations cross the boundary $\partial A$. Accordingly, $g_x$ cannot have support on the surface of $\partial A$, meaning the two-body Ising terms between classical spins living on the surface must be positive. In other words, the spins on the boundary $\partial A$ must all be aligned (see fig. \ref{fig:3dtc-negativity}a). This generically gains energy $\abs{\partial A}$. In the paramagnetic phase, spins are free to fluctuate above the correlation length $\xi$, so fixing the spins on the boundary removes entropy of order $\abs{\partial A}/\xi$. There is a residual freedom to choose which way the aligned spins point, however, giving a sub-leading correction $-\log 2$. In the ferromagnetic phase, spins can fluctuate below the scale $\xi$, so fixing them also removes $\abs{\partial A}/\xi$ degrees of freedom, since we can fit $\abs{\partial A}/\xi$ patches of fluctuating spins in the area $\abs{\partial A}$. There is no sub-leading correction, since the aligned surface spins must match up with the bulk spins.
 
 It is interesting to note that these calculations show a finite phase error transition for the 3D toric code entanglement negativity; previous studies have considered a finite-temperature transition and demonstrated the negativity disappears for any $T>0$ \cite{Lu_2020}. Intuitively, the decoherence due to local error channel is ``weaker'' than the finite-temperature channel, and so the 3D toric code is able to self-correct up to some finite-error rate. 

In the case of bit-flip errors, we now want to enforce the constraint that no $g_z$ (2D loop) crosses the boundary $\partial A$. This means $g_z$ cannot have support on any of the edges normal to the boundary. This forces the gauge interactions assigned to these edges to be positive, i.e. all the plaquette interactions on the dual lattice surface $\partial \tilde{A}$ must be +1 (see fig. \ref{fig:3dtc-negativity}b). Again, the energetic contribution to the free energy is always $\sim \abs{\partial A}$. In the high-temperature phase, a Wilson loop living on this surface should have an exponentially decaying correlation function $\sim e^{-A/\xi}$--- on scales larger than the correlation length $\xi$, the plaquettes are free to fluctuate, as so by pinning them we are removing $\abs{\partial A/\xi}$ degrees of freedom, as in the 3D Ising model. However, also like the 3D Ising model, the spins on the surface are free to fluctuate together. The two possibilities in this case can be seen by performing a ``temporal'' gauge transformation such that all spins along one direction in the plane are pointing up \cite{kogutIntroductionLatticeGauge1979}. The residual freedom leftover is whether the spins in the other direction point up or down, and so the sub-leading, topological, correction is $-\log 2$.

In the low-temperature phase, a Wilson loop on the surface $\partial A$ has correlation function $\sim e^{-L/\xi}$, where $L$ is the perimeter. Plaquette violations (magnetic monopoles) are confined, but free to proliferate below the correlation length, $\xi$. By pinning, we lose $\abs{\partial A/\xi}$ degrees of freedom, as before. However, there is no sub-leading correction, since the spins left free after the temporal gauge transformation must be aligned with the bulk spins away from the surface.

\subsection{X-Cube Model}

\begin{figure}
    \includegraphics[width=\linewidth]{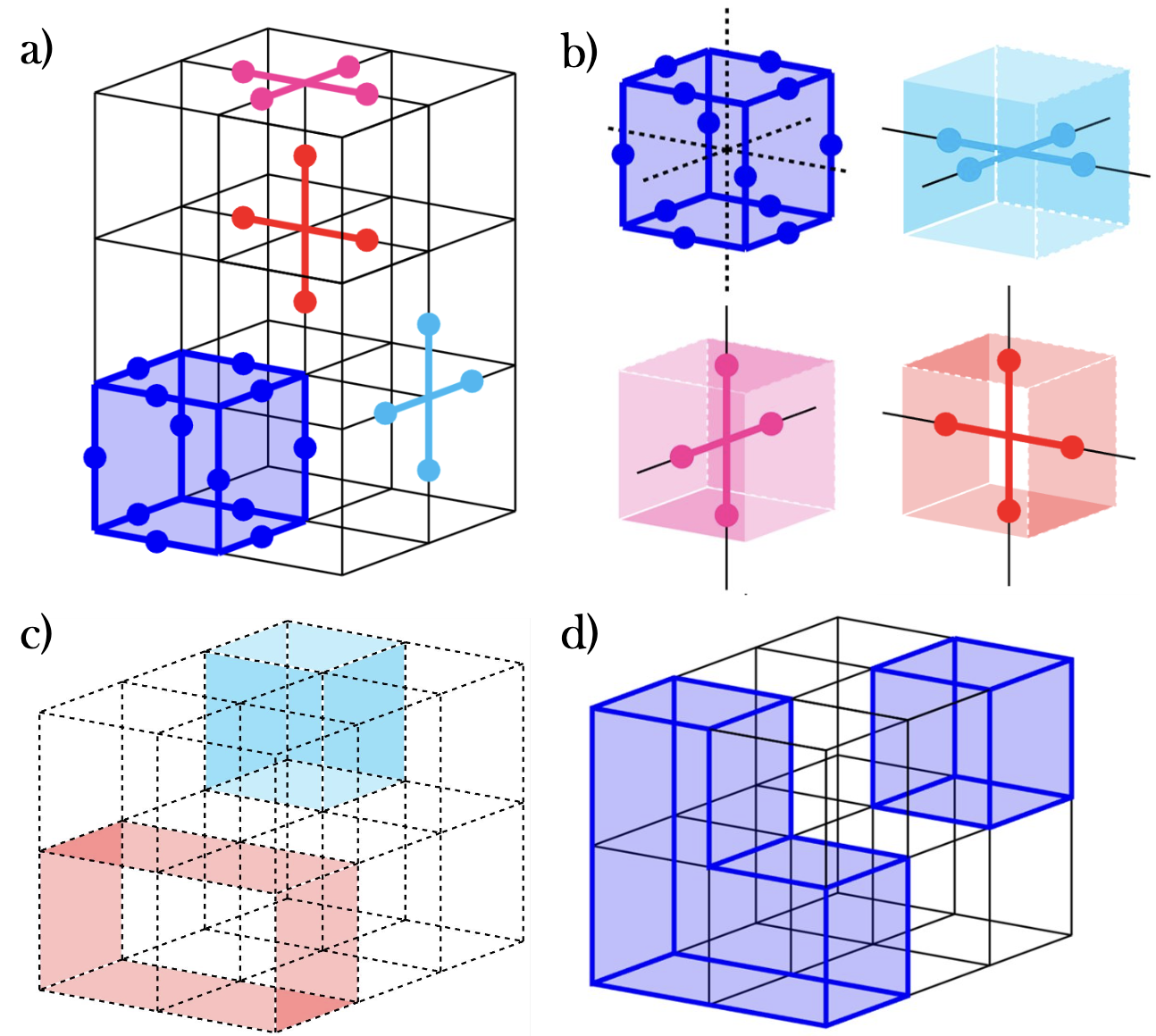}
    \caption{\textbf{X-Cube Model:} a) The stabilizers as defined on the cubic lattice for the X-Cube model. $A_c$ is pictured in dark blue, while the three flavors of vertex operator are pictured in light blue, red, and pink, respectively. b) The mapping from stabilizers on the direct lattice to stabilizers on the dual lattice. The cube term becomes a vertex on the dual lattice, while the three vertex terms are mapped to four plaquettes each, forming an ``open box''. c) An example stabilizer configuration $g_z$, which consists of ``open boxes'' on the dual lattice. d) An example stabilizer configuration $g_x$, which is a set of closed wireframes on the direct lattice.}
    \label{fig:xcube}
\end{figure}

We consider the X-Cube model, first introduced in \cite{vijayFractonTopologicalOrder2016}. This is a type-I fracton order, meaning it has immobile fractonic excitations as well as excitations mobile in a plane or along a line. We consider the X-Cube model defined on the cubic lattice with vertices $v \in V$, edges $e \in E$, plaquettes $p \in P$, and cubes $c \in C$. The Hamiltonian is the following:
\begin{equation}
    H_{\text{XCube}} = - \sum_{v, \mu = \hat{x}, \hat{y}, \hat{x}} B^{\mu}_v - \sum_c A_c
\end{equation}
where the stabilizers are given by:
\begin{equation}
    \begin{aligned}
        A_c &= \prod_{e \in \partial c} X_e \\
        B^{\mu}_v &= \prod_{\partial e \ni v, ~e \perp \mu} Z_e
    \end{aligned}
\end{equation}

Here, $e \in \partial c$ denotes all edges making up the boundary of a cube $c$, while $e \perp \mu$ indicates the edge $e$ lies in the plane normal to the unit vector $\mu$. See Fig. \ref{fig:xcube}a, b for a visualization of the stabilizers on the cubic lattice. The lowest energy excitations of the X-Cube model are the \emph{fractons}, which are completely immobile, and the \emph{lineons}, which are allowed to move along a line. Fractons and lineons can be combined in pairs to form fracton- or lineon-dipoles which are free to move in the 2D plane normal to their dipole moment. These are sometimes called \emph{planons}. A single Pauli $Z$ creates four fractons (or equivalently, a pair of fracton dipoles) on the neighboring cubes, while a single Pauli $X$ creates a pair of lineons.

We construct the stabilizer and excitation maps from the X-Cube Hamiltonian:
\begin{widetext}
\begin{equation}
    \mathsf{S}_{\text{XCube}} = \begin{pmatrix}
        1+\bar{x} & 1+\bar{x} & 0 & 0\\
        1+\bar{y} & 0 & 1+\bar{y} & 0\\
        0 & 1+\bar{z} & 1+\bar{z} & 0\\
        0 & 0 & 0 & 1 + y + z + yz \\
        0 & 0 & 0 & 1 + x + z + xz\\
        0 & 0 & 0 & 1 + x + y + xy
    \end{pmatrix}
\end{equation}

\begin{equation}
    \mathsf{E}_{\text{XCube}} = \begin{pmatrix}
        0 & 0 & 0 & 1+\bar{y}+\bar{z}+\bar{y}\bar{z} & 1+\bar{x}+\bar{z}+\bar{x}\bar{z} & 1+\bar{x}+\bar{y}+\bar{x}\bar{y}\\
        1+x & 1+y & 0 & 0 & 0 & 0 \\
        1+x & 0 & 1+z & 0 & 0 & 0  \\
        0 & 1+y & 1+z & 0 & 0 & 0
    \end{pmatrix}
\end{equation}
\end{widetext}

\begin{figure}
    \centering
    \includegraphics[width=1\linewidth]{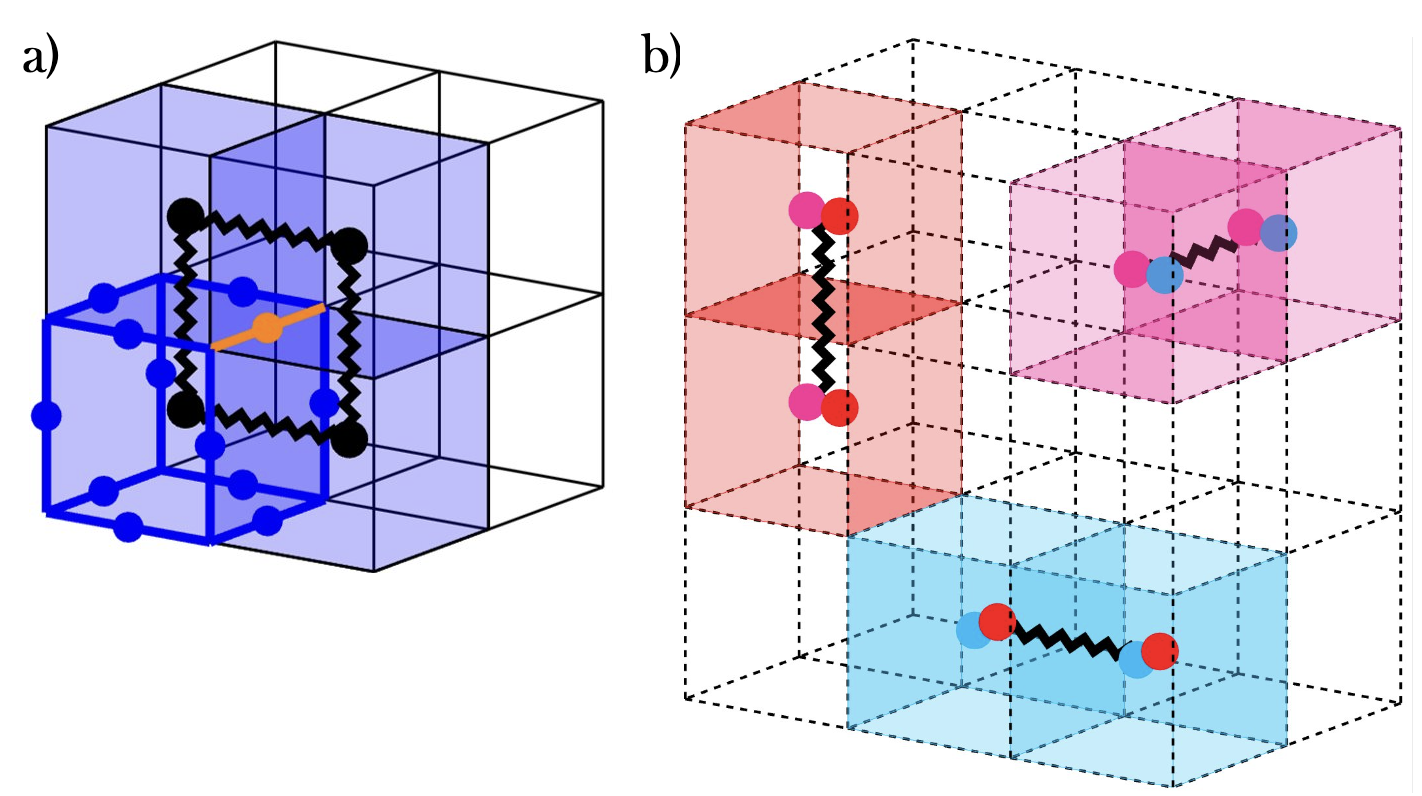}
    \caption{\textbf{SM Models for the X-Cube Model:} a) The output of $E_{\text{XCube}}$ for a single Pauli $Z$ error is a plaquette-type interaction. The four cubes that share an edge are shaded in--- only one stabilizer is shown to reduce clutter. The orange edge indicated where the error has occurred, and the black zigzag line denotes the induced interaction between the classical spins for the four shaded cubes. This form of interaction corresponds to the plaquette-Ising model. b) The classical interactions induced by $X$ errors. Here, we are working on the dual lattice, and $X$ errors act on dual plaquettes rather than edges. Classical spins live at the centers of dual cubes, and are colored according to their corresponding stabilizer. Each $X$ error in the three basis directions of the lattice couples four classical spins, two for each vertex type with support on the affected dual plaquette (here we have only drawn one color stabilizer, again to reduce clutter). All three interaction types are shown. Using the fact that the product of two differently colored vertex terms gives the third color, these interaction terms can be reduced to the anisotropically-coupled Ashkin-Teller model described in the main text.}
    \label{fig:xcube-SM}
\end{figure}

\subsubsection{Phase Errors}

We can directly read off the form of interactions in the SM models by looking at the columns of $E_{\text{XCube}}$. A single site $Z$ error along the $\hat{x}$-direction produces interaction $\begin{pmatrix} 1 + \bar{y} + \bar{z} + \bar{y}\bar{z} & 0 & 0 & 0\end{pmatrix}$. This a four-body interaction between classical spins assigned to cubes sharing the affected edge; mapping to the dual lattice, it corresponds to a four-body interaction between the classical spins living on the vertices of a plaquette lying in the y-z plane. $Z$ errors in the other directions yield the other two kinds of plaquette interactions, and we see that the SM model for phase errors in the X-Cube model is the plaquette Ising model (PIM) \cite{Savvidy_1994, CAPPI1992659, johnstonPlaquetteIsingModels2017, muellerExactSolutionsPlaquette2017, johnstonGonihedric3DIsing1996} (see Fig. \ref{fig:xcube-SM}a): 
\begin{equation}
\begin{aligned}
    \mathcal{H}^{(2)}_{\text{XCube}, Z} = -\sum_{i} (&\sigma_i \sigma_{i-\hat{y}} \sigma_{i - \hat{z}} \sigma_{i - \hat{y} - \hat{z}}\\
    +& \sigma_i \sigma_{i-\hat{x}} \sigma_{i - \hat{z}} \sigma_{i - \hat{x} - \hat{z}} \\
    +& \sigma_i \sigma_{i-\hat{x}} \sigma_{i - \hat{y}} \sigma_{i - \hat{x} - \hat{y}})
\end{aligned}
\end{equation}
where $i$ indexes sites (vertices) of the cubic lattice, $\sigma_i = \pm 1$ is a classical spin. Numerical studies have estimated the inverse critical temperature for the PIM at $\beta_c = 0.554$ \cite{johnstonGonihedric3DIsing1996}, which means $p^{(2)}_{c, Z} = 0.213$. 

For open boundary conditions, the PIM has a subextensive ground state degeneracy of $2^{3L}$, where $L$ is the linear dimension of the cubic lattice \cite{johnstonPlaquetteIsingModels2017}. The large number of degenerate ground states stems from the subsystem symmetry present in the model; any 2D plane of spins may be flipped without changing the energy of the spin configuration. This means that the ordered phase of the PIM will not be distinguished by typical ferromagnetic order parameters. For example, the magnetization $\langle \sigma_i \rangle$ and the two-spin correlator $\langle \sigma_i \sigma_j \rangle$ will generically vanish even at zero temperature, given the freedom to flip planes of spins. Instead, we must consider order parameters that account for the planar ordering present in the model. Numerics have indicated that the following ``pseudo''- magnetizations are good order parameters for the PIM \cite{johnstonPlaquetteIsingModels2017}: 
\begin{equation}
\begin{aligned}
    m_x &= \left \langle \frac{1}{L^3} \sum_{x=1}^L \abs{\sum_{y,z=1}^L \sigma_{x,y,z} \sigma_{x+1, y, z}} \right\rangle\\
    m_y &= \left \langle \frac{1}{L^3} \sum_{y=1}^L \abs{\sum_{x,z=1}^L \sigma_{x,y,z} \sigma_{x, y+1, z}} \right\rangle \\
    m_z &= \left \langle \frac{1}{L^3} \sum_{z=1}^L \abs{\sum_{x,y=1}^L \sigma_{x,y,z} \sigma_{x, y, z+1}} \right\rangle \\
\end{aligned}
\end{equation}
where the absolute value serves to prevent cancellations between planes that have been flipped relative to each other. If this is the analogue of the magnetization, we can construct an analogue spin-spin correlator:
\begin{equation}
\begin{aligned}
    s_x(y_1, z_1, y_2, z_2) &= \langle \sigma_{x, y_1, z_1} \sigma_{x + 1, y_1, z_1} \sigma_{x, y_2, z_2} \sigma_{x+1, y_2, z_2}\rangle\\
    s_y(x_1, z_1, x_2, z_2) &= \langle \sigma_{x_1, y, z_1} \sigma_{x_1, y+1, z_1} \sigma_{x_2, y, z_2} \sigma_{x_2, y+1, z_2}\rangle\\
    s_z(x_1, y_1, x_2, y_2) &= \langle \sigma_{x_1, y_1, z} \sigma_{x_1, y_1, z+1} \sigma_{x_2, y_2, z} \sigma_{x_2, y_2, z+1}\rangle\\
\label{eq:dipole-correlation}
\end{aligned}
\end{equation}
which measures the correlation between pairs of spin-``dipoles''. Intuitively, this ``dipole''-``dipole'' correlator should detect the unusual ordering of the PIM ground states, since they are sensitive to whether the relative orientation between two spins persists along the parallel planes they live in. 

These correlators are directly related to the relative entropy (Eq. \ref{eq:rel-entropy}) in the X-Cube model. Consider creating a pair of fracton dipoles with a single Pauli $Z$ on an edge $k$ belonging to unit cell $s = (x_1+1, y_1+1, z_1)$. Assume $k$ is in the $\hat{z}$ direction:
\begin{equation}
\begin{aligned}
    D^{(2)}(\rho \vert \vert \rho_\epsilon) &= \frac{1}{1-n}\log \left\langle \mathsf{E} \cdot\mathsf{\Omega}(Z_k)\right\rangle \\
    &= \frac{1}{1-n}\log \left\langle \sigma_{k} \sigma_{k - \hat{x}} \sigma_{k - \hat{y}} \sigma_{k - \hat{x} - \hat{y}} \right\rangle \\
    &= \frac{1}{1-n}\log s_{x_1}(y_1, z_1, y_1 +1, z_1)\\
\end{aligned}
\end{equation}
We can create pairs of farther-separated fracton dipoles, which will correspond to longer-distance correlators. But we can see that the relative entropy is likely a good witness for the X-Cube information transitions, given that the corresponding classical correlation function is likely a good witness for the SM model finite-temperature transition. We note that, as far as we are aware, numerical studies on the PIM have not yet considered the correlators in Eq. \ref{eq:dipole-correlation}. In this way, knowledge of quantum CSS codes can give inspiration for better understanding their classical SM model counterparts.

\subsubsection{Bit-Flip Errors}

Our SM model will have three flavors of spin, corresponding to the three types of vertex term. Each $X$ error will excite two kinds of vertex terms, whose colors depend on the direction of the edge in question (see Fig. \ref{fig:xcube-SM}b). Indeed, if we plug in error pattern $\begin{pmatrix}1 & 0 & 0 & 0 & 0 & 0 \end{pmatrix}^T$ into the excitation map, we get the interaction term $\begin{pmatrix} 1 + x & 1 + x & 0 \end{pmatrix}^T$. Let the vertex term lying in the $x-y$ plane have classical spins $\sigma$, the vertex term lying in the $x-z$ plane have classical spins $\tau$, and the vertex term lying in the $y-z$ plane have classical spins $\eta$. The the SM model for Réyni-2 quantities is given by:
\begin{equation}
    \begin{aligned}
        \mathcal{H}^{(2)}_{XCube, X} = \sum_i (&\sigma_i \sigma_{i+\hat{x}} \tau_i \tau_{i +\hat{x}} + \sigma_i \sigma_{i+\hat{y}} \eta_i \eta_{i + \hat{y}} \\
        &\qquad + \tau_i \tau_{i+\hat{z}} \eta_i \eta_{i + \hat{z}} )
    \end{aligned}
\end{equation}

Using the fact that the product of two of the vertex terms gives the third kind, we have the relationship $\sigma \tau = \eta$. So we can simplify the above SM model:
\begin{equation}
        \mathcal{H}^{(2)}_{XCube, X} = \sum_i \sigma_i \sigma_{i+\hat{z}} + \tau_i \tau_{i+\hat{y}} + \sigma_i \sigma_{i+\hat{x}} \tau_i \tau_{i+\hat{x}}
    \label{eq:ACAT-ham}
\end{equation}

This is known as the anisotropically-coupled Ashkin-Teller (ACAT) model, and it is dual model to the plaquette Ising model \cite{Savvidy_1994, johnstonDualGonihedric3D2011, Johnston_2020}. It has a finite-temperature transition at $\beta_c = 1.313$ \cite{johnstonPlaquetteIsingModels2017}, which corresponds to a threshold error rate of $p^{(2)}_{c, X} \approx 0.336$. We can compare with the maximum-likelihood threshold of $p^{(1)}_{c, X} = 0.152$ obtained from the random-bond version of the ACAT model studied in \cite{songOptimalThresholdsFracton2022}; as expected, the threshold for Réyni-2 quantities is larger than their von-Neumann counterparts.

Similarly to the PIM, the ACAT has a subextensive ground state degeneracy, which numerics suggest persists to finite-temperatures \cite{johnstonDualGonihedric3D2011, mueller2014}. This ground state degeneracy is also due to the presence of subsystem symmetries; here, we are free to flip any plane of $\sigma$ spins normal to the $\hat{y}$ direction, and plane of $\tau$ spins normal to the $\hat{z}$ direction, or we can flip both $\sigma$ and $\tau$ simultaneously in a plane normal to the $\hat{x}$ direction. The ability to flip planes of spins relative to each other without an energy cost means the ground states do not have conventional ferromagnetic order. As with the PIM, we need a modified version of the magnetization or usual spin-spin correlators to detect the ordered phase of the ACAT. So far, the existing numerical studies of this model have relied on the energy and energy cumulants to detect the critical point \cite{johnstonDualGonihedric3D2011}. As with the PIM, perhaps we can derive some possible order parameters using inspiration from the X-Cube model. 

The correlation function that maps onto the relative entropy is given by:
\begin{equation}
\begin{aligned}
    \langle \mathsf{E} \cdot \mathsf{\Omega}(X^s) \rangle &= \begin{cases}
        \langle \sigma_i \sigma_{j} \tau_i \tau_{j}\rangle & \text{if } s \parallel \hat{x}\\
        \langle \tau_i \tau_j \rangle & \text{if } s \parallel \hat{y} \\
        \langle \sigma_i \sigma_j \rangle & \text{if } s \parallel \hat{z}
    \end{cases}
\end{aligned}
\end{equation}
Where $s$ is a line, and $i$, $j$ are the vertices at the two ends of $s$. If we introduce any bends into $s$, the resulting correlator will have spins located at the bends as well, since bends create extra lineons in the X-Cube model. For simplicity, we consider the straight-line case. Taking inspiration from these correlators, we propose the following potential magnetizations for the ACAT:
\begin{equation}
    \begin{aligned}
        m_x &= \left \langle \frac{1}{L^3} \sum_{x=1}^L \abs{\sum_{y,z=1}^L \sigma_{x,y,z} \tau_{x,y,z} \sigma_{x+1, y, z} \tau_{x+1, y, z} }\right\rangle\\
        m_y &= \left \langle \frac{1}{L^3} \sum_{y=1}^L \abs{\sum_{x,z=1}^L \tau_{x,y,z} \tau_{x, y+1, z}} \right\rangle \\
        m_z &= \left \langle \frac{1}{L^3} \sum_{z=1}^L \abs{\sum_{x,y=1}^L \sigma_{x,y,z} \sigma_{x, y, z+1}} \right\rangle \\
    \end{aligned}
\end{equation}
These are similar to the ``fuki-nuke'' magnetizations that have been shown to be good order parameters for the PIM phase transition, but have an anisotropy reflecting the anisotropy of the ACAT itself. The subsystem symmetries of the ACAT and the PIM are connected to stabilizer constraints in the X-Cube model--- since a stabilizer configuration $g$ is unchanged by adding a set of stabilizers $\hat{g}$ that multiply to the identity, when we assign a classical spin configuration to $g$, we should be able to flip the spins corresponding to $\hat{g}$ without changing the energy of the configuration. Alternately, we can view the stabilizer constraints as being the result of the pre-existing subsystem symmetries when we gauge the classical model to get the quantum one. The ability to flip any plane of spins in the PIM corresponds to the constraint that $\prod B_c = \mathbb{I}$ for all $c$ in a plane. The ability to flip certain flavors of spins in certain planes in the ACAT corresponds to the constraint that $\prod A_{v}^\mu =\mathbb{I}$ for all $v$ in a plane for fixed $\mu$. Going back to the 3D toric code, we can now see why the SM model for bit-flip errors is a lattice gauge theory; this stems from the fact that the product of plaquette stabilizers around a single cube must be the identity. The bit-flip SM model is constrained to have a local spin-flip symmetry, which means it must be a gauge theory-like model.

\subsubsection{Coherent Information}

\begin{figure}
    \centering
    \includegraphics[width=0.9\linewidth]{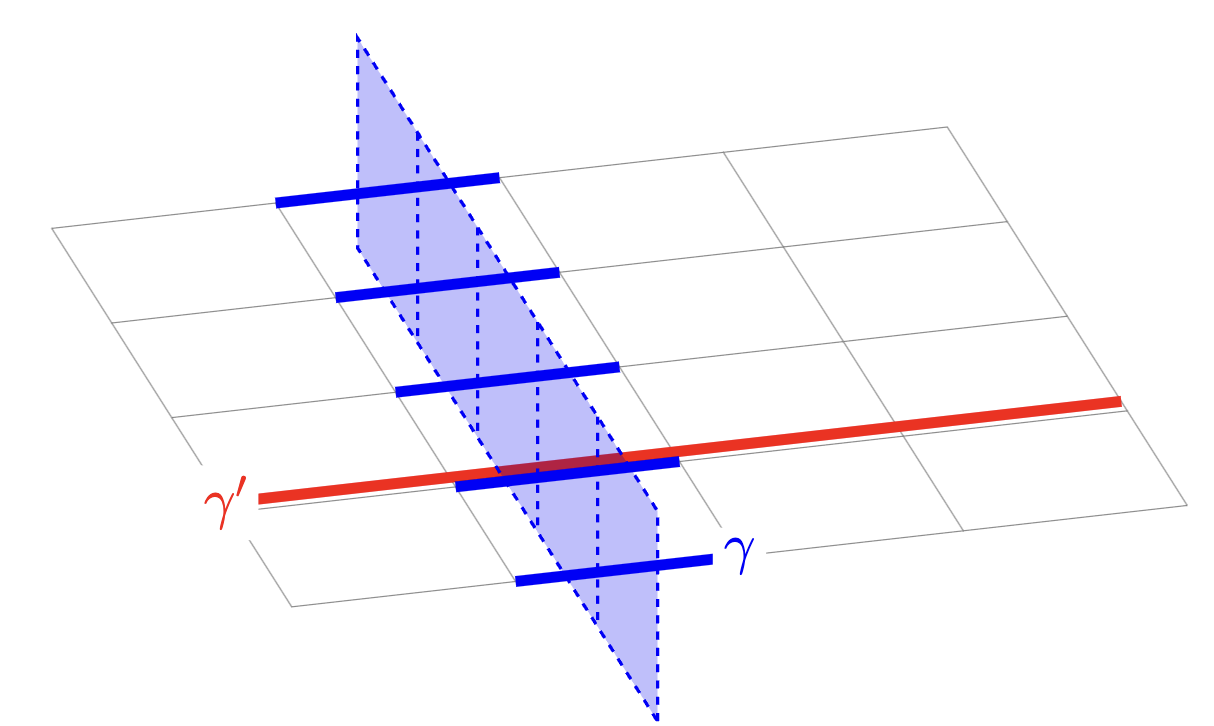}
    \caption{\textbf{X-Cube Logical Operators: } The X-Cube model has 6$L$ pairs of logical operators--- one for each plane of the 3D cubic lattice. An example pair of logicals $\gamma$, $\gamma'$ is pictured for an arbitrary 2D plane. These logicals are not all independent, which can be seen from the X-Cube GSD of $6L-3$. The $Z$-type logical operator ($\gamma$) is a string operator on the dual lattice; the coherent information under $Z$ errors is then connected to the excess free energy of inserting a line of antiferromagnetic plaquettes along this line. The $X$-type logical ($\gamma'$) is a string on the direct lattice, and so the coherent information under bit-flip errors is connected to the free energy of inserting antiferromagnetic couplings along this line in the ACAT. }
    \label{fig:xcube-info}
\end{figure}

What is the behavior of the Rényi-2 coherent information, given our two SM models? We know that in general, it will map onto the excess free energy of flipping signs of classical interactions along a the path of a logical operator in the original quantum code. $Z$ type logical operators correspond to a fracton-dipole string operator $Z_\gamma$ wrapped around a non-contractible loop $\gamma$; $X$-type logicals correspond to a lineon string operator $X_{\gamma'}$ wrapped around a non-contractible loop $\gamma'$. In the PIM, we flip interactions along a line of in-plane plaquettes (see Fig. \ref{fig:xcube-info}). In the ACAT, we flip the sign of interactions along a line in the same direction as the $X$ logical in question. 

First we consider the PIM with defect. In the paramagnetic phase, plaquettes are free to fluctuate, and so there is no free energy gain to introducing the antiferromagnetic interactions. In the ferromagnetic phase, however, the energy cost of the flipped plaquette interactions in the PIM scales with the length of the defect. This is a result of frustration; we can choose to satisfy the antiferromagnetic interactions along the defect, at the cost of violating ferromagnetic terms elsewhere, or we can violate the flipped terms and satisfy the unflipped terms everywhere else. The entropic gain from the defects, on the other hand, is constant, and does not compete with the energy cost in the thermodynamic limit: $\Delta \mathcal{F}_z \rightarrow \infty$. 

For the ACAT, the situation is similar. In the paramagnetic phase, both flavors of spins are free to fluctuate, and inserting defects along a line costs nothing. In the ferromagnetic phase, we pay a linear energy cost, as we must choose between satisfying the defect interactions or the clean ones. So we arrive at the same conclusion as for the 3D toric code: if both SM models are in the paramagnetic phase, we have maximum channel capacity. If only one is in the paramagnetic phase, we reduce the coherent information to zero, indicating we can protect \emph{classical} information. When both models are ordered, we lose all information capacity.

\subsubsection{Entanglement Negativity}

\begin{figure}
     \centering
     \includegraphics[width=\linewidth]{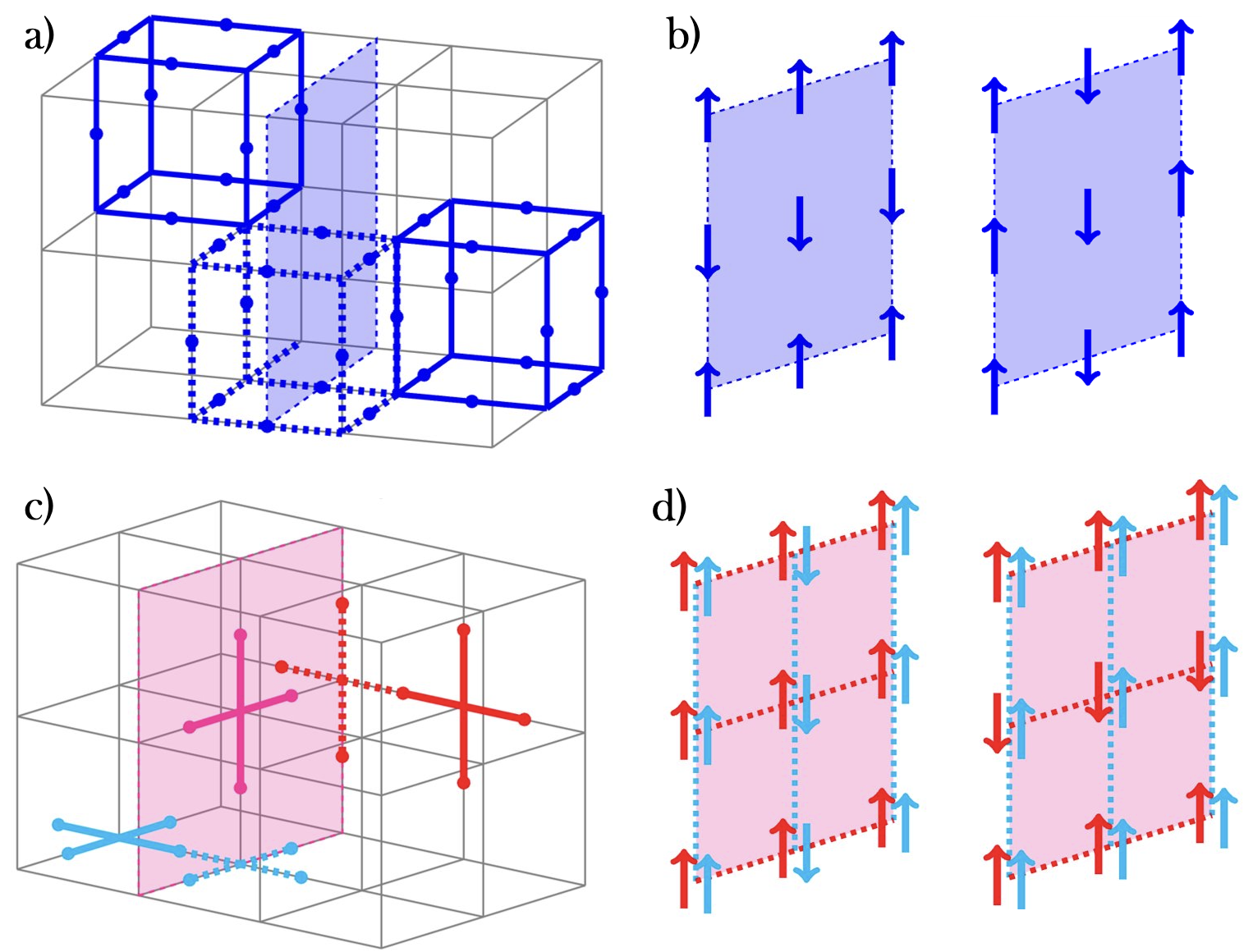}
     \caption{\textbf{Entanglement Negativity in the X-Cube Model: }a) $A$-separability prohibits $g_x$ configurations that cross $\partial A$; for example, the cube pictured in dashed lines is not allowed. b) The separability condition implies that all plaquette interactions on $\partial A$ must be $+1$. Two possible spin configurations satifying this requirement are shown. These configurations are related by a spin flip along a horizontal or vertical line in-plane. c) $A$-separability for $g_z$ configurations prevents any vertices not parallel to $\partial A$ from crossing the bipartition. d) Allowed classical configurations on the bipartition. The two independent flavors of spins are each coupled only along one direction, so spin flips along these directions are allowed.}
     \label{fig:xcube-negativity}
 \end{figure}

We consider the same setup as for the 3D toric code, where the boundary of our region $A$, $\partial A$, is a bipartition of the cubic lattice. For phase errors, we consider a ``rough'' boundary (see Fig. \ref{fig:xcube-negativity}a). No $g_x$ stabilizers are allowed to cross this boundary, which means we must pin all the PIM interactions on the surface dual to the rough boundary. Pinning these plaquette interactions always leads to an energy gain $\abs{\partial A}$, but the entropic considerations are different in the paramagnetic vs. ferromagnetic regime. In the high-temperature, paramagnet phase, beyond some correlation length $\xi$, the sign of plaquettes will fluctuate, so pinning reduces the entropy by $\abs{\partial A}/ \xi$. However, there is still freedom for the configuration of spins on the plane to fluctuate together; the subsystem symmetry of the PIM, when projected onto $\partial A$, means we are free to flip any 1D line of spins in the plane. This leads to a linear correction to the negativity of $-(2L +1)\log 2$, since there are $2(L+1)-1$ independent lines to flip in the plane. The entanglement entropy for fracton phases has been shown to contain a geometry-dependent linear correction term \cite{Ma_2018, Shi_2018, shirleyUniversalEntanglementSignatures2019}, so it is plausible that the negativity should display similar scaling. The linear term disappears in the ferromagnetic phase of the PIM, however--- the in-plane spins have to align with out-of-plane spins, so no independent fluctuations on the surface are possible except for a spin flip everywhere in the plane. The only correction left is $-\log 2$.

For bit-flip errors, we take $\partial A$ to be smooth (see Fig. \ref{fig:xcube-negativity}b). The separability requirement for $g_z$ configurations means the only allowable vertex terms on the surface $\partial A$ are ones that lie completely in-plane, or alternately, the matchbox with its open faces in-plane. The classical spins assigned to the other two vertex terms will be forced to be aligned. Suppose that $\partial A$ lies in the $y-z$ plane; according to Eq. \ref{eq:ACAT-ham}, this means $\sigma$ and $\tau$ spins will be pinned. Imporantly, due to the anisotropic coupling of the ACAT, $\sigma$-$\sigma$ and $\tau-\tau$ interactions lie in plane, while the four-body coupling is out-of-plane. In the paramagnetic phase, using the same arguments as above, the pinning leads to the loss of $\abs{\partial A}/\xi$ degrees of freedom. However, similarly to the PIM, the subsystem symmetry of the ACAT means the in-plane spins can still fluctuate collectively. In particular, since the out-of-plane $\sigma \sigma \tau \tau$ bonds are able to fluctuate, we can freely flip lines of $\sigma$ and $\tau$ spins in-plane along the direction of their coupling. This leads to a linear correction to the negativity of $-(2L+1)\log 2$, the same counting as for the phase errors. This freedom is partially lost in the ferromagnetic phase, as now the out-of-plane four-body term removes the ability to flip $\tau$ and $\sigma$ independently. However, we can still flip the entire plane of spins, leaving a residual correction of $-\log 2$. 

The residual $-\log 2$ correction to the negativities in the case of $X$ and $Z$ errors indicates that some nontrivial entanglement structure remains in the system even when we have lost all ability to encode information in the ground state manifold. A similar phenomenon has been demonstrated in the 2D toric code under correlated $X$ and $Z$ errors \cite{wang2023intrinsic}, which we will discuss in section \ref{sec:corr-errs}. It would be interesting to explore the nature of this memoryless phase further; we leave this to future work.

\subsubsection{Optimal Thresholds}

Recently, the random-bond versions of both the PIM and the ACAT have been studied numerically \cite{songOptimalThresholdsFracton2022}. The critical error rates along the Nishimori line were located at $p_{c, X} = 0.152$ and $p_{c, Z} = 0.075$. As expected, these are lower than the thresholds for Réyni-2 quantities.

\section{Correlated Errors}
\label{sec:corr-errs}

The analysis so far has focused on single-site Pauli errors--- however, the SM model we have outlined works just as well for error channels involving correlated Pauli errors, including error channels that span multiple physical sites. All that needs to be done is to input the error $P_i$ (which now may have support on multiple sites) into the excitation map $\mathsf{E}$ to derive the form of interactions in the classical model. We note that a formalism for analyzing the effect of general correlated errors in stabilizer codes has been developed in \cite{chubbStatisticalMechanicalModels2021}. However they focus specifically on SM models for the Kitaev-Preskill decoder, whereas we are considering intrinsic behavior of the density  matrix. 

Correlated errors spanning multiple sites will in general produce longer-range interactions in the corresponding SM models, as a single error can affect farther separated stabilizers. Depending on the structure of the SM model, this may or may not effect the critical behavior. Correlated errors involving both $X$ and $Z$ will also in general couple different types of stabilizers together, meaning the partition function of the SM model no longer factorizes into two decoupled $X$- and $Z$-error models. 

\subsection{Example: Correlated X and Z Errors in the 2D Toric Code}

\begin{figure}
    \centering
    \includegraphics[width=\linewidth]{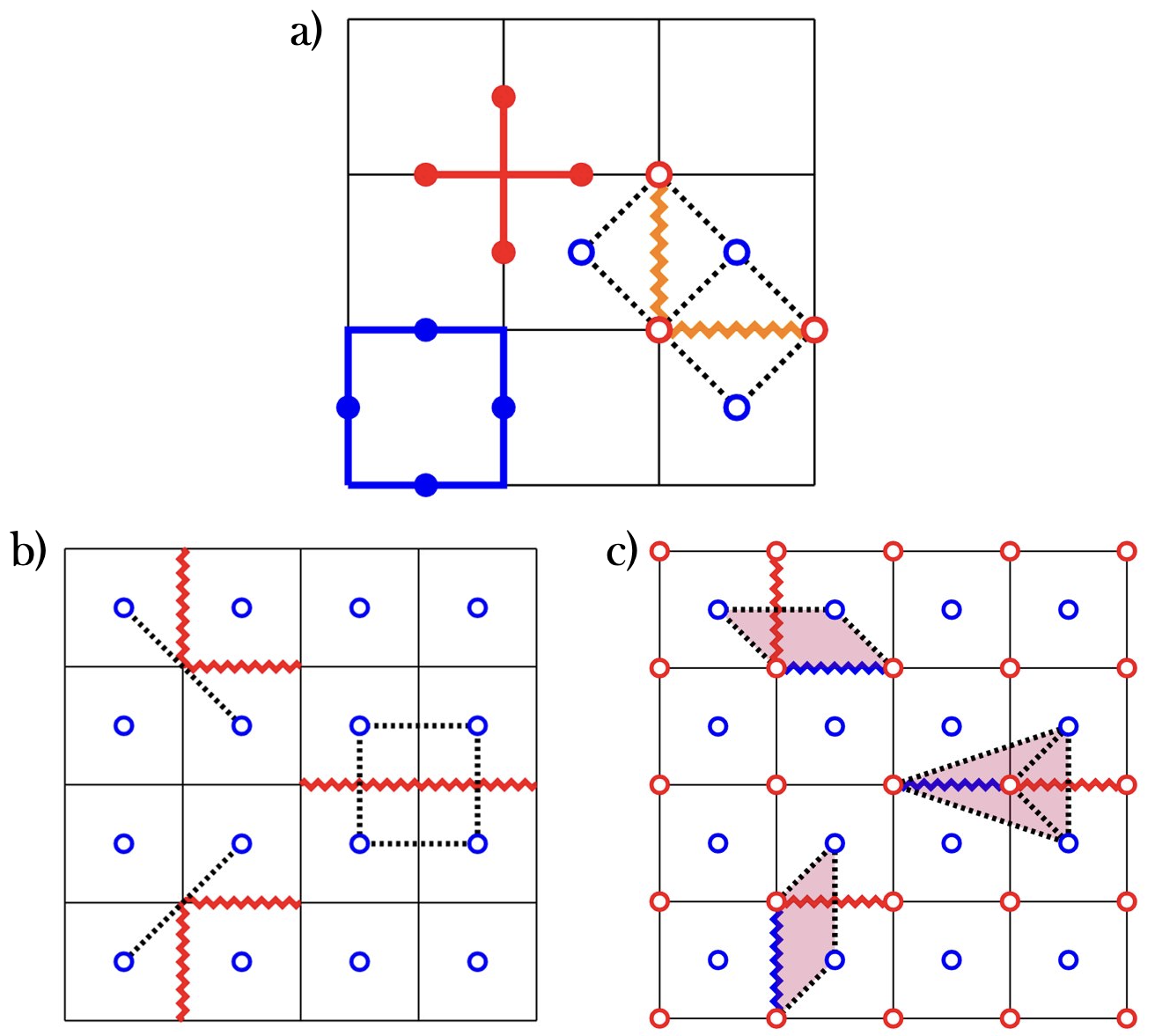}
    \caption{\textbf{2D Toric Code with Correlated Errors:} a) The 2D toric code stabilizers are pictured on the left, with $X$ operators in red and $Z$ operators in blue. Classical spins corresponding to vertex terms and plaquette terms are denoted by open circles, in red and blue respectively. The orange zig-zag lines denote the location of $Y$ errors and the dashed lines show the induced interaction between the classical spins. b) Induced classical interactions between plaquette spins for correlated $X$ errors. Different possible correlated errors are pictured as red zig-zag lines, and the corresponding classical interaction is represented by the dashed black lines. c) Induced classical interactions between plaquette and vertex spins for correlated two-site $X$ and $Z$ errors (red and blue zig-zag lines, respectively). }
    \label{fig:2dtc-corr}
\end{figure}

We consider various kinds of correlated and multi-site error channels in the 2D toric code. A simple possibility is that $X$ and $Z$ errors always occur together; otherwise known as a Pauli-$Y$ channel. A single Pauli $Y_e$ on an edge $e$ excites two vertices and two plaquettes, inducing a four-body interaction between their corresponding classical spins (see Fig. \ref{fig:2dtc-corr}a). If we work on the rotated lattice, these interactions are the same as the 2D plaquette Ising model \cite{SAVVIDY199472, muellerExactSolutionsPlaquette2017}, which is disordered at all finite-temperatures. This is easily understood from the fact that there exists a change of variables that maps this model onto a set of decoupled 1D Ising chains. This gives us the somewhat irrelevant result that the threshold for the 2D toric code under $Y$ errors is bounded from above by $p^{(2)}_c =0.5$. However, we know that the random-bond version of the 2D plaquette Ising model should describe the density matrix in the replica limit $n \rightarrow 1$--- this model is also paramagnetic at all temperatures, since we can still map it onto a set of decoupled (random-bond) 1D Ising chains. Now, however, the paramagnet phase corresponds to the high-error rate phase, and we find that $p_c = 0$. The $Y$ error channel trivializes the 2D toric code for any error rate. 

Consider a scenario where we can have single-site bit-flip errors with some probability $p_x$, and a correlated bit-flip on two neighboring sites with probability $p_{xx}$. The two-site correlated bit flip can induce several new types of interactions on top of the Ising-type interactions generated by the single-site errors (see Fig. \ref{fig:2dtc-corr}b). By tuning the relative probability of $p_x$ and $p_{xx}$, we can tune the relative strength of these longer-range interaction terms. 

Still further interaction terms can be produced by correlated $X$ and $Z$ errors on neighboring sites; we can think of this type of error-channel as a $\psi$-channel, since it will create pairs of fermions. These $\psi$-errors will couple two plaquette and two vertex spins together, leading to a four-body interaction. We can change the geometry of this four-body interaction by changing the relative positions of the $X$ and $Z$ error (see Fig. \ref{fig:2dtc-corr}c). It has been shown that such the $\psi$-channel can destroy the memory properties of the toric code, without trivializing the topological order \cite{wang2023intrinsic}. This is captured by a topological entanglement entropy that does not depend on error rate $p_\psi$. We can understand how this comes about from the form of the SM model. The particular $\psi$ channel considered in \cite{wang2023intrinsic} included only some of the possible $X$, $Z$ correlated errors (see Fig. \ref{fig:2dtc-corr}d). The SM model with only these interactions is a single-layer anisotropic plaquette Ising model, sometimes known as a ``fuki-nuke'' model \cite{suzuki1972, Johnston_2012}:
\begin{equation}
    \mathcal{H}^{(2)}_{\psi} = -\sum_{j=1}^N \sigma_j \sigma_{j + \hat{z}} \sigma_{j + \hat{x}} \sigma_{j + \hat{x} + \hat{z}} + \sigma_j \sigma_{j + \hat{z}} \sigma_{j + \hat{y}} \sigma_{j + \hat{y} + \hat{z}} 
\end{equation}
where the $z$-direction only extends for one layer. This model can be mapped onto the 2D Ising model using the transformation $\tau_j = \sigma_j \sigma_{j+\hat{z}}$, and so for the purposes of calculating the relative entropy and coherent information, the results will be the same as for $X$ or $Z$ errors. The additional spin-flip symmetries due to the underlying plaquette interaction become important when calculating the negativity; it can be shown these spin-flip symmetry remain even in the ferromagnetic phase since we are free to take $\sigma_j \rightarrow -\sigma_j$ and $\sigma_{j + \hat{z}} \rightarrow -\sigma_{j + \hat{z}}$ simultaneously. This leads to a constant negativity independent of error rate.  

%\zx{[I think it's important to comment on the $\psi$ channel decoherence. It's like whenever the Z error happens, there is a nearby X error such that you are always exciting a pair of fermions. People have found nontrivial TEN even after the topological memory is lost. See https://arxiv.org/abs/2307.13758]} %\zx{[Actually now I'm confused - did Ruihua's paper say that the TEN is trivial in this case??]}

\section{Non-CSS Codes}
\label{sec:non-CSS}

We have focused on CSS-type codes up to this point, but the SM mapping developed in section \ref{sec:generic-mapping} can be generalized to deal with non-CSS stabilizer codes as well. As with correlated errors, when we consider a non-CSS code, the SM models for $X$ and $Z$ errors will not necessarily factorize any more. This is now due to the fact a single $X$ or $Z$ error might excite the same stabilizer. Consider a generic, non-CSS stabilizer Hamiltonian:
\begin{equation}
    H = -\sum_{\alpha=1}^M \sum_{i_{\alpha}} S^{(\alpha)}_{i_{\alpha}}
    \label{eq:non-CSS-stabilizer-Hamiltonian}
\end{equation}
Here, $\alpha = 1, \dots, M$ simply indexes distinct kinds of stabilizers, which are not specifically $X$- or $Z$-type. Stabilizers of type $\alpha$ are centered at locations $i_\alpha$ on the lattice. The completely mixed ground state can be written as a sum over all stabilizer configurations $g$:
\begin{equation}
    \rho_0 = \frac{1}{2^N} \sum_{g} g
\end{equation}

If we apply $\mathcal{E} = \mathcal{E}_X \circ \mathcal{E}_Z$ to this density matrix, with error probabilities $p_x$ and $p_z$, we obtain:
\begin{equation}
    \begin{aligned}
        \mathcal{E}[\rho_0] &= \frac{1}{2^N} \sum_g e^{-\sum_P \beta_P \sum_{j=1}^N \mathsf{E} \cdot\mathsf{\Omega}(P_j) \vert_g} g \\
    \end{aligned}
\end{equation}
where $\beta_P = \mu_P/2 = -1/2 \log (1-2p_P)$, $P = X, Z$. Letting $\rho = \mathcal{E}[\rho_0]$, the $n$th moment of $\rho$ is then described by the following SM model:
\begin{equation}
\begin{aligned}
    \tr \rho^n &= \frac{1}{2^{(n-1)N}} \sum_{\{g^{(m)}\}} e^{-\beta \sum_P \mathcal{H}^{(n)}_P} \\
    \mathcal{H}^{(n)}_P &= \frac{\beta_P}{\beta} \sum\limits_{j=1}^N \left(\sum\limits_{m=1}^{n-1} \mathsf{E}^{(m)}\cdot\mathsf{\Omega}(P^{(m)}_j)\right. \\
    &\left.\qquad\qquad\qquad\qquad\qquad+ \prod\limits_{m=1}^{n-1} \mathsf{E}^{(m)}\cdot\mathsf{\Omega}(P^{(m)}_j)\right)
\end{aligned}
\end{equation}
We now have to consider a combined SM model that includes interaction terms derived from both $X$ and $Z$ errors. We will discuss non-CSS models in more depth in an upcoming work \cite{upcoming}. As an example, however, we will derive the statistical mechanics model for the CBLT code. 

\subsection{Example: CBLT Model}

\begin{figure}
\centering
    \includegraphics[width=\linewidth]{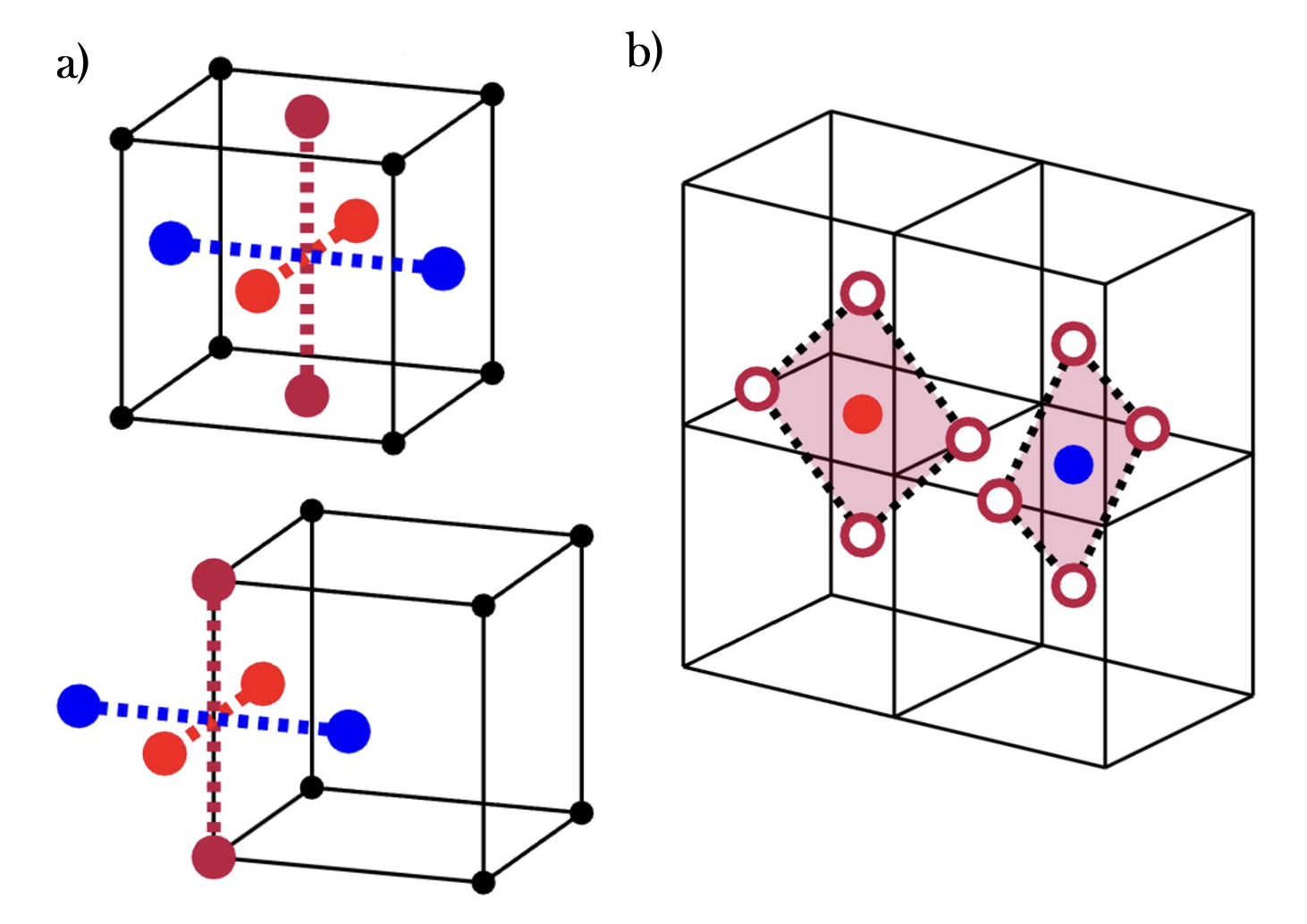}
    \caption{\textbf{CBLT Model:} a) Stabilizers for the CBLT model. Qubits live on the sites of the FCC lattice; $X$ operators are represented by red dots, $Z$ operators by blue dots, and $Y$ operators by maroon dots. b) Two kinds of classical interactions induced by $X$ and $Z$ errors respectively. The SM model is a sum over all such terms, centered at every face of the cubic lattice. The open circles denote classical spins assigned to each stabilizer. The filled red and blue circle represent respectively an $X$ or $Z$ error applied to the qubit living at that site.}
    \label{fig:cblt}
\end{figure}

The Chamon-Bravyi-Leemhuis-Terhal (CBLT) code \cite{chamonQuantumGlassiness2005, bravyiTopologicalOrderExactly2011} is a non-CSS type-I fracton stabilizer model. It is defined on the FCC lattice, with one type of stabilizer (see Fig. \ref{fig:cblt}a). 

The stabilizer and excitation maps are given by \cite{vijayFractonTopologicalOrder2016}:
\begin{equation}
    \begin{aligned}
        \mathsf{S}_{CBLT} &= \begin{pmatrix}
            1 + x + z + x\bar{z} \\
            1 + x + y + x\bar{y} 
        \end{pmatrix} \\
        \mathsf{E}_{CBLT} &= \begin{pmatrix}
            1 + \bar{x} + \bar{y} + \bar{x}y \\
            1 + \bar{x} + \bar{z} + \bar{x}z 
         \end{pmatrix}
    \end{aligned}
\end{equation}

So the SM model includes two flavors of interaction, with relative coupling strengths determined by the relative rates of $X$ and $Z$ errors (see Fig. \ref{fig:cblt}b). As far as we are aware, statistical mechanics models with these interactions have not been studied. However, if we only have one kind of error, we can simplify the SM model so it only includes one flavor of interaction. In this case, the SM model for Réyni-2 quantities maps onto a stack of decoupled 2D plaquette Ising models, which, as discussed in the previous section, is paramagnetic at all temperatures. The random-bond version of the 2D PIM is also paramagnetic at all temperatures, indicating that any amount of $X$ errors will trivialize the CBLT code. The result is the same if we only have $Z$ errors. It would be interesting to see if the CBLT model is more robust to correlated errors--- we leave this for future work.

\section{Discussion and Outlook}
\label{sec:discussion}
% \begin{itemize}
%     \item extend to qudits
%     \item consider qLDPC codes 
%     \item move past Ising type stat-mech models
%     \item physical implications of the phase transition (implications of first order vs. continuous transitions?)
%     \item novel decoherence-induced phase transitions 
% \end{itemize}

We have introduced a general prescription for mapping the information-theoretic properties of stabilizer codes under decoherence to the thermodynamic quantities of the corresponding classical statistical mechanics models. This allows the derivation of error thresholds that are intrinsic to a given code, rather than dependent on a specific decoder.  Our mapping is applicable to both CSS and non-CSS codes, as well as generic Pauli error channels (including correlated errors). Using the general mapping, we have examined in depth the properties of the 3D toric code and the X-Cube model under local Pauli errors, and also commented on the CBLT model.  

The statistical mechanics mapping provides a useful tool for understanding the information properties of quantum codes; we are able to provide estimates for quantities like the entanglement negativity by analyzing the classical SM models. It would be interesting to verify the scaling estimates calculated for the X-Cube model, since as far as we are aware, the entanglement negativity has not been calculated for this model or for other fracton phases. Additionally, given the mapping, it would be interesting to try to construct novel quantum codes from classical models with desirable properties. For instance, robust quantum codes can be built by choosing a classical model with a critical point corresponding to a high-error threshold. 

The SM mapping can also shed light on the classical statistical mechanics models themselves; not only can we generate interesting SM models by starting with known quantum codes, but the quantum codes can also point us in the direction of relevant order parameters that have not been considered in the literature previously. For instance, we have proposed novel correlation functions that should be able to probe the unconventional magnetic ordering of the plaquette Ising model and its dual, the anisotropically-coupled Askhin-Teller model. 

While we have considered mainly local stabilizer Hamiltonians in this work, the SM mapping is not restricted to local codes. It would be interesting to apply the homological perspective on the mapping presented in section \ref{sec:homology} to important \emph{non-local} codes, such as the recently discovered good qLDPC codes \cite{tillichQuantumLDPCCodes2014, panteleevAsymptoticallyGoodQuantum2022, panteleevQuantumLDPCCodes2022}. To what extent can these codes be considered phases of matter, and how can we understand their intrinsic error correcting properties through their corresponding classical SM models? 

Another interesting avenue for further study is exploring novel decoherence-induced phases of matter--- we have demonstrated that single-site Pauli $X$ and $Z$ operators can drive a quantum memory into a phase only able to store classical information. Are there more exotic phases we can reach via decoherence? This will be the focus of an upcoming work \cite{upcoming}. Some examples of novel decoherence-induced phases have been studied recently--- for instance,  average-symmetry protected topological orders can be obtained from conventional symmetry-protected topological phases \cite{lee2024symmetry, Ma_2023, ma2023topological}.

These questions are all intimately related to understanding mixed phases of matter and the behavior of quantum codes under decoherence. The tools developed in this paper provide a valuable starting point from which to delve deeper into these areas, and highlight the deep connections between quantum information and condensed matter. 

\emph{Note Added:} While preparing this paper, we became aware of the following work \cite{su2024tapestry}, which also introduces a general statistical mechanics mapping for stabilizer codes under Pauli errors. The focus of their paper is on proving various duality relationships between the different SM models, and so they give more rigorous proofs of the various dualities mentioned in this paper and prove others we did not discuss. We have focused on the applicability of this mapping to studying the information properties of stabilizer codes. In particular, we provide the general mapping for the relative entropy, coherent information, and entanglement negativity. Our work also presents a different perspective on the SM mapping, through the polynomial and homological perspectives. Additionally, our mapping is generally applicable to correlated Pauli errors. Where we study the same models, our derived thresholds agree. 

%sources: 
% \begin{itemize}
%     \item Stabilizers: \cite{gottesman1997stabilizer}
%     \item toric code: \cite{Kitaev_2003}
%     %\item plaquette ising: \cite{Savvidy_1994, CAPPI1992659, johnstonPlaquetteIsingModels2017, muellerExactSolutionsPlaquette2017, johnstonGonihedric3DIsing1996}
%     \item anisotropic ashkin-teller: \cite{johnstonDualGonihedric3D2011}
%     \item random bond plaquette + ashkin-teller: \cite{songOptimalThresholdsFracton2022}
%     \item CBLT: \cite{chamonQuantumGlassiness2005, bravyiTopologicalOrderExactly2011}
%     \item Kitaev Honeycomb: \cite{Kitaev_2006}
%     \item Kitaev-Preskill decoder: \cite{Dennis_2002}
%     \item qLDPC codes: \cite{bravyiHomologicalProductCodes2013, panteleevAsymptoticallyGoodQuantum2022, panteleevQuantumLDPCCodes2022, tillichQuantumLDPCCodes2014}
%     \item LDPC codes: \cite{gallagher1962}
%     \item cohomology: 
% \end{itemize}

\section{Acknowledgements}
The author thanks Chiu Fan Bowen (Leo) Lo, Ruihua Fan, Ashvin Vishwanath, and Zhu-Xi Luo for insightful discussions, and Zhu-Xi Luo in particular for very helpful comments on the manuscript. A.L. acknowledges support by the National Science Foundation Graduate Student Fellowship Program (NSF GRFP). 

\bibliography{literature.bib}

\begin{thebibliography}{90}%
\makeatletter
\providecommand \@ifxundefined [1]{%
 \@ifx{#1\undefined}
}%
\providecommand \@ifnum [1]{%
 \ifnum #1\expandafter \@firstoftwo
 \else \expandafter \@secondoftwo
 \fi
}%
\providecommand \@ifx [1]{%
 \ifx #1\expandafter \@firstoftwo
 \else \expandafter \@secondoftwo
 \fi
}%
\providecommand \natexlab [1]{#1}%
\providecommand \enquote  [1]{``#1''}%
\providecommand \bibnamefont  [1]{#1}%
\providecommand \bibfnamefont [1]{#1}%
\providecommand \citenamefont [1]{#1}%
\providecommand \href@noop [0]{\@secondoftwo}%
\providecommand \href [0]{\begingroup \@sanitize@url \@href}%
\providecommand \@href[1]{\@@startlink{#1}\@@href}%
\providecommand \@@href[1]{\endgroup#1\@@endlink}%
\providecommand \@sanitize@url [0]{\catcode `\\12\catcode `\$12\catcode
  `\&12\catcode `\#12\catcode `\^12\catcode `\_12\catcode `\%12\relax}%
\providecommand \@@startlink[1]{}%
\providecommand \@@endlink[0]{}%
\providecommand \url  [0]{\begingroup\@sanitize@url \@url }%
\providecommand \@url [1]{\endgroup\@href {#1}{\urlprefix }}%
\providecommand \urlprefix  [0]{URL }%
\providecommand \Eprint [0]{\href }%
\providecommand \doibase [0]{https://doi.org/}%
\providecommand \selectlanguage [0]{\@gobble}%
\providecommand \bibinfo  [0]{\@secondoftwo}%
\providecommand \bibfield  [0]{\@secondoftwo}%
\providecommand \translation [1]{[#1]}%
\providecommand \BibitemOpen [0]{}%
\providecommand \bibitemStop [0]{}%
\providecommand \bibitemNoStop [0]{.\EOS\space}%
\providecommand \EOS [0]{\spacefactor3000\relax}%
\providecommand \BibitemShut  [1]{\csname bibitem#1\endcsname}%
\let\auto@bib@innerbib\@empty
%</preamble>
\bibitem [{\citenamefont {Haah}(2013)}]{haahCommutingPauliHamiltonians2013}%
  \BibitemOpen
  \bibfield  {author} {\bibinfo {author} {\bibfnamefont {J.}~\bibnamefont
  {Haah}},\ }\href {https://doi.org/10.1007/s00220-013-1810-2} {\bibfield
  {journal} {\bibinfo  {journal} {Communications in Mathematical Physics}\
  }\textbf {\bibinfo {volume} {324}},\ \bibinfo {pages} {351} (\bibinfo {year}
  {2013})},\ \Eprint {https://arxiv.org/abs/1204.1063} {arxiv:1204.1063
  [cond-mat, physics:math-ph, physics:quant-ph]} \BibitemShut {NoStop}%
\bibitem [{\citenamefont {Kitaev}(2003)}]{Kitaev_2003}%
  \BibitemOpen
  \bibfield  {author} {\bibinfo {author} {\bibfnamefont {A.}~\bibnamefont
  {Kitaev}},\ }\href {https://doi.org/10.1016/s0003-4916(02)00018-0} {\bibfield
   {journal} {\bibinfo  {journal} {Annals of Physics}\ }\textbf {\bibinfo
  {volume} {303}},\ \bibinfo {pages} {2–30} (\bibinfo {year}
  {2003})}\BibitemShut {NoStop}%
\bibitem [{\citenamefont {Dennis}\ \emph {et~al.}(2002)\citenamefont {Dennis},
  \citenamefont {Kitaev}, \citenamefont {Landahl},\ and\ \citenamefont
  {Preskill}}]{Dennis_2002}%
  \BibitemOpen
  \bibfield  {author} {\bibinfo {author} {\bibfnamefont {E.}~\bibnamefont
  {Dennis}}, \bibinfo {author} {\bibfnamefont {A.}~\bibnamefont {Kitaev}},
  \bibinfo {author} {\bibfnamefont {A.}~\bibnamefont {Landahl}},\ and\ \bibinfo
  {author} {\bibfnamefont {J.}~\bibnamefont {Preskill}},\ }\href
  {https://doi.org/10.1063/1.1499754} {\bibfield  {journal} {\bibinfo
  {journal} {Journal of Mathematical Physics}\ }\textbf {\bibinfo {volume}
  {43}},\ \bibinfo {pages} {4452–4505} (\bibinfo {year} {2002})}\BibitemShut
  {NoStop}%
\bibitem [{\citenamefont {Mochon}(2003)}]{Mochon_2003}%
  \BibitemOpen
  \bibfield  {author} {\bibinfo {author} {\bibfnamefont {C.}~\bibnamefont
  {Mochon}},\ }\bibfield  {journal} {\bibinfo  {journal} {Physical Review A}\
  }\textbf {\bibinfo {volume} {67}},\ \href
  {https://doi.org/10.1103/physreva.67.022315} {10.1103/physreva.67.022315}
  (\bibinfo {year} {2003})\BibitemShut {NoStop}%
\bibitem [{\citenamefont {Das~Sarma}\ \emph {et~al.}(2005)\citenamefont
  {Das~Sarma}, \citenamefont {Freedman},\ and\ \citenamefont
  {Nayak}}]{Das_Sarma_2005}%
  \BibitemOpen
  \bibfield  {author} {\bibinfo {author} {\bibfnamefont {S.}~\bibnamefont
  {Das~Sarma}}, \bibinfo {author} {\bibfnamefont {M.}~\bibnamefont
  {Freedman}},\ and\ \bibinfo {author} {\bibfnamefont {C.}~\bibnamefont
  {Nayak}},\ }\bibfield  {journal} {\bibinfo  {journal} {Physical Review
  Letters}\ }\textbf {\bibinfo {volume} {94}},\ \href
  {https://doi.org/10.1103/physrevlett.94.166802}
  {10.1103/physrevlett.94.166802} (\bibinfo {year} {2005})\BibitemShut
  {NoStop}%
\bibitem [{\citenamefont {Satzinger}\ \emph {et~al.}(2021)\citenamefont
  {Satzinger}, \citenamefont {Liu}, \citenamefont {Smith}, \citenamefont
  {Knapp}, \citenamefont {Newman}, \citenamefont {Jones}, \citenamefont {Chen},
  \citenamefont {Quintana}, \citenamefont {Mi}, \citenamefont {Dunsworth},
  \citenamefont {Gidney}, \citenamefont {Aleiner}, \citenamefont {Arute},
  \citenamefont {Arya}, \citenamefont {Atalaya} \emph
  {et~al.}}]{satzingerRealizingTopologicallyOrdered2021}%
  \BibitemOpen
  \bibfield  {author} {\bibinfo {author} {\bibfnamefont {K.~J.}\ \bibnamefont
  {Satzinger}}, \bibinfo {author} {\bibfnamefont {Y.-J.}\ \bibnamefont {Liu}},
  \bibinfo {author} {\bibfnamefont {A.}~\bibnamefont {Smith}}, \bibinfo
  {author} {\bibfnamefont {C.}~\bibnamefont {Knapp}}, \bibinfo {author}
  {\bibfnamefont {M.}~\bibnamefont {Newman}}, \bibinfo {author} {\bibfnamefont
  {C.}~\bibnamefont {Jones}}, \bibinfo {author} {\bibfnamefont
  {Z.}~\bibnamefont {Chen}}, \bibinfo {author} {\bibfnamefont {C.}~\bibnamefont
  {Quintana}}, \bibinfo {author} {\bibfnamefont {X.}~\bibnamefont {Mi}},
  \bibinfo {author} {\bibfnamefont {A.}~\bibnamefont {Dunsworth}}, \bibinfo
  {author} {\bibfnamefont {C.}~\bibnamefont {Gidney}}, \bibinfo {author}
  {\bibfnamefont {I.}~\bibnamefont {Aleiner}}, \bibinfo {author} {\bibfnamefont
  {F.}~\bibnamefont {Arute}}, \bibinfo {author} {\bibfnamefont
  {K.}~\bibnamefont {Arya}}, \bibinfo {author} {\bibfnamefont {J.}~\bibnamefont
  {Atalaya}}, \emph {et~al.},\ }\href {https://doi.org/10.1126/science.abi8378}
  {\bibfield  {journal} {\bibinfo  {journal} {Science}\ }\textbf {\bibinfo
  {volume} {374}},\ \bibinfo {pages} {1237} (\bibinfo {year}
  {2021})}\BibitemShut {NoStop}%
\bibitem [{\citenamefont {Semeghini}\ \emph {et~al.}(2021)\citenamefont
  {Semeghini}, \citenamefont {Levine}, \citenamefont {Keesling}, \citenamefont
  {Ebadi}, \citenamefont {Wang}, \citenamefont {Bluvstein}, \citenamefont
  {Verresen}, \citenamefont {Pichler}, \citenamefont {Kalinowski},
  \citenamefont {Samajdar}, \citenamefont {Omran}, \citenamefont {Sachdev},
  \citenamefont {Vishwanath}, \citenamefont {Greiner}, \citenamefont
  {Vuletić},\ and\ \citenamefont {Lukin}}]{Semeghini_2021}%
  \BibitemOpen
  \bibfield  {author} {\bibinfo {author} {\bibfnamefont {G.}~\bibnamefont
  {Semeghini}}, \bibinfo {author} {\bibfnamefont {H.}~\bibnamefont {Levine}},
  \bibinfo {author} {\bibfnamefont {A.}~\bibnamefont {Keesling}}, \bibinfo
  {author} {\bibfnamefont {S.}~\bibnamefont {Ebadi}}, \bibinfo {author}
  {\bibfnamefont {T.~T.}\ \bibnamefont {Wang}}, \bibinfo {author}
  {\bibfnamefont {D.}~\bibnamefont {Bluvstein}}, \bibinfo {author}
  {\bibfnamefont {R.}~\bibnamefont {Verresen}}, \bibinfo {author}
  {\bibfnamefont {H.}~\bibnamefont {Pichler}}, \bibinfo {author} {\bibfnamefont
  {M.}~\bibnamefont {Kalinowski}}, \bibinfo {author} {\bibfnamefont
  {R.}~\bibnamefont {Samajdar}}, \bibinfo {author} {\bibfnamefont
  {A.}~\bibnamefont {Omran}}, \bibinfo {author} {\bibfnamefont
  {S.}~\bibnamefont {Sachdev}}, \bibinfo {author} {\bibfnamefont
  {A.}~\bibnamefont {Vishwanath}}, \bibinfo {author} {\bibfnamefont
  {M.}~\bibnamefont {Greiner}}, \bibinfo {author} {\bibfnamefont
  {V.}~\bibnamefont {Vuletić}},\ and\ \bibinfo {author} {\bibfnamefont
  {M.~D.}\ \bibnamefont {Lukin}},\ }\href
  {https://doi.org/10.1126/science.abi8794} {\bibfield  {journal} {\bibinfo
  {journal} {Science}\ }\textbf {\bibinfo {volume} {374}},\ \bibinfo {pages}
  {1242–1247} (\bibinfo {year} {2021})}\BibitemShut {NoStop}%
\bibitem [{\citenamefont {Andersen}\ \emph {et~al.}(2023)\citenamefont
  {Andersen}, \citenamefont {Lensky}, \citenamefont {Kechedzhi}, \citenamefont
  {Drozdov}, \citenamefont {Bengtsson}, \citenamefont {Hong}, \citenamefont
  {Morvan}, \citenamefont {Mi}, \citenamefont {Opremcak}, \citenamefont
  {Acharya}, \citenamefont {Allen}, \citenamefont {Ansmann}, \citenamefont
  {Arute}, \citenamefont {Arya}, \citenamefont {Asfaw} \emph
  {et~al.}}]{andersen2023nonabelian}%
  \BibitemOpen
  \bibfield  {author} {\bibinfo {author} {\bibfnamefont {T.~I.}\ \bibnamefont
  {Andersen}}, \bibinfo {author} {\bibfnamefont {Y.~D.}\ \bibnamefont
  {Lensky}}, \bibinfo {author} {\bibfnamefont {K.}~\bibnamefont {Kechedzhi}},
  \bibinfo {author} {\bibfnamefont {I.}~\bibnamefont {Drozdov}}, \bibinfo
  {author} {\bibfnamefont {A.}~\bibnamefont {Bengtsson}}, \bibinfo {author}
  {\bibfnamefont {S.}~\bibnamefont {Hong}}, \bibinfo {author} {\bibfnamefont
  {A.}~\bibnamefont {Morvan}}, \bibinfo {author} {\bibfnamefont
  {X.}~\bibnamefont {Mi}}, \bibinfo {author} {\bibfnamefont {A.}~\bibnamefont
  {Opremcak}}, \bibinfo {author} {\bibfnamefont {R.}~\bibnamefont {Acharya}},
  \bibinfo {author} {\bibfnamefont {R.}~\bibnamefont {Allen}}, \bibinfo
  {author} {\bibfnamefont {M.}~\bibnamefont {Ansmann}}, \bibinfo {author}
  {\bibfnamefont {F.}~\bibnamefont {Arute}}, \bibinfo {author} {\bibfnamefont
  {K.}~\bibnamefont {Arya}}, \bibinfo {author} {\bibfnamefont {A.}~\bibnamefont
  {Asfaw}}, \emph {et~al.},\ }\href@noop {} {\bibinfo {title} {Non-abelian
  braiding of graph vertices in a superconducting processor}} (\bibinfo {year}
  {2023}),\ \Eprint {https://arxiv.org/abs/2210.10255} {arXiv:2210.10255
  [quant-ph]} \BibitemShut {NoStop}%
\bibitem [{\citenamefont {Iqbal}\ \emph {et~al.}(2023)\citenamefont {Iqbal},
  \citenamefont {Tantivasadakarn}, \citenamefont {Gatterman}, \citenamefont
  {Gerber}, \citenamefont {Gilmore}, \citenamefont {Gresh}, \citenamefont
  {Hankin}, \citenamefont {Hewitt}, \citenamefont {Horst}, \citenamefont
  {Matheny}, \citenamefont {Mengle}, \citenamefont {Neyenhuis}, \citenamefont
  {Vishwanath}, \citenamefont {Foss-Feig}, \citenamefont {Verresen},\ and\
  \citenamefont {Dreyer}}]{iqbal2023topological}%
  \BibitemOpen
  \bibfield  {author} {\bibinfo {author} {\bibfnamefont {M.}~\bibnamefont
  {Iqbal}}, \bibinfo {author} {\bibfnamefont {N.}~\bibnamefont
  {Tantivasadakarn}}, \bibinfo {author} {\bibfnamefont {T.~M.}\ \bibnamefont
  {Gatterman}}, \bibinfo {author} {\bibfnamefont {J.~A.}\ \bibnamefont
  {Gerber}}, \bibinfo {author} {\bibfnamefont {K.}~\bibnamefont {Gilmore}},
  \bibinfo {author} {\bibfnamefont {D.}~\bibnamefont {Gresh}}, \bibinfo
  {author} {\bibfnamefont {A.}~\bibnamefont {Hankin}}, \bibinfo {author}
  {\bibfnamefont {N.}~\bibnamefont {Hewitt}}, \bibinfo {author} {\bibfnamefont
  {C.~V.}\ \bibnamefont {Horst}}, \bibinfo {author} {\bibfnamefont
  {M.}~\bibnamefont {Matheny}}, \bibinfo {author} {\bibfnamefont
  {T.}~\bibnamefont {Mengle}}, \bibinfo {author} {\bibfnamefont
  {B.}~\bibnamefont {Neyenhuis}}, \bibinfo {author} {\bibfnamefont
  {A.}~\bibnamefont {Vishwanath}}, \bibinfo {author} {\bibfnamefont
  {M.}~\bibnamefont {Foss-Feig}}, \bibinfo {author} {\bibfnamefont
  {R.}~\bibnamefont {Verresen}},\ and\ \bibinfo {author} {\bibfnamefont
  {H.}~\bibnamefont {Dreyer}},\ }\href@noop {} {\bibinfo {title} {Topological
  order from measurements and feed-forward on a trapped ion quantum computer}}
  (\bibinfo {year} {2023}),\ \Eprint {https://arxiv.org/abs/2302.01917}
  {arXiv:2302.01917 [quant-ph]} \BibitemShut {NoStop}%
\bibitem [{\citenamefont {Iqbal}\ \emph {et~al.}(2024)\citenamefont {Iqbal},
  \citenamefont {Tantivasadakarn}, \citenamefont {Verresen}, \citenamefont
  {Campbell}, \citenamefont {Dreiling}, \citenamefont {Figgatt}, \citenamefont
  {Gaebler}, \citenamefont {Johansen}, \citenamefont {Mills}, \citenamefont
  {Moses}, \citenamefont {Pino}, \citenamefont {Ransford}, \citenamefont
  {Rowe}, \citenamefont {Siegfried}, \citenamefont {Stutz}, \citenamefont
  {Foss-Feig}, \citenamefont {Vishwanath},\ and\ \citenamefont
  {Dreyer}}]{Iqbal_2024}%
  \BibitemOpen
  \bibfield  {author} {\bibinfo {author} {\bibfnamefont {M.}~\bibnamefont
  {Iqbal}}, \bibinfo {author} {\bibfnamefont {N.}~\bibnamefont
  {Tantivasadakarn}}, \bibinfo {author} {\bibfnamefont {R.}~\bibnamefont
  {Verresen}}, \bibinfo {author} {\bibfnamefont {S.~L.}\ \bibnamefont
  {Campbell}}, \bibinfo {author} {\bibfnamefont {J.~M.}\ \bibnamefont
  {Dreiling}}, \bibinfo {author} {\bibfnamefont {C.}~\bibnamefont {Figgatt}},
  \bibinfo {author} {\bibfnamefont {J.~P.}\ \bibnamefont {Gaebler}}, \bibinfo
  {author} {\bibfnamefont {J.}~\bibnamefont {Johansen}}, \bibinfo {author}
  {\bibfnamefont {M.}~\bibnamefont {Mills}}, \bibinfo {author} {\bibfnamefont
  {S.~A.}\ \bibnamefont {Moses}}, \bibinfo {author} {\bibfnamefont {J.~M.}\
  \bibnamefont {Pino}}, \bibinfo {author} {\bibfnamefont {A.}~\bibnamefont
  {Ransford}}, \bibinfo {author} {\bibfnamefont {M.}~\bibnamefont {Rowe}},
  \bibinfo {author} {\bibfnamefont {P.}~\bibnamefont {Siegfried}}, \bibinfo
  {author} {\bibfnamefont {R.~P.}\ \bibnamefont {Stutz}}, \bibinfo {author}
  {\bibfnamefont {M.}~\bibnamefont {Foss-Feig}}, \bibinfo {author}
  {\bibfnamefont {A.}~\bibnamefont {Vishwanath}},\ and\ \bibinfo {author}
  {\bibfnamefont {H.}~\bibnamefont {Dreyer}},\ }\href
  {https://doi.org/10.1038/s41586-023-06934-4} {\bibfield  {journal} {\bibinfo
  {journal} {Nature}\ }\textbf {\bibinfo {volume} {626}},\ \bibinfo {pages}
  {505–511} (\bibinfo {year} {2024})}\BibitemShut {NoStop}%
\bibitem [{\citenamefont {Gottesman}(1997)}]{gottesman1997stabilizer}%
  \BibitemOpen
  \bibfield  {author} {\bibinfo {author} {\bibfnamefont {D.}~\bibnamefont
  {Gottesman}},\ }\href@noop {} {\bibinfo {title} {Stabilizer codes and quantum
  error correction}} (\bibinfo {year} {1997}),\ \Eprint
  {https://arxiv.org/abs/quant-ph/9705052} {arXiv:quant-ph/9705052 [quant-ph]}
  \BibitemShut {NoStop}%
\bibitem [{\citenamefont {Bombín}(2014)}]{Bomb_n_2014}%
  \BibitemOpen
  \bibfield  {author} {\bibinfo {author} {\bibfnamefont {H.}~\bibnamefont
  {Bombín}},\ }\href {https://doi.org/10.1007/s00220-014-1893-4} {\bibfield
  {journal} {\bibinfo  {journal} {Communications in Mathematical Physics}\
  }\textbf {\bibinfo {volume} {327}},\ \bibinfo {pages} {387–432} (\bibinfo
  {year} {2014})}\BibitemShut {NoStop}%
\bibitem [{\citenamefont {Haah}(2021)}]{haah_2021}%
  \BibitemOpen
  \bibfield  {author} {\bibinfo {author} {\bibfnamefont {J.}~\bibnamefont
  {Haah}},\ }\href {https://doi.org/10.1063/5.0021068} {\bibfield  {journal}
  {\bibinfo  {journal} {Journal of Mathematical Physics}\ }\textbf {\bibinfo
  {volume} {62}},\ \bibinfo {pages} {012201} (\bibinfo {year} {2021})},\
  \Eprint
  {https://arxiv.org/abs/https://pubs.aip.org/aip/jmp/article-pdf/doi/10.1063/5.0021068/15978399/012201{\textbackslash}\_1{\textbackslash}\_online.pdf}
  {https://pubs.aip.org/aip/jmp/article-pdf/doi/10.1063/5.0021068/15978399/012201{\textbackslash}\_1{\textbackslash}\_online.pdf}
  \BibitemShut {NoStop}%
\bibitem [{\citenamefont {Ellison}\ \emph {et~al.}(2022)\citenamefont
  {Ellison}, \citenamefont {Chen}, \citenamefont {Dua}, \citenamefont
  {Shirley}, \citenamefont {Tantivasadakarn},\ and\ \citenamefont
  {Williamson}}]{Ellison_2022}%
  \BibitemOpen
  \bibfield  {author} {\bibinfo {author} {\bibfnamefont {T.~D.}\ \bibnamefont
  {Ellison}}, \bibinfo {author} {\bibfnamefont {Y.-A.}\ \bibnamefont {Chen}},
  \bibinfo {author} {\bibfnamefont {A.}~\bibnamefont {Dua}}, \bibinfo {author}
  {\bibfnamefont {W.}~\bibnamefont {Shirley}}, \bibinfo {author} {\bibfnamefont
  {N.}~\bibnamefont {Tantivasadakarn}},\ and\ \bibinfo {author} {\bibfnamefont
  {D.~J.}\ \bibnamefont {Williamson}},\ }\bibfield  {journal} {\bibinfo
  {journal} {PRX Quantum}\ }\textbf {\bibinfo {volume} {3}},\ \href
  {https://doi.org/10.1103/prxquantum.3.010353} {10.1103/prxquantum.3.010353}
  (\bibinfo {year} {2022})\BibitemShut {NoStop}%
\bibitem [{\citenamefont {Chamon}(2005)}]{chamonQuantumGlassiness2005}%
  \BibitemOpen
  \bibfield  {author} {\bibinfo {author} {\bibfnamefont {C.}~\bibnamefont
  {Chamon}},\ }\href {https://doi.org/10.1103/PhysRevLett.94.040402} {\bibfield
   {journal} {\bibinfo  {journal} {Physical Review Letters}\ }\textbf {\bibinfo
  {volume} {94}},\ \bibinfo {pages} {040402} (\bibinfo {year} {2005})},\
  \Eprint {https://arxiv.org/abs/cond-mat/0404182} {arxiv:cond-mat/0404182}
  \BibitemShut {NoStop}%
\bibitem [{\citenamefont {Haah}(2011)}]{Haah_2011}%
  \BibitemOpen
  \bibfield  {author} {\bibinfo {author} {\bibfnamefont {J.}~\bibnamefont
  {Haah}},\ }\bibfield  {journal} {\bibinfo  {journal} {Physical Review A}\
  }\textbf {\bibinfo {volume} {83}},\ \href
  {https://doi.org/10.1103/physreva.83.042330} {10.1103/physreva.83.042330}
  (\bibinfo {year} {2011})\BibitemShut {NoStop}%
\bibitem [{\citenamefont {Yoshida}(2013)}]{Yoshida_2013}%
  \BibitemOpen
  \bibfield  {author} {\bibinfo {author} {\bibfnamefont {B.}~\bibnamefont
  {Yoshida}},\ }\bibfield  {journal} {\bibinfo  {journal} {Physical Review B}\
  }\textbf {\bibinfo {volume} {88}},\ \href
  {https://doi.org/10.1103/physrevb.88.125122} {10.1103/physrevb.88.125122}
  (\bibinfo {year} {2013})\BibitemShut {NoStop}%
\bibitem [{\citenamefont {Vijay}\ \emph {et~al.}(2015)\citenamefont {Vijay},
  \citenamefont {Haah},\ and\ \citenamefont
  {Fu}}]{vijayNewKindTopological2015}%
  \BibitemOpen
  \bibfield  {author} {\bibinfo {author} {\bibfnamefont {S.}~\bibnamefont
  {Vijay}}, \bibinfo {author} {\bibfnamefont {J.}~\bibnamefont {Haah}},\ and\
  \bibinfo {author} {\bibfnamefont {L.}~\bibnamefont {Fu}},\ }\href
  {https://doi.org/10.1103/PhysRevB.92.235136} {\bibfield  {journal} {\bibinfo
  {journal} {Physical Review B}\ }\textbf {\bibinfo {volume} {92}},\ \bibinfo
  {pages} {235136} (\bibinfo {year} {2015})}\BibitemShut {NoStop}%
\bibitem [{\citenamefont {Vijay}\ \emph {et~al.}(2016)\citenamefont {Vijay},
  \citenamefont {Haah},\ and\ \citenamefont
  {Fu}}]{vijayFractonTopologicalOrder2016}%
  \BibitemOpen
  \bibfield  {author} {\bibinfo {author} {\bibfnamefont {S.}~\bibnamefont
  {Vijay}}, \bibinfo {author} {\bibfnamefont {J.}~\bibnamefont {Haah}},\ and\
  \bibinfo {author} {\bibfnamefont {L.}~\bibnamefont {Fu}},\ }\href
  {https://doi.org/10.1103/PhysRevB.94.235157} {\bibfield  {journal} {\bibinfo
  {journal} {Physical Review B}\ }\textbf {\bibinfo {volume} {94}},\ \bibinfo
  {pages} {235157} (\bibinfo {year} {2016})}\BibitemShut {NoStop}%
\bibitem [{\citenamefont {Panteleev}\ and\ \citenamefont
  {Kalachev}(2022{\natexlab{a}})}]{panteleevAsymptoticallyGoodQuantum2022}%
  \BibitemOpen
  \bibfield  {author} {\bibinfo {author} {\bibfnamefont {P.}~\bibnamefont
  {Panteleev}}\ and\ \bibinfo {author} {\bibfnamefont {G.}~\bibnamefont
  {Kalachev}},\ }\href@noop {} {\bibinfo {title} {Asymptotically {{Good
  Quantum}} and {{Locally Testable Classical LDPC Codes}}}} (\bibinfo {year}
  {2022}{\natexlab{a}}),\ \Eprint {https://arxiv.org/abs/2111.03654}
  {arxiv:2111.03654 [quant-ph]} \BibitemShut {NoStop}%
\bibitem [{\citenamefont {Panteleev}\ and\ \citenamefont
  {Kalachev}(2022{\natexlab{b}})}]{panteleevQuantumLDPCCodes2022}%
  \BibitemOpen
  \bibfield  {author} {\bibinfo {author} {\bibfnamefont {P.}~\bibnamefont
  {Panteleev}}\ and\ \bibinfo {author} {\bibfnamefont {G.}~\bibnamefont
  {Kalachev}},\ }\href {https://doi.org/10.1109/TIT.2021.3119384} {\bibfield
  {journal} {\bibinfo  {journal} {IEEE Transactions on Information Theory}\
  }\textbf {\bibinfo {volume} {68}},\ \bibinfo {pages} {213} (\bibinfo {year}
  {2022}{\natexlab{b}})},\ \Eprint {https://arxiv.org/abs/2012.04068}
  {arxiv:2012.04068 [quant-ph]} \BibitemShut {NoStop}%
\bibitem [{\citenamefont {Wang}\ \emph {et~al.}(2003)\citenamefont {Wang},
  \citenamefont {Harrington},\ and\ \citenamefont
  {Preskill}}]{wangConfinementHiggsTransitionDisordered2003}%
  \BibitemOpen
  \bibfield  {author} {\bibinfo {author} {\bibfnamefont {C.}~\bibnamefont
  {Wang}}, \bibinfo {author} {\bibfnamefont {J.}~\bibnamefont {Harrington}},\
  and\ \bibinfo {author} {\bibfnamefont {J.}~\bibnamefont {Preskill}},\ }\href
  {https://doi.org/10.1016/S0003-4916(02)00019-2} {\bibfield  {journal}
  {\bibinfo  {journal} {Annals of Physics}\ }\textbf {\bibinfo {volume}
  {303}},\ \bibinfo {pages} {31} (\bibinfo {year} {2003})},\ \Eprint
  {https://arxiv.org/abs/quant-ph/0207088} {arxiv:quant-ph/0207088}
  \BibitemShut {NoStop}%
\bibitem [{\citenamefont {Bombin}\ \emph {et~al.}(2012)\citenamefont {Bombin},
  \citenamefont {Andrist}, \citenamefont {Ohzeki}, \citenamefont {Katzgraber},\
  and\ \citenamefont {Martin-Delgado}}]{Bombin_2012}%
  \BibitemOpen
  \bibfield  {author} {\bibinfo {author} {\bibfnamefont {H.}~\bibnamefont
  {Bombin}}, \bibinfo {author} {\bibfnamefont {R.~S.}\ \bibnamefont {Andrist}},
  \bibinfo {author} {\bibfnamefont {M.}~\bibnamefont {Ohzeki}}, \bibinfo
  {author} {\bibfnamefont {H.~G.}\ \bibnamefont {Katzgraber}},\ and\ \bibinfo
  {author} {\bibfnamefont {M.~A.}\ \bibnamefont {Martin-Delgado}},\ }\bibfield
  {journal} {\bibinfo  {journal} {Physical Review X}\ }\textbf {\bibinfo
  {volume} {2}},\ \href {https://doi.org/10.1103/physrevx.2.021004}
  {10.1103/physrevx.2.021004} (\bibinfo {year} {2012})\BibitemShut {NoStop}%
\bibitem [{\citenamefont {Kovalev}\ \emph {et~al.}(2018)\citenamefont
  {Kovalev}, \citenamefont {Prabhakar}, \citenamefont {Dumer},\ and\
  \citenamefont {Pryadko}}]{kovalevNumericalAnalyticalBounds2018}%
  \BibitemOpen
  \bibfield  {author} {\bibinfo {author} {\bibfnamefont {A.~A.}\ \bibnamefont
  {Kovalev}}, \bibinfo {author} {\bibfnamefont {S.}~\bibnamefont {Prabhakar}},
  \bibinfo {author} {\bibfnamefont {I.}~\bibnamefont {Dumer}},\ and\ \bibinfo
  {author} {\bibfnamefont {L.~P.}\ \bibnamefont {Pryadko}},\ }\href
  {https://doi.org/10.1103/PhysRevA.97.062320} {\bibfield  {journal} {\bibinfo
  {journal} {Physical Review A}\ }\textbf {\bibinfo {volume} {97}},\ \bibinfo
  {pages} {062320} (\bibinfo {year} {2018})}\BibitemShut {NoStop}%
\bibitem [{\citenamefont {Chubb}\ and\ \citenamefont
  {Flammia}(2021)}]{chubbStatisticalMechanicalModels2021}%
  \BibitemOpen
  \bibfield  {author} {\bibinfo {author} {\bibfnamefont {C.~T.}\ \bibnamefont
  {Chubb}}\ and\ \bibinfo {author} {\bibfnamefont {S.~T.}\ \bibnamefont
  {Flammia}},\ }\href {https://doi.org/10.4171/AIHPD/105} {\bibfield  {journal}
  {\bibinfo  {journal} {Annales de l'Institut Henri Poincar{\'e} D}\ }\textbf
  {\bibinfo {volume} {8}},\ \bibinfo {pages} {269} (\bibinfo {year} {2021})},\
  \Eprint {https://arxiv.org/abs/1809.10704} {arxiv:1809.10704 [cond-mat,
  physics:quant-ph]} \BibitemShut {NoStop}%
\bibitem [{\citenamefont {Chen}\ and\ \citenamefont
  {Grover}(2023)}]{chen2023separability}%
  \BibitemOpen
  \bibfield  {author} {\bibinfo {author} {\bibfnamefont {Y.-H.}\ \bibnamefont
  {Chen}}\ and\ \bibinfo {author} {\bibfnamefont {T.}~\bibnamefont {Grover}},\
  }\href@noop {} {\bibinfo {title} {Separability transitions in topological
  states induced by local decoherence}} (\bibinfo {year} {2023}),\ \Eprint
  {https://arxiv.org/abs/2309.11879} {arXiv:2309.11879 [quant-ph]} \BibitemShut
  {NoStop}%
\bibitem [{\citenamefont {Fan}\ \emph {et~al.}(2023)\citenamefont {Fan},
  \citenamefont {Bao}, \citenamefont {Altman},\ and\ \citenamefont
  {Vishwanath}}]{fanDiagnosticsMixedstateTopological2023}%
  \BibitemOpen
  \bibfield  {author} {\bibinfo {author} {\bibfnamefont {R.}~\bibnamefont
  {Fan}}, \bibinfo {author} {\bibfnamefont {Y.}~\bibnamefont {Bao}}, \bibinfo
  {author} {\bibfnamefont {E.}~\bibnamefont {Altman}},\ and\ \bibinfo {author}
  {\bibfnamefont {A.}~\bibnamefont {Vishwanath}},\ }\href@noop {} {\bibinfo
  {title} {Diagnostics of mixed-state topological order and breakdown of
  quantum memory}} (\bibinfo {year} {2023}),\ \Eprint
  {https://arxiv.org/abs/2301.05689} {arxiv:2301.05689 [cond-mat,
  physics:quant-ph]} \BibitemShut {NoStop}%
\bibitem [{\citenamefont
  {Kitaev}(1997)}]{kitaevQuantumComputationsAlgorithms1997a}%
  \BibitemOpen
  \bibfield  {author} {\bibinfo {author} {\bibfnamefont {A.~Y.}\ \bibnamefont
  {Kitaev}},\ }\href {https://doi.org/10.1070/RM1997v052n06ABEH002155}
  {\bibfield  {journal} {\bibinfo  {journal} {Russian Mathematical Surveys}\
  }\textbf {\bibinfo {volume} {52}},\ \bibinfo {pages} {1191} (\bibinfo {year}
  {1997})}\BibitemShut {NoStop}%
\bibitem [{\citenamefont {Bombin}\ and\ \citenamefont
  {Martin-Delgado}(2007)}]{Bombin_2007}%
  \BibitemOpen
  \bibfield  {author} {\bibinfo {author} {\bibfnamefont {H.}~\bibnamefont
  {Bombin}}\ and\ \bibinfo {author} {\bibfnamefont {M.~A.}\ \bibnamefont
  {Martin-Delgado}},\ }\bibfield  {journal} {\bibinfo  {journal} {Journal of
  Mathematical Physics}\ }\textbf {\bibinfo {volume} {48}},\ \href
  {https://doi.org/10.1063/1.2731356} {10.1063/1.2731356} (\bibinfo {year}
  {2007})\BibitemShut {NoStop}%
\bibitem [{\citenamefont {Bravyi}\ and\ \citenamefont
  {Hastings}(2013)}]{bravyiHomologicalProductCodes2013}%
  \BibitemOpen
  \bibfield  {author} {\bibinfo {author} {\bibfnamefont {S.}~\bibnamefont
  {Bravyi}}\ and\ \bibinfo {author} {\bibfnamefont {M.~B.}\ \bibnamefont
  {Hastings}},\ }\href@noop {} {\bibinfo {title} {Homological {{Product
  Codes}}}} (\bibinfo {year} {2013}),\ \Eprint
  {https://arxiv.org/abs/1311.0885} {arxiv:1311.0885 [quant-ph]} \BibitemShut
  {NoStop}%
\bibitem [{\citenamefont {Hamma}\ \emph {et~al.}(2005)\citenamefont {Hamma},
  \citenamefont {Zanardi},\ and\ \citenamefont {Wen}}]{Hamma_2005}%
  \BibitemOpen
  \bibfield  {author} {\bibinfo {author} {\bibfnamefont {A.}~\bibnamefont
  {Hamma}}, \bibinfo {author} {\bibfnamefont {P.}~\bibnamefont {Zanardi}},\
  and\ \bibinfo {author} {\bibfnamefont {X.-G.}\ \bibnamefont {Wen}},\
  }\bibfield  {journal} {\bibinfo  {journal} {Physical Review B}\ }\textbf
  {\bibinfo {volume} {72}},\ \href {https://doi.org/10.1103/physrevb.72.035307}
  {10.1103/physrevb.72.035307} (\bibinfo {year} {2005})\BibitemShut {NoStop}%
\bibitem [{\citenamefont {Castelnovo}\ and\ \citenamefont
  {Chamon}(2008)}]{Castelnovo_2008}%
  \BibitemOpen
  \bibfield  {author} {\bibinfo {author} {\bibfnamefont {C.}~\bibnamefont
  {Castelnovo}}\ and\ \bibinfo {author} {\bibfnamefont {C.}~\bibnamefont
  {Chamon}},\ }\bibfield  {journal} {\bibinfo  {journal} {Physical Review B}\
  }\textbf {\bibinfo {volume} {78}},\ \href
  {https://doi.org/10.1103/physrevb.78.155120} {10.1103/physrevb.78.155120}
  (\bibinfo {year} {2008})\BibitemShut {NoStop}%
\bibitem [{\citenamefont
  {Michnicki}(2014)}]{michnicki3DTopologicalQuantum2014}%
  \BibitemOpen
  \bibfield  {author} {\bibinfo {author} {\bibfnamefont {K.~P.}\ \bibnamefont
  {Michnicki}},\ }\href {https://doi.org/10.1103/PhysRevLett.113.130501}
  {\bibfield  {journal} {\bibinfo  {journal} {Physical Review Letters}\
  }\textbf {\bibinfo {volume} {113}},\ \bibinfo {pages} {130501} (\bibinfo
  {year} {2014})}\BibitemShut {NoStop}%
\bibitem [{\citenamefont {Wegner}(1971)}]{wegnerDualityGeneralizedIsing1971}%
  \BibitemOpen
  \bibfield  {author} {\bibinfo {author} {\bibfnamefont {F.~J.}\ \bibnamefont
  {Wegner}},\ }\href {https://doi.org/10.1063/1.1665530} {\bibfield  {journal}
  {\bibinfo  {journal} {Journal of Mathematical Physics}\ }\textbf {\bibinfo
  {volume} {12}},\ \bibinfo {pages} {2259} (\bibinfo {year}
  {1971})}\BibitemShut {NoStop}%
\bibitem [{\citenamefont {Kogut}(1979)}]{kogutIntroductionLatticeGauge1979}%
  \BibitemOpen
  \bibfield  {author} {\bibinfo {author} {\bibfnamefont {J.~B.}\ \bibnamefont
  {Kogut}},\ }\href {https://doi.org/10.1103/RevModPhys.51.659} {\bibfield
  {journal} {\bibinfo  {journal} {Reviews of Modern Physics}\ }\textbf
  {\bibinfo {volume} {51}},\ \bibinfo {pages} {659} (\bibinfo {year}
  {1979})}\BibitemShut {NoStop}%
\bibitem [{\citenamefont {Haah}(2017)}]{Haah_2017}%
  \BibitemOpen
  \bibfield  {author} {\bibinfo {author} {\bibfnamefont {J.}~\bibnamefont
  {Haah}},\ }\href {https://doi.org/10.15446/recolma.v50n2.62214} {\bibfield
  {journal} {\bibinfo  {journal} {Revista Colombiana de Matemáticas}\ }\textbf
  {\bibinfo {volume} {50}},\ \bibinfo {pages} {299} (\bibinfo {year}
  {2017})}\BibitemShut {NoStop}%
\bibitem [{\citenamefont {R{\'e}nyi}(1961)}]{renyi1961measures}%
  \BibitemOpen
  \bibfield  {author} {\bibinfo {author} {\bibfnamefont {A.}~\bibnamefont
  {R{\'e}nyi}},\ }in\ \href@noop {} {\emph {\bibinfo {booktitle} {Proceedings
  of the Fourth Berkeley Symposium on Mathematical Statistics and Probability,
  Volume 1: Contributions to the Theory of Statistics}}},\ Vol.~\bibinfo
  {volume} {4}\ (\bibinfo {organization} {University of California Press},\
  \bibinfo {year} {1961})\ pp.\ \bibinfo {pages} {547--562}\BibitemShut
  {NoStop}%
\bibitem [{\citenamefont {Müller-Lennert}\ \emph {et~al.}(2013)\citenamefont
  {Müller-Lennert}, \citenamefont {Dupuis}, \citenamefont {Szehr},
  \citenamefont {Fehr},\ and\ \citenamefont
  {Tomamichel}}]{Muller_Lennert_2013}%
  \BibitemOpen
  \bibfield  {author} {\bibinfo {author} {\bibfnamefont {M.}~\bibnamefont
  {Müller-Lennert}}, \bibinfo {author} {\bibfnamefont {F.}~\bibnamefont
  {Dupuis}}, \bibinfo {author} {\bibfnamefont {O.}~\bibnamefont {Szehr}},
  \bibinfo {author} {\bibfnamefont {S.}~\bibnamefont {Fehr}},\ and\ \bibinfo
  {author} {\bibfnamefont {M.}~\bibnamefont {Tomamichel}},\ }\bibfield
  {journal} {\bibinfo  {journal} {Journal of Mathematical Physics}\ }\textbf
  {\bibinfo {volume} {54}},\ \href {https://doi.org/10.1063/1.4838856}
  {10.1063/1.4838856} (\bibinfo {year} {2013})\BibitemShut {NoStop}%
\bibitem [{\citenamefont {Umegaki}(1962)}]{ume1962}%
  \BibitemOpen
  \bibfield  {author} {\bibinfo {author} {\bibfnamefont {H.}~\bibnamefont
  {Umegaki}},\ }\href {https://doi.org/10.2996/kmj/1138844604} {\bibfield
  {journal} {\bibinfo  {journal} {Kodai Mathematical Seminar Reports}\ }\textbf
  {\bibinfo {volume} {14}},\ \bibinfo {pages} {59 } (\bibinfo {year}
  {1962})}\BibitemShut {NoStop}%
\bibitem [{\citenamefont {Vedral}(2002)}]{Vedral_2002}%
  \BibitemOpen
  \bibfield  {author} {\bibinfo {author} {\bibfnamefont {V.}~\bibnamefont
  {Vedral}},\ }\href {https://doi.org/10.1103/revmodphys.74.197} {\bibfield
  {journal} {\bibinfo  {journal} {Reviews of Modern Physics}\ }\textbf
  {\bibinfo {volume} {74}},\ \bibinfo {pages} {197–234} (\bibinfo {year}
  {2002})}\BibitemShut {NoStop}%
\bibitem [{\citenamefont {Petz}(1986)}]{PETZ198657}%
  \BibitemOpen
  \bibfield  {author} {\bibinfo {author} {\bibfnamefont {D.}~\bibnamefont
  {Petz}},\ }\href {https://doi.org/10.1016/0034-4877(86)90067-4} {\bibfield
  {journal} {\bibinfo  {journal} {Reports on Mathematical Physics}\ }\textbf
  {\bibinfo {volume} {23}},\ \bibinfo {pages} {57} (\bibinfo {year}
  {1986})}\BibitemShut {NoStop}%
\bibitem [{\citenamefont {Schumacher}\ and\ \citenamefont
  {Nielsen}(1996)}]{Schumacher_1996}%
  \BibitemOpen
  \bibfield  {author} {\bibinfo {author} {\bibfnamefont {B.}~\bibnamefont
  {Schumacher}}\ and\ \bibinfo {author} {\bibfnamefont {M.~A.}\ \bibnamefont
  {Nielsen}},\ }\href {https://doi.org/10.1103/physreva.54.2629} {\bibfield
  {journal} {\bibinfo  {journal} {Physical Review A}\ }\textbf {\bibinfo
  {volume} {54}},\ \bibinfo {pages} {2629–2635} (\bibinfo {year}
  {1996})}\BibitemShut {NoStop}%
\bibitem [{\citenamefont {Lloyd}(1997)}]{Lloyd_1997}%
  \BibitemOpen
  \bibfield  {author} {\bibinfo {author} {\bibfnamefont {S.}~\bibnamefont
  {Lloyd}},\ }\href {https://doi.org/10.1103/physreva.55.1613} {\bibfield
  {journal} {\bibinfo  {journal} {Physical Review A}\ }\textbf {\bibinfo
  {volume} {55}},\ \bibinfo {pages} {1613–1622} (\bibinfo {year}
  {1997})}\BibitemShut {NoStop}%
\bibitem [{\citenamefont {Peres}(1996)}]{Peres_1996}%
  \BibitemOpen
  \bibfield  {author} {\bibinfo {author} {\bibfnamefont {A.}~\bibnamefont
  {Peres}},\ }\href {https://doi.org/10.1103/physrevlett.77.1413} {\bibfield
  {journal} {\bibinfo  {journal} {Physical Review Letters}\ }\textbf {\bibinfo
  {volume} {77}},\ \bibinfo {pages} {1413–1415} (\bibinfo {year}
  {1996})}\BibitemShut {NoStop}%
\bibitem [{\citenamefont {Horodecki}\ \emph {et~al.}(1996)\citenamefont
  {Horodecki}, \citenamefont {Horodecki},\ and\ \citenamefont
  {Horodecki}}]{Horodecki_1996}%
  \BibitemOpen
  \bibfield  {author} {\bibinfo {author} {\bibfnamefont {M.}~\bibnamefont
  {Horodecki}}, \bibinfo {author} {\bibfnamefont {P.}~\bibnamefont
  {Horodecki}},\ and\ \bibinfo {author} {\bibfnamefont {R.}~\bibnamefont
  {Horodecki}},\ }\href {https://doi.org/10.1016/s0375-9601(96)00706-2}
  {\bibfield  {journal} {\bibinfo  {journal} {Physics Letters A}\ }\textbf
  {\bibinfo {volume} {223}},\ \bibinfo {pages} {1–8} (\bibinfo {year}
  {1996})}\BibitemShut {NoStop}%
\bibitem [{\citenamefont {\ifmmode~\dot{Z}\else \.{Z}\fi{}yczkowski}\ \emph
  {et~al.}(1998)\citenamefont {\ifmmode~\dot{Z}\else \.{Z}\fi{}yczkowski},
  \citenamefont {Horodecki}, \citenamefont {Sanpera},\ and\ \citenamefont
  {Lewenstein}}]{horodecki1998}%
  \BibitemOpen
  \bibfield  {author} {\bibinfo {author} {\bibfnamefont {K.}~\bibnamefont
  {\ifmmode~\dot{Z}\else \.{Z}\fi{}yczkowski}}, \bibinfo {author}
  {\bibfnamefont {P.}~\bibnamefont {Horodecki}}, \bibinfo {author}
  {\bibfnamefont {A.}~\bibnamefont {Sanpera}},\ and\ \bibinfo {author}
  {\bibfnamefont {M.}~\bibnamefont {Lewenstein}},\ }\href
  {https://doi.org/10.1103/PhysRevA.58.883} {\bibfield  {journal} {\bibinfo
  {journal} {Phys. Rev. A}\ }\textbf {\bibinfo {volume} {58}},\ \bibinfo
  {pages} {883} (\bibinfo {year} {1998})}\BibitemShut {NoStop}%
\bibitem [{\citenamefont {Vidal}\ and\ \citenamefont
  {Werner}(2002)}]{Vidal_2002}%
  \BibitemOpen
  \bibfield  {author} {\bibinfo {author} {\bibfnamefont {G.}~\bibnamefont
  {Vidal}}\ and\ \bibinfo {author} {\bibfnamefont {R.~F.}\ \bibnamefont
  {Werner}},\ }\bibfield  {journal} {\bibinfo  {journal} {Physical Review A}\
  }\textbf {\bibinfo {volume} {65}},\ \href
  {https://doi.org/10.1103/physreva.65.032314} {10.1103/physreva.65.032314}
  (\bibinfo {year} {2002})\BibitemShut {NoStop}%
\bibitem [{\citenamefont {Lee}\ and\ \citenamefont {Vidal}(2013)}]{lee2013}%
  \BibitemOpen
  \bibfield  {author} {\bibinfo {author} {\bibfnamefont {Y.~A.}\ \bibnamefont
  {Lee}}\ and\ \bibinfo {author} {\bibfnamefont {G.}~\bibnamefont {Vidal}},\
  }\href {https://doi.org/10.1103/PhysRevA.88.042318} {\bibfield  {journal}
  {\bibinfo  {journal} {Phys. Rev. A}\ }\textbf {\bibinfo {volume} {88}},\
  \bibinfo {pages} {042318} (\bibinfo {year} {2013})}\BibitemShut {NoStop}%
\bibitem [{\citenamefont {Ma}\ \emph {et~al.}(2018)\citenamefont {Ma},
  \citenamefont {Schmitz}, \citenamefont {Parameswaran}, \citenamefont
  {Hermele},\ and\ \citenamefont {Nandkishore}}]{Ma_2018}%
  \BibitemOpen
  \bibfield  {author} {\bibinfo {author} {\bibfnamefont {H.}~\bibnamefont
  {Ma}}, \bibinfo {author} {\bibfnamefont {A.~T.}\ \bibnamefont {Schmitz}},
  \bibinfo {author} {\bibfnamefont {S.~A.}\ \bibnamefont {Parameswaran}},
  \bibinfo {author} {\bibfnamefont {M.}~\bibnamefont {Hermele}},\ and\ \bibinfo
  {author} {\bibfnamefont {R.~M.}\ \bibnamefont {Nandkishore}},\ }\bibfield
  {journal} {\bibinfo  {journal} {Physical Review B}\ }\textbf {\bibinfo
  {volume} {97}},\ \href {https://doi.org/10.1103/physrevb.97.125101}
  {10.1103/physrevb.97.125101} (\bibinfo {year} {2018})\BibitemShut {NoStop}%
\bibitem [{\citenamefont {Shi}\ and\ \citenamefont {Lu}(2018)}]{Shi_2018}%
  \BibitemOpen
  \bibfield  {author} {\bibinfo {author} {\bibfnamefont {B.}~\bibnamefont
  {Shi}}\ and\ \bibinfo {author} {\bibfnamefont {Y.-M.}\ \bibnamefont {Lu}},\
  }\bibfield  {journal} {\bibinfo  {journal} {Physical Review B}\ }\textbf
  {\bibinfo {volume} {97}},\ \href {https://doi.org/10.1103/physrevb.97.144106}
  {10.1103/physrevb.97.144106} (\bibinfo {year} {2018})\BibitemShut {NoStop}%
\bibitem [{\citenamefont {Shirley}\ \emph {et~al.}(2019)\citenamefont
  {Shirley}, \citenamefont {Slagle},\ and\ \citenamefont
  {Chen}}]{shirleyUniversalEntanglementSignatures2019}%
  \BibitemOpen
  \bibfield  {author} {\bibinfo {author} {\bibfnamefont {W.}~\bibnamefont
  {Shirley}}, \bibinfo {author} {\bibfnamefont {K.}~\bibnamefont {Slagle}},\
  and\ \bibinfo {author} {\bibfnamefont {X.}~\bibnamefont {Chen}},\ }\href
  {https://doi.org/10.21468/SciPostPhys.6.1.015} {\bibfield  {journal}
  {\bibinfo  {journal} {SciPost Physics}\ }\textbf {\bibinfo {volume} {6}},\
  \bibinfo {pages} {015} (\bibinfo {year} {2019})},\ \Eprint
  {https://arxiv.org/abs/1803.10426} {arxiv:1803.10426 [cond-mat,
  physics:quant-ph]} \BibitemShut {NoStop}%
\bibitem [{\citenamefont {Calabrese}\ \emph {et~al.}(2012)\citenamefont
  {Calabrese}, \citenamefont {Cardy},\ and\ \citenamefont
  {Tonni}}]{calabrese2012}%
  \BibitemOpen
  \bibfield  {author} {\bibinfo {author} {\bibfnamefont {P.}~\bibnamefont
  {Calabrese}}, \bibinfo {author} {\bibfnamefont {J.}~\bibnamefont {Cardy}},\
  and\ \bibinfo {author} {\bibfnamefont {E.}~\bibnamefont {Tonni}},\ }\href
  {https://doi.org/10.1103/PhysRevLett.109.130502} {\bibfield  {journal}
  {\bibinfo  {journal} {Phys. Rev. Lett.}\ }\textbf {\bibinfo {volume} {109}},\
  \bibinfo {pages} {130502} (\bibinfo {year} {2012})}\BibitemShut {NoStop}%
\bibitem [{\citenamefont {Levin}\ and\ \citenamefont {Gu}(2012)}]{Levin_2012}%
  \BibitemOpen
  \bibfield  {author} {\bibinfo {author} {\bibfnamefont {M.}~\bibnamefont
  {Levin}}\ and\ \bibinfo {author} {\bibfnamefont {Z.-C.}\ \bibnamefont {Gu}},\
  }\bibfield  {journal} {\bibinfo  {journal} {Physical Review B}\ }\textbf
  {\bibinfo {volume} {86}},\ \href {https://doi.org/10.1103/physrevb.86.115109}
  {10.1103/physrevb.86.115109} (\bibinfo {year} {2012})\BibitemShut {NoStop}%
\bibitem [{\citenamefont {Haegeman}\ \emph {et~al.}(2015)\citenamefont
  {Haegeman}, \citenamefont {Van~Acoleyen}, \citenamefont {Schuch},
  \citenamefont {Cirac},\ and\ \citenamefont {Verstraete}}]{Haegeman_2015}%
  \BibitemOpen
  \bibfield  {author} {\bibinfo {author} {\bibfnamefont {J.}~\bibnamefont
  {Haegeman}}, \bibinfo {author} {\bibfnamefont {K.}~\bibnamefont
  {Van~Acoleyen}}, \bibinfo {author} {\bibfnamefont {N.}~\bibnamefont
  {Schuch}}, \bibinfo {author} {\bibfnamefont {J.~I.}\ \bibnamefont {Cirac}},\
  and\ \bibinfo {author} {\bibfnamefont {F.}~\bibnamefont {Verstraete}},\
  }\bibfield  {journal} {\bibinfo  {journal} {Physical Review X}\ }\textbf
  {\bibinfo {volume} {5}},\ \href {https://doi.org/10.1103/physrevx.5.011024}
  {10.1103/physrevx.5.011024} (\bibinfo {year} {2015})\BibitemShut {NoStop}%
\bibitem [{\citenamefont {Williamson}(2016)}]{Williamson_2016}%
  \BibitemOpen
  \bibfield  {author} {\bibinfo {author} {\bibfnamefont {D.~J.}\ \bibnamefont
  {Williamson}},\ }\bibfield  {journal} {\bibinfo  {journal} {Physical Review
  B}\ }\textbf {\bibinfo {volume} {94}},\ \href
  {https://doi.org/10.1103/physrevb.94.155128} {10.1103/physrevb.94.155128}
  (\bibinfo {year} {2016})\BibitemShut {NoStop}%
\bibitem [{Note1()}]{Note1}%
  \BibitemOpen
  \bibinfo {note} {Intuitively, $\protect \mathsf {G}$ generates the collection
  of subsets of classical spins that can be simultaneously flipped without
  incurring an energy penalty, i.e., symmetries of the classical model. When we
  gauge the classical model to obtain the quantum one, these symmetries become
  allowed gauge transformations}\BibitemShut {NoStop}%
\bibitem [{\citenamefont {Kramers}\ and\ \citenamefont
  {Wannier}(1941)}]{kramers-wannier}%
  \BibitemOpen
  \bibfield  {author} {\bibinfo {author} {\bibfnamefont {H.~A.}\ \bibnamefont
  {Kramers}}\ and\ \bibinfo {author} {\bibfnamefont {G.~H.}\ \bibnamefont
  {Wannier}},\ }\href {https://doi.org/10.1103/PhysRev.60.252} {\bibfield
  {journal} {\bibinfo  {journal} {Phys. Rev.}\ }\textbf {\bibinfo {volume}
  {60}},\ \bibinfo {pages} {252} (\bibinfo {year} {1941})}\BibitemShut
  {NoStop}%
\bibitem [{\citenamefont {Kardar}(2007)}]{Kardar_2007}%
  \BibitemOpen
  \bibfield  {author} {\bibinfo {author} {\bibfnamefont {M.}~\bibnamefont
  {Kardar}},\ }\href@noop {} {\emph {\bibinfo {title} {Statistical Physics of
  Particles}}}\ (\bibinfo  {publisher} {Cambridge University Press},\ \bibinfo
  {year} {2007})\BibitemShut {NoStop}%
\bibitem [{\citenamefont {Browne}(2014)}]{browne-notes}%
  \BibitemOpen
  \bibfield  {author} {\bibinfo {author} {\bibfnamefont {D.}~\bibnamefont
  {Browne}},\ }\href
  {https://sites.google.com/site/danbrowneucl/teaching/lectures-on-topological-codes-and-quantum-computation}
  {\bibinfo {title} {Lectures on topological codes and quantum computation}}
  (\bibinfo {year} {2014})\BibitemShut {NoStop}%
\bibitem [{\citenamefont {Mittal}(2022)}]{mittal-notes}%
  \BibitemOpen
  \bibfield  {author} {\bibinfo {author} {\bibfnamefont {T.}~\bibnamefont
  {Mittal}},\ }\href {https://mittaltushant.github.io/readings/qldpc.pdf}
  {\bibinfo {title} {Quantum {{LDPC}} codes: an exposition of recent results}}
  (\bibinfo {year} {2022})\BibitemShut {NoStop}%
\bibitem [{\citenamefont {Rakovszky}\ and\ \citenamefont
  {Khemani}(2023)}]{rakovszkyPhysicsGoodLDPC2023}%
  \BibitemOpen
  \bibfield  {author} {\bibinfo {author} {\bibfnamefont {T.}~\bibnamefont
  {Rakovszky}}\ and\ \bibinfo {author} {\bibfnamefont {V.}~\bibnamefont
  {Khemani}},\ }\href@noop {} {\bibinfo {title} {The {{Physics}} of (good)
  {{LDPC Codes I}}. {{Gauging}} and dualities}} (\bibinfo {year} {2023}),\
  \Eprint {https://arxiv.org/abs/2310.16032} {arxiv:2310.16032 [cond-mat,
  physics:hep-th, physics:quant-ph]} \BibitemShut {NoStop}%
\bibitem [{\citenamefont {Pal}(2019)}]{palPhysicistIntroductionAlgebraic2019}%
  \BibitemOpen
  \bibfield  {author} {\bibinfo {author} {\bibfnamefont {P.~B.}\ \bibnamefont
  {Pal}},\ }\href {https://doi.org/10.1017/9781108679114} {\emph {\bibinfo
  {title} {A {{Physicist}}'s {{Introduction}} to {{Algebraic Structures}}:
  {{Vector Spaces}}, {{Groups}}, {{Topological Spaces}} and {{More}}}}},\
  \bibinfo {edition} {1st}\ ed.\ (\bibinfo  {publisher} {{Cambridge University
  Press}},\ \bibinfo {year} {2019})\BibitemShut {NoStop}%
\bibitem [{Note2()}]{Note2}%
  \BibitemOpen
  \bibinfo {note} {Here, $\protect \mathbb {F}_2 = \{0, 1\}$ is the field over
  two elements.}\BibitemShut {Stop}%
\bibitem [{\citenamefont {Hasenbusch}(2001)}]{HASENBUSCHMARTIN2001MCSO}%
  \BibitemOpen
  \bibfield  {author} {\bibinfo {author} {\bibfnamefont {M.}~\bibnamefont
  {Hasenbusch}},\ }\href@noop {} {\bibfield  {journal} {\bibinfo  {journal}
  {International journal of modern physics. C, Computational physics, physical
  computation}\ }\textbf {\bibinfo {volume} {12}},\ \bibinfo {pages} {911}
  (\bibinfo {year} {2001})}\BibitemShut {NoStop}%
\bibitem [{\citenamefont {Johnston}\ and\ \citenamefont
  {Ranasinghe}(2011)}]{johnstonDualGonihedric3D2011}%
  \BibitemOpen
  \bibfield  {author} {\bibinfo {author} {\bibfnamefont {D.~A.}\ \bibnamefont
  {Johnston}}\ and\ \bibinfo {author} {\bibfnamefont {R.~P. K. C.~M.}\
  \bibnamefont {Ranasinghe}},\ }\href
  {https://doi.org/10.1088/1751-8113/44/29/295004} {\bibfield  {journal}
  {\bibinfo  {journal} {Journal of Physics A: Mathematical and Theoretical}\
  }\textbf {\bibinfo {volume} {44}},\ \bibinfo {pages} {295004} (\bibinfo
  {year} {2011})},\ \Eprint {https://arxiv.org/abs/1104.3224} {arxiv:1104.3224
  [cond-mat]} \BibitemShut {NoStop}%
\bibitem [{\citenamefont {Johnston}\ and\ \citenamefont
  {Malmini}(1996)}]{johnstonGonihedric3DIsing1996}%
  \BibitemOpen
  \bibfield  {author} {\bibinfo {author} {\bibfnamefont {D.}~\bibnamefont
  {Johnston}}\ and\ \bibinfo {author} {\bibfnamefont {R.~P.}\ \bibnamefont
  {Malmini}},\ }\href {https://doi.org/10.1016/0370-2693(96)00391-7} {\bibfield
   {journal} {\bibinfo  {journal} {Physics Letters B}\ }\textbf {\bibinfo
  {volume} {378}},\ \bibinfo {pages} {87} (\bibinfo {year} {1996})}\BibitemShut
  {NoStop}%
\bibitem [{\citenamefont {Ozeki}\ and\ \citenamefont
  {Nishimori}(1987)}]{Ozeki1987PhaseDA}%
  \BibitemOpen
  \bibfield  {author} {\bibinfo {author} {\bibfnamefont {Y.}~\bibnamefont
  {Ozeki}}\ and\ \bibinfo {author} {\bibfnamefont {H.}~\bibnamefont
  {Nishimori}},\ }\href {https://api.semanticscholar.org/CorpusID:123092112}
  {\bibfield  {journal} {\bibinfo  {journal} {Journal of the Physical Society
  of Japan}\ }\textbf {\bibinfo {volume} {56}},\ \bibinfo {pages} {1568}
  (\bibinfo {year} {1987})}\BibitemShut {NoStop}%
\bibitem [{\citenamefont {Hasenbusch}\ \emph {et~al.}(2007)\citenamefont
  {Hasenbusch}, \citenamefont {Toldin}, \citenamefont {Pelissetto},\ and\
  \citenamefont {Vicari}}]{Hasenbusch_2007}%
  \BibitemOpen
  \bibfield  {author} {\bibinfo {author} {\bibfnamefont {M.}~\bibnamefont
  {Hasenbusch}}, \bibinfo {author} {\bibfnamefont {F.~P.}\ \bibnamefont
  {Toldin}}, \bibinfo {author} {\bibfnamefont {A.}~\bibnamefont {Pelissetto}},\
  and\ \bibinfo {author} {\bibfnamefont {E.}~\bibnamefont {Vicari}},\
  }\bibfield  {journal} {\bibinfo  {journal} {Physical Review B}\ }\textbf
  {\bibinfo {volume} {76}},\ \href {https://doi.org/10.1103/physrevb.76.094402}
  {10.1103/physrevb.76.094402} (\bibinfo {year} {2007})\BibitemShut {NoStop}%
\bibitem [{\citenamefont {Ohno}\ \emph {et~al.}(2004)\citenamefont {Ohno},
  \citenamefont {Arakawa}, \citenamefont {Ichinose},\ and\ \citenamefont
  {Matsui}}]{ohnoPhaseStructureRandomPlaquette2004}%
  \BibitemOpen
  \bibfield  {author} {\bibinfo {author} {\bibfnamefont {T.}~\bibnamefont
  {Ohno}}, \bibinfo {author} {\bibfnamefont {G.}~\bibnamefont {Arakawa}},
  \bibinfo {author} {\bibfnamefont {I.}~\bibnamefont {Ichinose}},\ and\
  \bibinfo {author} {\bibfnamefont {T.}~\bibnamefont {Matsui}},\ }\href
  {https://doi.org/10.1016/j.nuclphysb.2004.07.003} {\bibfield  {journal}
  {\bibinfo  {journal} {Nuclear Physics B}\ }\textbf {\bibinfo {volume}
  {697}},\ \bibinfo {pages} {462} (\bibinfo {year} {2004})},\ \Eprint
  {https://arxiv.org/abs/quant-ph/0401101} {arxiv:quant-ph/0401101}
  \BibitemShut {NoStop}%
\bibitem [{\citenamefont {Ashkin}\ and\ \citenamefont
  {Teller}(1943)}]{ashkin-teller}%
  \BibitemOpen
  \bibfield  {author} {\bibinfo {author} {\bibfnamefont {J.}~\bibnamefont
  {Ashkin}}\ and\ \bibinfo {author} {\bibfnamefont {E.}~\bibnamefont
  {Teller}},\ }\href {https://doi.org/10.1103/PhysRev.64.178} {\bibfield
  {journal} {\bibinfo  {journal} {Phys. Rev.}\ }\textbf {\bibinfo {volume}
  {64}},\ \bibinfo {pages} {178} (\bibinfo {year} {1943})}\BibitemShut
  {NoStop}%
\bibitem [{\citenamefont {Fan}(1972)}]{FAN1972136}%
  \BibitemOpen
  \bibfield  {author} {\bibinfo {author} {\bibfnamefont {C.}~\bibnamefont
  {Fan}},\ }\href {https://doi.org/10.1016/0375-9601(72)91051-1} {\bibfield
  {journal} {\bibinfo  {journal} {Physics Letters A}\ }\textbf {\bibinfo
  {volume} {39}},\ \bibinfo {pages} {136} (\bibinfo {year} {1972})}\BibitemShut
  {NoStop}%
\bibitem [{\citenamefont {Lu}\ \emph {et~al.}(2020)\citenamefont {Lu},
  \citenamefont {Hsieh},\ and\ \citenamefont {Grover}}]{Lu_2020}%
  \BibitemOpen
  \bibfield  {author} {\bibinfo {author} {\bibfnamefont {T.-C.}\ \bibnamefont
  {Lu}}, \bibinfo {author} {\bibfnamefont {T.~H.}\ \bibnamefont {Hsieh}},\ and\
  \bibinfo {author} {\bibfnamefont {T.}~\bibnamefont {Grover}},\ }\bibfield
  {journal} {\bibinfo  {journal} {Physical Review Letters}\ }\textbf {\bibinfo
  {volume} {125}},\ \href {https://doi.org/10.1103/physrevlett.125.116801}
  {10.1103/physrevlett.125.116801} (\bibinfo {year} {2020})\BibitemShut
  {NoStop}%
\bibitem [{\citenamefont {Savvidy}\ and\ \citenamefont
  {Wegner}(1994)}]{Savvidy_1994}%
  \BibitemOpen
  \bibfield  {author} {\bibinfo {author} {\bibfnamefont {G.}~\bibnamefont
  {Savvidy}}\ and\ \bibinfo {author} {\bibfnamefont {F.}~\bibnamefont
  {Wegner}},\ }\href {https://doi.org/10.1016/0550-3213(94)90003-5} {\bibfield
  {journal} {\bibinfo  {journal} {Nuclear Physics B}\ }\textbf {\bibinfo
  {volume} {413}},\ \bibinfo {pages} {605–613} (\bibinfo {year}
  {1994})}\BibitemShut {NoStop}%
\bibitem [{\citenamefont {Cappi}\ \emph {et~al.}(1992)\citenamefont {Cappi},
  \citenamefont {Colangelo}, \citenamefont {Gonnella},\ and\ \citenamefont
  {Maritan}}]{CAPPI1992659}%
  \BibitemOpen
  \bibfield  {author} {\bibinfo {author} {\bibfnamefont {A.}~\bibnamefont
  {Cappi}}, \bibinfo {author} {\bibfnamefont {P.}~\bibnamefont {Colangelo}},
  \bibinfo {author} {\bibfnamefont {G.}~\bibnamefont {Gonnella}},\ and\
  \bibinfo {author} {\bibfnamefont {A.}~\bibnamefont {Maritan}},\ }\href
  {https://doi.org/10.1016/0550-3213(92)90427-D} {\bibfield  {journal}
  {\bibinfo  {journal} {Nuclear Physics B}\ }\textbf {\bibinfo {volume}
  {370}},\ \bibinfo {pages} {659} (\bibinfo {year} {1992})}\BibitemShut
  {NoStop}%
\bibitem [{\citenamefont {Johnston}\ \emph {et~al.}(2017)\citenamefont
  {Johnston}, \citenamefont {Mueller},\ and\ \citenamefont
  {Janke}}]{johnstonPlaquetteIsingModels2017}%
  \BibitemOpen
  \bibfield  {author} {\bibinfo {author} {\bibfnamefont {D.~A.}\ \bibnamefont
  {Johnston}}, \bibinfo {author} {\bibfnamefont {M.}~\bibnamefont {Mueller}},\
  and\ \bibinfo {author} {\bibfnamefont {W.}~\bibnamefont {Janke}},\ }\href
  {https://doi.org/10.1140/epjst/e2016-60329-4} {\bibfield  {journal} {\bibinfo
   {journal} {The European Physical Journal Special Topics}\ }\textbf {\bibinfo
  {volume} {226}},\ \bibinfo {pages} {749} (\bibinfo {year} {2017})},\ \Eprint
  {https://arxiv.org/abs/1612.00060} {arxiv:1612.00060 [cond-mat]} \BibitemShut
  {NoStop}%
\bibitem [{\citenamefont {Mueller}\ \emph {et~al.}(2017)\citenamefont
  {Mueller}, \citenamefont {Johnston},\ and\ \citenamefont
  {Janke}}]{muellerExactSolutionsPlaquette2017}%
  \BibitemOpen
  \bibfield  {author} {\bibinfo {author} {\bibfnamefont {M.}~\bibnamefont
  {Mueller}}, \bibinfo {author} {\bibfnamefont {D.~A.}\ \bibnamefont
  {Johnston}},\ and\ \bibinfo {author} {\bibfnamefont {W.}~\bibnamefont
  {Janke}},\ }\href {https://doi.org/10.1016/j.nuclphysb.2016.11.005}
  {\bibfield  {journal} {\bibinfo  {journal} {Nuclear Physics B}\ }\textbf
  {\bibinfo {volume} {914}},\ \bibinfo {pages} {388} (\bibinfo {year}
  {2017})}\BibitemShut {NoStop}%
\bibitem [{\citenamefont {Johnston}\ and\ \citenamefont
  {Ranasinghe}(2020)}]{Johnston_2020}%
  \BibitemOpen
  \bibfield  {author} {\bibinfo {author} {\bibfnamefont {D.~A.}\ \bibnamefont
  {Johnston}}\ and\ \bibinfo {author} {\bibfnamefont {R.~P. K. C.~M.}\
  \bibnamefont {Ranasinghe}},\ }\href {https://doi.org/10.3390/e22060633}
  {\bibfield  {journal} {\bibinfo  {journal} {Entropy}\ }\textbf {\bibinfo
  {volume} {22}},\ \bibinfo {pages} {633} (\bibinfo {year} {2020})}\BibitemShut
  {NoStop}%
\bibitem [{\citenamefont {Song}\ \emph {et~al.}(2022)\citenamefont {Song},
  \citenamefont {{Sch{\"o}nmeier-Kromer}}, \citenamefont {Liu}, \citenamefont
  {Viyuela}, \citenamefont {Pollet},\ and\ \citenamefont
  {{Martin-Delgado}}}]{songOptimalThresholdsFracton2022}%
  \BibitemOpen
  \bibfield  {author} {\bibinfo {author} {\bibfnamefont {H.}~\bibnamefont
  {Song}}, \bibinfo {author} {\bibfnamefont {J.}~\bibnamefont
  {{Sch{\"o}nmeier-Kromer}}}, \bibinfo {author} {\bibfnamefont
  {K.}~\bibnamefont {Liu}}, \bibinfo {author} {\bibfnamefont {O.}~\bibnamefont
  {Viyuela}}, \bibinfo {author} {\bibfnamefont {L.}~\bibnamefont {Pollet}},\
  and\ \bibinfo {author} {\bibfnamefont {M.~A.}\ \bibnamefont
  {{Martin-Delgado}}},\ }\href {https://doi.org/10.1103/PhysRevLett.129.230502}
  {\bibfield  {journal} {\bibinfo  {journal} {Physical Review Letters}\
  }\textbf {\bibinfo {volume} {129}},\ \bibinfo {pages} {230502} (\bibinfo
  {year} {2022})}\BibitemShut {NoStop}%
\bibitem [{\citenamefont {Mueller}\ \emph {et~al.}(2014)\citenamefont
  {Mueller}, \citenamefont {Janke},\ and\ \citenamefont
  {Johnston}}]{mueller2014}%
  \BibitemOpen
  \bibfield  {author} {\bibinfo {author} {\bibfnamefont {M.}~\bibnamefont
  {Mueller}}, \bibinfo {author} {\bibfnamefont {W.}~\bibnamefont {Janke}},\
  and\ \bibinfo {author} {\bibfnamefont {D.~A.}\ \bibnamefont {Johnston}},\
  }\href {https://doi.org/10.1103/PhysRevLett.112.200601} {\bibfield  {journal}
  {\bibinfo  {journal} {Phys. Rev. Lett.}\ }\textbf {\bibinfo {volume} {112}},\
  \bibinfo {pages} {200601} (\bibinfo {year} {2014})}\BibitemShut {NoStop}%
\bibitem [{\citenamefont {Wang}\ \emph {et~al.}(2023)\citenamefont {Wang},
  \citenamefont {Wu},\ and\ \citenamefont {Wang}}]{wang2023intrinsic}%
  \BibitemOpen
  \bibfield  {author} {\bibinfo {author} {\bibfnamefont {Z.}~\bibnamefont
  {Wang}}, \bibinfo {author} {\bibfnamefont {Z.}~\bibnamefont {Wu}},\ and\
  \bibinfo {author} {\bibfnamefont {Z.}~\bibnamefont {Wang}},\ }\href@noop {}
  {\bibinfo {title} {Intrinsic mixed-state topological order without quantum
  memory}} (\bibinfo {year} {2023}),\ \Eprint
  {https://arxiv.org/abs/2307.13758} {arXiv:2307.13758 [quant-ph]} \BibitemShut
  {NoStop}%
\bibitem [{\citenamefont {Savvidy}\ and\ \citenamefont
  {Savvidy}(1994)}]{SAVVIDY199472}%
  \BibitemOpen
  \bibfield  {author} {\bibinfo {author} {\bibfnamefont {G.}~\bibnamefont
  {Savvidy}}\ and\ \bibinfo {author} {\bibfnamefont {K.}~\bibnamefont
  {Savvidy}},\ }\href {https://doi.org/10.1016/0370-2693(94)00114-6} {\bibfield
   {journal} {\bibinfo  {journal} {Physics Letters B}\ }\textbf {\bibinfo
  {volume} {324}},\ \bibinfo {pages} {72} (\bibinfo {year} {1994})}\BibitemShut
  {NoStop}%
\bibitem [{\citenamefont {Suzuki}(1972)}]{suzuki1972}%
  \BibitemOpen
  \bibfield  {author} {\bibinfo {author} {\bibfnamefont {M.}~\bibnamefont
  {Suzuki}},\ }\href {https://doi.org/10.1103/PhysRevLett.28.507} {\bibfield
  {journal} {\bibinfo  {journal} {Phys. Rev. Lett.}\ }\textbf {\bibinfo
  {volume} {28}},\ \bibinfo {pages} {507} (\bibinfo {year} {1972})}\BibitemShut
  {NoStop}%
\bibitem [{\citenamefont {Johnston}(2012)}]{Johnston_2012}%
  \BibitemOpen
  \bibfield  {author} {\bibinfo {author} {\bibfnamefont {D.~A.}\ \bibnamefont
  {Johnston}},\ }\href {https://doi.org/10.1088/1751-8113/45/40/405001}
  {\bibfield  {journal} {\bibinfo  {journal} {Journal of Physics A:
  Mathematical and Theoretical}\ }\textbf {\bibinfo {volume} {45}},\ \bibinfo
  {pages} {405001} (\bibinfo {year} {2012})}\BibitemShut {NoStop}%
\bibitem [{\citenamefont {Lyons}\ and\ \citenamefont {Luo}(2024)}]{upcoming}%
  \BibitemOpen
  \bibfield  {author} {\bibinfo {author} {\bibfnamefont {A.}~\bibnamefont
  {Lyons}}\ and\ \bibinfo {author} {\bibfnamefont {Z.-X.}\ \bibnamefont {Luo}}}
  (\bibinfo {year} {2024}),\ \bibinfo {note} {in preparation}\BibitemShut
  {NoStop}%
\bibitem [{\citenamefont {Bravyi}\ \emph {et~al.}(2011)\citenamefont {Bravyi},
  \citenamefont {Leemhuis},\ and\ \citenamefont
  {Terhal}}]{bravyiTopologicalOrderExactly2011}%
  \BibitemOpen
  \bibfield  {author} {\bibinfo {author} {\bibfnamefont {S.}~\bibnamefont
  {Bravyi}}, \bibinfo {author} {\bibfnamefont {B.}~\bibnamefont {Leemhuis}},\
  and\ \bibinfo {author} {\bibfnamefont {B.~M.}\ \bibnamefont {Terhal}},\
  }\href {https://doi.org/10.1016/j.aop.2010.11.002} {\bibfield  {journal}
  {\bibinfo  {journal} {Annals of Physics}\ }\textbf {\bibinfo {volume}
  {326}},\ \bibinfo {pages} {839} (\bibinfo {year} {2011})},\ \Eprint
  {https://arxiv.org/abs/1006.4871} {arxiv:1006.4871 [quant-ph]} \BibitemShut
  {NoStop}%
\bibitem [{\citenamefont {Tillich}\ and\ \citenamefont
  {Zemor}(2014)}]{tillichQuantumLDPCCodes2014}%
  \BibitemOpen
  \bibfield  {author} {\bibinfo {author} {\bibfnamefont {J.-P.}\ \bibnamefont
  {Tillich}}\ and\ \bibinfo {author} {\bibfnamefont {G.}~\bibnamefont
  {Zemor}},\ }\href {https://doi.org/10.1109/TIT.2013.2292061} {\bibfield
  {journal} {\bibinfo  {journal} {IEEE Transactions on Information Theory}\
  }\textbf {\bibinfo {volume} {60}},\ \bibinfo {pages} {1193} (\bibinfo {year}
  {2014})},\ \Eprint {https://arxiv.org/abs/0903.0566} {arxiv:0903.0566
  [quant-ph]} \BibitemShut {NoStop}%
\bibitem [{\citenamefont {Lee}\ \emph {et~al.}(2024)\citenamefont {Lee},
  \citenamefont {You},\ and\ \citenamefont {Xu}}]{lee2024symmetry}%
  \BibitemOpen
  \bibfield  {author} {\bibinfo {author} {\bibfnamefont {J.~Y.}\ \bibnamefont
  {Lee}}, \bibinfo {author} {\bibfnamefont {Y.-Z.}\ \bibnamefont {You}},\ and\
  \bibinfo {author} {\bibfnamefont {C.}~\bibnamefont {Xu}},\ }\href@noop {}
  {\bibinfo {title} {Symmetry protected topological phases under decoherence}}
  (\bibinfo {year} {2024}),\ \Eprint {https://arxiv.org/abs/2210.16323}
  {arXiv:2210.16323 [cond-mat.str-el]} \BibitemShut {NoStop}%
\bibitem [{\citenamefont {Ma}\ and\ \citenamefont {Wang}(2023)}]{Ma_2023}%
  \BibitemOpen
  \bibfield  {author} {\bibinfo {author} {\bibfnamefont {R.}~\bibnamefont
  {Ma}}\ and\ \bibinfo {author} {\bibfnamefont {C.}~\bibnamefont {Wang}},\
  }\bibfield  {journal} {\bibinfo  {journal} {Physical Review X}\ }\textbf
  {\bibinfo {volume} {13}},\ \href {https://doi.org/10.1103/physrevx.13.031016}
  {10.1103/physrevx.13.031016} (\bibinfo {year} {2023})\BibitemShut {NoStop}%
\bibitem [{\citenamefont {Ma}\ \emph {et~al.}(2023)\citenamefont {Ma},
  \citenamefont {Zhang}, \citenamefont {Bi}, \citenamefont {Cheng},\ and\
  \citenamefont {Wang}}]{ma2023topological}%
  \BibitemOpen
  \bibfield  {author} {\bibinfo {author} {\bibfnamefont {R.}~\bibnamefont
  {Ma}}, \bibinfo {author} {\bibfnamefont {J.-H.}\ \bibnamefont {Zhang}},
  \bibinfo {author} {\bibfnamefont {Z.}~\bibnamefont {Bi}}, \bibinfo {author}
  {\bibfnamefont {M.}~\bibnamefont {Cheng}},\ and\ \bibinfo {author}
  {\bibfnamefont {C.}~\bibnamefont {Wang}},\ }\href@noop {} {\bibinfo {title}
  {Topological phases with average symmetries: the decohered, the disordered,
  and the intrinsic}} (\bibinfo {year} {2023}),\ \Eprint
  {https://arxiv.org/abs/2305.16399} {arXiv:2305.16399 [cond-mat.str-el]}
  \BibitemShut {NoStop}%
\bibitem [{\citenamefont {Su}\ \emph {et~al.}(2024)\citenamefont {Su},
  \citenamefont {Yang},\ and\ \citenamefont {Jian}}]{su2024tapestry}%
  \BibitemOpen
  \bibfield  {author} {\bibinfo {author} {\bibfnamefont {K.}~\bibnamefont
  {Su}}, \bibinfo {author} {\bibfnamefont {Z.}~\bibnamefont {Yang}},\ and\
  \bibinfo {author} {\bibfnamefont {C.-M.}\ \bibnamefont {Jian}},\ }\href@noop
  {} {\bibinfo {title} {Tapestry of dualities in decohered quantum error
  correction codes}} (\bibinfo {year} {2024}),\ \Eprint
  {https://arxiv.org/abs/2401.17359} {arXiv:2401.17359 [cond-mat.str-el]}
  \BibitemShut {NoStop}%
\end{thebibliography}%

\end{document}